\documentclass[12pt]{article}

\usepackage{amssymb}
\usepackage{epsf}
\usepackage{epsfig}
\advance \topmargin by -2cm
\advance \oddsidemargin by -1cm
\advance \evensidemargin by -3cm

\begin{document}
\def\be{\begin{equation}}
\def\ee{\end{equation}}
\def\bea{\begin{eqnarray}}
\def\eea{\end{eqnarray}}

%%%\debugon

\title{\bf Statistics of energy levels and eigenfunctions in disordered
and chaotic systems: Supersymmetry approach}
\author{Alexander D. Mirlin \thanks{Also at Petersburg Nuclear Physics
Institute, 188350 Gatchina, St. Petersburg, Russia} \\
Institut f\"ur Theorie der kondensierten Materie,\\
Universit\"at Karlsruhe, 76128 Karlsruhe, Germany}
\date{October 5, 1999}
\maketitle

\bigskip

%{\it Running head:} 
%Supersymmetry approach to level and eigenfunction statistics.

\tableofcontents

\section{Introduction}
\label{s1}

The supersymmetry method pioneered by Efetov  \cite{efetov83}
has proven to be a very powerful tool of study of the statistical
properties of energy levels and eigenfunctions in disordered systems. 
The aim of these lectures is to present a tutorial introduction
to the method, as well as an overview of the recent developments. 
One of the major achievements of Efetov \cite{efetov83} was a proof of
applicability of the random matrix theory (RMT) to a disordered
metallic sample. More recently, the focus of the research interest 
shifted to the study of system-specific deviations from the universal
(RMT) behavior. This will be the central topic of the present
lectures. Since these deviations are determined by the underlying
classical dynamics, this issue is closely related to the subject of 
quantum chaos. 

We begin by introducing the
basic notions and ideas of the supermatrix $\sigma$-model method
in Sec.~\ref{s2},
considering the level statistics in the Gaussian unitary
ensemble (GUE) as the simplest example. In Sec.~\ref{s3} we turn to
the problem of level correlations in a diffusive sample. We
outline a derivation of the diffusive $\sigma$-model and discuss
deviations of the level statistics from RMT. Section \ref{s4}
is devoted to the eigenfunction statistics. Finally, in Sec.~\ref{s5} 
we discuss very recent ideas of application of the method to chaotic
ballistic systems.

It is appropriate to give  here  references to a few recent review
articles and books which are close in their topics to the present
lectures. These are the Efetov's book \cite{ef-book}, the reviews
by Guhr, M\"uller-Groeling, and Weidenm\"uller \cite{gmw-rev} and by
the author \cite{m-rev}, and the proceedings volume of the NATO
Advanced Study Institute \cite{cambridge} (in particular the review by 
Altland, Offer, and Simons \cite{aos-rev} there).
Other references to the relevant literature will be given in 
appropriate places below.  

\section{Introduction to the supersymmetry method and application to
RMT}
\label{s2}

\subsection{Green's function approach}
\label{s2.1}

We consider the Gaussian unitary ensemble (GUE) defined
\cite{mehta,bohigas} as an ensemble
of $N\times N$ ($N\to\infty$) Hermitian matrices
$\hat{H}=\hat{H}^\dagger$ with probability density
\be
\label{e1.1}
{\cal P}(\hat{H})={\cal N}\exp\left\{-{N\over 2} \mbox{Tr}\,
\hat{H}^2\right\}\ ,
\ee
where ${\cal N}$ is a normalization factor. 
According to (\ref{e1.1}), all the matrix elements of $\hat{H}$ have a
Gaussian distribution with variance
\be
\label{e1.2}
\langle H_{ij}H_{i'j'}^*\rangle={1\over N}\delta_{ii'}\delta_{jj'}\ .
\ee
The factor $N$ in the exponent of (\ref{e1.1}) allows one to keep the
eigenvalues $E_i$ of $\hat{H}$ finite in the limit $N\to\infty$, with
$\langle E_i^2\rangle=1$. We demonstrate below how the well-known
results \cite{mehta,bohigas} for the average density of states and the
two-level correlation function are derived within the supermatrix
$\sigma$-model method.

The density of states (DOS) is defined as
\be
\label{e1.3}
\nu(E)={1\over N} \,\mbox{Tr}\,\delta(E-\hat{H}).
\ee
The first step of the strategy is to express the quantity of interest
in terms of the retarded and/or advanced Green's functions 
\be
\label{e1.4}
\hat{G}_{\rm R,A}=(E-\hat{H}\pm i\eta)^{-1}\ .
\ee
We have
\be
\label{e1.5}
\nu(E)=-{1\over\pi N}\,{\rm Im}\, {\rm Tr}\,(E-\hat{H}+
i\eta)^{-1}|_{\eta\to +0}\ .
\ee
The second step is to write the Green's function (in a general case, a
product of Green's functions) as a functional integral. More
precisely, in the case of RMT this will be an integral over a vector
field with the number of components proportional to $N$; for the problem
of a particle in a random potential (Sec.~\ref{s3}) the discrete index
will be replaced by a continuously changing spatial coordinate, which
will result in a functional integral. The simplest way to do it is to
introduce an $N$-component complex vector $\phi_i$, $i=1,2,\ldots,N$,
so that
\be
\label{e1.6}
(E-\hat{H}+i\eta)^{-1}_{kl}=-i
{\int[{\rm d}\phi^*{\rm d}\phi]\phi_k\phi_l^*
\exp\{i\sum_{ij}\phi_i^*[(E+i\eta)\delta_{ij}-H_{ij}]\phi_j\} 
\over
\int[{\rm d}\phi^*{\rm d}\phi]
\exp\{i\sum_{ij}\phi_i^*[(E+i\eta)\delta_{ij}-H_{ij}]\phi_j\}}\ ,
\ee
where $[{\rm d}\phi^*{\rm d}\phi]=\prod_{i=1}^N {\rm d}\phi_i^*{\rm
d}\phi_i$. We have  to put 
the imaginary unit
$i$ in front of the quadratic form in the exponent of Eq.~(\ref{e1.6})
to get a convergent integral; the convergence being guaranteed by
$\eta>0$. 

The next step is to average over the ensemble of
random matrices $\hat{H}$. However, direct averaging of
Eq.~(\ref{e1.6}) is complicated by the fact that $\hat{H}$ enters not
only the numerator but also the denominator. If there were no
denominator, the averaging over $\hat{H}$ with the probability density
(\ref{e1.1}) would be straightforward (a Gaussian integral over
$\hat{H}$). One possible way to get rid of the denominator is the
replica trick first introduced by Edwards and Anderson
\cite{edwards75} in the context of the spin glass theory. 
The idea of the method is to introduce $n$ species of the field,
$\phi_i^{(\alpha)}$, $\alpha=1,2,\ldots,n$. Then the denominator $Z$ of 
Eq.~(\ref{e1.6}) is transformed to $Z^n$ and disappears in the limit
$n\to 0$. However, the replica trick turns out to be ill-founded
\cite{verbzirn} for the
$\sigma$-model approach to the problem of the level and 
eigenfunction statistics reviewed in these lectures.\footnote{Note
that if the $\sigma$-model is treated perturbatively  
(including the renormalization group treatment which is a resummation
of the perturbative expansion),
then the replica trick is completely equivalent to the supersymmetry
method.}.  An alternative method, which uses a combination of 
commuting and anticommuting \cite{berezin66}
variables instead of the $n\to 0$ replica limit, was
proposed by McKane \cite{mckane80} and by Parisi and Sourlas
\cite{parsour81}. The effective theories which are obtained in this
way are invariant with respect to a transformation mixing commuting
(``bosonic'') and anticommuting (``fermionic'') degrees of freedom,
which is conventionally referred to as
supersymmetry.\footnote{Mathematicians also use the term ``graded''
(or, more specifically, $\mathbb{Z}_2$-graded) to characterize the
arising algebraic structures.} 

A number of publications containing an introduction to the
supersymmetry approach are available; we mention, in addition to the
Efetov's review \cite{efetov83}  and his more recent book
\cite{ef-book}, the papers \cite{vwz,fyodorov95,zuk94}. In particular,
in Refs.~\cite{fyodorov95,zuk94} a detailed exposition of the
calculation of the GUE level statistics with the supersymmetry is
given. In our presentation we will concentrate on
``ideological'' aspects of the method and skip some technical details
of calculations, which can be found in the literature cited above.

\subsection{Supermathematics}
\label{s2.2}

We describe now briefly the basic properties of anticommuting
(Grassmannian) 
variables and introduce the notions of the supermathematics which we
will use. For a detailed exposition of the Grassmannian
mathematics and of the superanalysis the reader is referred to the
book \cite{berezin87} (see also \cite{ef-book} for a physicist's
summary of the most important properties). We introduce the
Grassmannian variables 
$\chi_k$, $\chi_k^*$, $k=1,\ldots,N$, which all anticommute to each other:
\be
\label{e1.7}
\chi_k \chi_l = -\chi_l\chi_k, \ \ \ 
\chi^*_k \chi_l = -\chi_l\chi^*_k, \ \ \ 
\chi^*_k \chi^*_l = -\chi^*_l\chi^*_k\ .
\ee
Note that $\chi_k$ and $\chi_k^*$ are to be considered as two
independent variables.
According to (\ref{e1.7}), the square of a Grassmannian variable is
zero. As a consequence, any function of Grassmanians, when expanded in
a power series, may contain only the terms of the zeroth and the first
order in each Grassmannian variable. For this reason, 
integration over Grassmannians is uniquely defined by the following
rules:
\be
\label{e1.8}
\int \chi_k {\rm d}\chi_k = \int \chi^*_k {\rm d}\chi^*_k 
={1\over\sqrt{2\pi}}\ ;\qquad 
\int {\rm d}\chi_k= \int {\rm d}\chi_k^*=0\ ,
\ee
the differentials ${\rm d}\chi_k$, ${\rm d}\chi_k^*$ anticommuting
with each other and with the Grassmannian variables. 
Note that the Grassmannian integration is just a formally defined
algebraic operation and the question ``What is the domain of
integration in Eqs.~(\ref{e1.8})?'' does not make sense. 

Using the rules (\ref{e1.7}), (\ref{e1.8}), one can calculate a
Gaussian integral over the Grassmannians,
\be
\label{e1.9}
\int {\rm d}\chi_1^* {\rm d}\chi_1\ldots {\rm d}\chi_N^* {\rm d}\chi_N 
\exp\{-\sum_{kl}\chi_k^*K_{kl}\chi_l\} 
= {\rm det}\left({K\over 2\pi}\right)\ .
\ee
An analogous integral over commuting variables would give 
${\rm det}^{-1}(K/ 2\pi)$. This property of the anticommuting
variables is  the reason for introducing them: it will allow us to replace
the Gaussian integral over the commuting variables in the denominator
of Eq.~(\ref{e1.6}) by an analogous Grassmannian integral in the
numerator, thus solving the denominator problem!

It is convenient  to define the ``complex conjugation'' for the
Grassmannians by the following rules\footnote{The notion of complex
conjugation of Grassmannian variables is introduced for notational
convenience only; it allows one to make the treatment of Grassmannians
similar to that of ordinary (commuting) variables and thus to
introduce compact supersymmetric notations. The fact that two
Grassmannian variables have been declared complex conjugate to each
other is, however, irrelevant for evaluation of integrals over
them, which are simply defined by the rules (\ref{e1.8}). }
\be
\label{e1.10}
(\chi_k)^*=\chi_k^*; \ \ (\chi_k^*)^*=-\chi_k;\ \
(\chi_k\chi_l)^*=\chi_k^*\chi_l^*\ .
\ee
Furthermore, we introduce the notion of a supervector
\be
\label{e1.11}
\Phi=\left\lgroup\begin{array}{c} S_1 \\ 
                            \vdots\\
                             S_n\\
                             \chi_1\\
                             \vdots\\
                              \chi_n
\end{array}\right\rgroup\ ;\qquad
\Phi^\dagger = (S_1^*,\ldots,S_n^*,\chi_1^*,\ldots,\chi_n^*),
\ee
where $S_i$ are the commuting and $\chi_i$ the anticommuting
components, and $n$ is an arbitrary positive integer. 
A supermatrix has the structure 
\be
\label{e1.12}
F=\left\lgroup\begin{array}{cc} a & \sigma \\ \rho & b \end{array}
\right\rgroup \ ,
\ee
where the boson-boson (bb) block $a$ and the fermion-fermion (ff)
block $b$ are 
ordinary $n\times n$ matrices, while the boson-fermion (bf) and
fermion-boson (fb) blocks $\sigma$, $\rho$ are $n\times n$ matrices with
anticommuting entries. For example, for any two supervectors $\Phi$,
$\Psi$, the tensor product $\Phi\otimes\Psi^\dagger$ is a supermatrix.
The supertrace and the superdeterminant of the supermatrix
(\ref{e1.12}) are defined as follows
\bea
&& {\rm Str}\, F = {\rm Tr}\, a - {\rm Tr}\, b\ ,  \label{e1.13}\\
&& {\rm Sdet}\, F = \exp {\rm Str} \ln F = {\rm
det}(a-\sigma b^{-1}\rho){\rm det}^{-1}b.
\label{e1.14}
\eea
Finally, hermitian conjugation of a supermatrix is defined by
\be
\label{e1.15}
F^\dagger =\left\lgroup\begin{array}{cc} a^\dagger & \rho^\dagger 
\\ -\sigma^\dagger & b^\dagger \end{array} \right\rgroup \ ,
\ee
where for the anticommuting (bf and fb) blocks the usual definition
holds, $\sigma^\dagger = (\sigma^*)^T$. 

It can be shown that with the above set of definitions, the usual
properties of the vector and matrix algebra are valid for
supervectors and supermatrices as well, in particular:
\begin{itemize}
\item if $F$ is a supermatrix and $\Phi$ a supervector, then
$\Psi=F\Phi$ is a supervector;
\item hermitian conjugation satisfies the usual properties
\bea
&& F=\Phi\otimes\Psi^\dagger \qquad \Longrightarrow \qquad
F^\dagger=\Psi\otimes\Phi^\dagger\ , \label{e1.16}\\
&& (\Phi^\dagger F\Psi)^\dagger=\Psi^\dagger F^\dagger\Phi\ ,
  \label{e1.17}\\
&&(F^\dagger)^\dagger=F\ ,\ \ {\rm
etc.} \label{e1.18}
\eea
\item supertrace and superdeterminant of a product:
\bea
&& {\rm Str} F_1F_2 = {\rm Str} F_2F_1\ , \label{e1.19}\\
&& {\rm Sdet} F_1F_2 = {\rm Sdet} F_1 \cdot {\rm Sdet} F_2\ .
\label{e1.20}
\eea
\item Gaussian integrals:
\bea
&& \int {\rm d}\Phi^\dagger {\rm d}\Phi \exp (-\Phi^\dagger K\Phi) =
{\rm Sdet} K^{-1} \label{e1.21}\\
&& \int {\rm d}\Phi^\dagger {\rm d}\Phi\, \Phi_\alpha\Phi_\beta^\dagger
\exp (-\Phi^\dagger K\Phi) = (K^{-1})_{\alpha\beta}\:{\rm Sdet}
K^{-1}\ . \label{e1.22}
\eea
It is assumed here that the boson-boson block $K_{\rm bb}$ of the
matrix $K$ defines a quadratic form with a positively defined real
part, ${\rm Re}\,   
{\bf S}^\dagger K_{\rm bb} {\bf S} > 0$, so that the integral over the
bosonic components ${\bf S}$ of the supervector $\Phi$ converges.
Integrals of the type (\ref{e1.22}) with a product of a larger number
of $\Phi_\alpha$'s in the preexponential factor can be evaluated via
the Wick theorem (taking into account that interchanging two
anticommuting variables produces a minus sign).
\end{itemize}

The notion of a supermanifold (including integration and change of
coordinates on it) will become important for us in the course of
calculation of the DOS-DOS correlation function (Sec.~\ref{s2.4}). We
restrict ourselves to saying that the corresponding
definitions are natural
extensions of those for ordinary analytic manifolds and refer the
reader to Berezin's book \cite{berezin87} (see also
\cite{zirnbauer99} for a summary). A discussion of
the structure of supermanifolds relevant to the supersymmetry
treatment of the random matrix ensembles and a more extended list of
the related mathematical literature can be found in \cite{zirn96}.

Now we return to the problem of the RMT level statistics.

\subsection{Average DOS from supersymmetry}
\label{s2.3}

We have, instead of Eq.~(\ref{e1.6}),
\be
\label{e1.23}
(E-\hat{H})^{-1}_{kl}=-i
\int[{\rm d}\Phi^*{\rm d}\Phi]S_k S_l^*
\exp\{i\sum_{ij}\Phi_i^\dagger[E\delta_{ij}-H_{ij}]\Phi_j\} \ ,
\ee
where 
$$
\Phi_i=\left\lgroup\begin{array}{cc} S_i \\ \chi_i
\end{array}\right\rgroup 
$$
is a two-component supervector, and we have included an infinitesimally
small imaginary part $+i\eta$ in the definition of $E$, so that ${\rm
Im}E>0$. The averaging over $\hat{H}$ is now straightforward,
\be
\label{e1.24}
\left\langle \exp(i\sum_{ij}\Phi_i^\dagger H_{ij}\Phi_j)\right\rangle
= \exp\left\{-{1\over
2N}\sum_{ij}(\Phi_i^\dagger\Phi_j)(\Phi_j^\dagger\Phi_i)\right\}\ .
\ee
The next step of our strategy is to decouple the $\Phi^4$ term in
Eq.~(\ref{e1.24}) by introducing an integration over a $2\times 2$
supermatrix $R$,
\be
\label{e1.25}
\exp\left\{-{1\over
2N}\sum_{ij}(\Phi_i^\dagger\Phi_j)(\Phi_j^\dagger\Phi_i)\right\}
=\int {\rm d}R \exp\left\{-{N\over 2}\,
{\rm Str}\,R^2-i\sum_i\Phi_i^\dagger R\Phi_i\right\}\ .
\ee
In order that the Gaussian integral (\ref{e1.25}) over the commuting
components of the matrix $R$ be convergent, the latter 
has to be chosen in the form\footnote{It does not play any role
whether $\rho$ and $\rho^*$ are considered to be complex conjugate of
each other or just two independent Grassmannian variables [see the
footnote preceding Eq.~(\ref{e1.10})].}
\be
\label{e1.26}
R=\left\lgroup \begin{array}{cc} q_{\rm b}  &  \rho^* \\
                            \rho &   iq_{\rm f}
          \end{array}
   \right\rgroup\ ; \qquad q_{\rm b}, q_{\rm f} \in \mathbb{R}\ .
\ee
In other words, integration over the ff-component of the matrix $R$
has to be performed along the imaginary axis. 
Indeed, due to the factor $i$ in the fermion-fermion element, the
quadratic form ${\rm Str}R^2 =q_{\rm b}^2+q_{\rm f}^2$ is positively
defined, ensuring the convergence. 
  
Substituting (\ref{e1.25}), (\ref{e1.24}) in (\ref{e1.23}), we get
\be
\label{e1.27}
\langle {\rm Tr} (E-\hat{H})^{-1}\rangle=-i
\int[{\rm d}\Phi^*{\rm d}\Phi]\sum_k S_k S_k^*
\exp\left\{iE\sum_i\Phi_i^\dagger\Phi_i-{N\over 2}\,
{\rm Str}\,R^2-i\sum_i\Phi_i^\dagger R\Phi_i\right\}\ .
\ee
Now we see what was the idea of the Hubbard-Stratonovich decoupling:
the integral over the $\Phi$-field, if taken first, 
is now Gaussian (and convergent due to ${\rm Im}E>0$). Furthermore,
since the matrix $R$ does not has any structure in the $N$-dimensional
space (where the matrices $\hat{H}$ act), the integral over each of
$\Phi_i$, $\Phi_i^\dagger$ ($i=1,2,\ldots,N$) produces the same factor
${\rm Sdet} (E-R)$. As a result, we find
\bea
\label{e1.28}
&&\langle {\rm Tr} (E-\hat{H})^{-1}\rangle=N\int {\rm d}R
(E-R)^{-1}_{\rm bb} \exp\{-NS[R]\}\ ,  \\
&& S[R]={1\over 2}{\rm Str}R^2+{\rm Str}\ln(E-R)\ .
\label{e1.29}
\eea
The presence of the factor $N\gg 1$ in the exponent of (\ref{e1.28})
allows us to apply the saddle-point method. The saddle point equation
for the action (\ref{e1.29}) has the form
\be
\label{e1.30}
R=(E-R)^{-1}\ .
\ee
We look first for diagonal solutions of (\ref{e1.30}). According to 
(\ref{e1.30}), the eigenvalues $r$ of the matrix $R$ can take 
two values $r={E\over 2}\pm i\sqrt{1-E^2/4}$, yielding four
diagonal solutions\footnote{it is implied in (\ref{e1.31}) that the
first term is a constant ($E/2$) times unit (super-)matrix. Likewise,
we omit the unit matrix symbol in other formulas; e.g., in the
l.h.s. (r.h.s.) of (\ref{e1.28})  $E$ implicitly includes the 
$N\times N$ (resp. $2\times 2$) unit matrix.}
\be
\label{e1.31}
R={E\over 2}-i\sqrt{1-E^2/4}\left\lgroup\begin{array}{cc} 
               s_{\rm b} &  \\  
                   & s_{\rm f}
\end{array}\right\rgroup\ ; \qquad s_{\rm b},s_{\rm f}=\pm 1\ .
\ee
Since these matrices do not belong to the original integration
manifold (\ref{e1.26}),  a shift of the integration contours over
$R_{\rm bb}=q_{\rm b}$ and $R_{\rm ff}=iq_{\rm f}$ is
necessary. However, at 
$q_{\rm b}=E$ the integrand of (\ref{e1.28}) is singular, 
\be
\label{e1.32}
\exp\{-N{\rm Str}\ln(E-R)\}\equiv {\rm Sdet}^{-N}(E-R)\propto
(E-q_{\rm b})^{-N} \to \infty\ . 
\ee
Recalling that ${\rm Im}E=+\eta >0$, we conclude that integration
contour over $q_{\rm b}$ can only be shifted to the half-plane ${\rm Im}
q_{\rm b}<0$, so that only the saddle points with $s_{\rm b}=+1$ are
relevant.  
For $q_{\rm f}$ this argumentation is not valid, since according to the
definition of the superdeterminant the  integrand has at $iq_{\rm f}=E$
a zero [$\propto (E-iq_{\rm f})^N$] rather than a
singularity. Nevertheless, 
it turns out that $s_{\rm f}=+1$ is the proper choice for the fermionic
sector as well. The reason is, however, different: the $s_{\rm f}=-1$
saddle-point, though being legitimate, gives a contribution suppressed
by $1/N$, as will be explained below. 

The leading (at $N\gg 1$) contribution is thus given by the vicinity of
the saddle point
\be
\label{e1.33}
R_0={E\over 2}-i\sqrt{1-E^2/4}\ .
\ee
To calculate the integral in the saddle-point approximation, we need
the action at the saddle-point and the quadratic form around it,
\bea
\label{e1.34}
&& S[R_0]=0\ ,\\
&& \delta^2 S[R_0] = C(\delta q_{\rm b}^2+\delta q_{\rm f}^2 +2
\rho^*\rho)\ ; 
\qquad C=2-E\left({E\over 2}-i\sqrt{1-E^2/4}\right).
\label{e1.35}
\eea
The Gaussian integration around $R_0$ yields 
\be
\label{e1.36}
\int {\rm d}q_{\rm b} {\rm d}q_{\rm f} {\rm d}\rho^*{\rm
d}\rho\exp\{-NC(\delta q_{\rm b}^2+\delta q_{\rm f}^2 +2 
\rho^*\rho)\} ={\rm Sdet} NC =1\ .
\ee
[Superdeterminant of the unit matrix multiplied by a number ($NC$)
is unity, since the contributions of bosons ($\pi/NC$) and fermions
($NC/\pi$) cancel each other.] 
Finally, the preexponential factor in (\ref{e1.28}) can be evaluated
at the saddle point,
\be
\label{e1.37}
(E-R_0)^{-1}_{\rm bb}=(R_0)_{\rm bb}={E\over 2}-i\sqrt{1-{E^2\over
4}}\ .
\ee
According to (\ref{e1.5}), we find thus the density of states
\be
\label{e1.38}
\nu(E)= \left\{\begin{array}{ll}
{1\over \pi}\sqrt{1-E^2/4}\ , & \ \ |E|\le 2\\
0\ ,                          & \ \ |E|\ge 2\ ,
\end{array}\right.
\ee
which is Wigner's semicircle law.

Now we return (as was promised in the paragraph below
Eq.~(\ref{e1.32})) to another possible choice of sign of $s_{\rm f}$ in
(\ref{e1.31}), i.e. $s_{\rm b}=-s_{\rm f}=1$. The corresponding
diagonal saddle point, 
$$
R_1= {E\over 2}-i\sqrt{1-{E^2\over 4}}\left(\begin{array}{cc}
1 &  \\  & -1 \end{array}\right)\ ,
$$
generates in fact, by means of rotations
with a Grassmannian generator, a whole manifold of saddle points, 
$$
R=UR_1U^{-1}\ ; \ \ U=\exp \left(\begin{array}{cc}
 & \alpha^*  \\ \alpha &  \end{array}\right)
=\left(\begin{array}{cc}
 1+{\alpha^*\alpha\over 2} & \alpha^*  \\ \alpha &  
1-{\alpha^*\alpha\over 2} \end{array}\right)\ .
$$
The quadratic form of the action around $R_1$ does not contain
Grassmannians, 
$$
\delta^2S[R_1]=C(\delta q_{\rm b}^2+\delta q_{\rm f}^2)\ ,
$$
since the action is the same on the whole manifold.
Therefore, the above compensation of the bosonic gaussian integral
(yielding a factor $\propto 1/N$) by the fermionic one (producing a
factor $\propto N$) does not take place, and the result is suppressed
by $1/N$ and can be neglected. We will see in Sec.~\ref{s2.4} that in
the problem of the level correlation function, a manifold of saddle
points emerges as well; in contrast to the present case,
however, the balance of fermionic and bosonic degrees of freedom will
be preserved, so that the result will be of the leading order in $1/N$.

\subsection{Level correlations.}
\label{s2.4}

The two-level correlation function characterizing the probability
density to have two levels separated by an energy interval $\omega$
is defined as
\be
\label{e1.39}
R_2(\omega)={\langle\nu(E-\omega/2)\nu(E+\omega/2)\rangle
\over \langle\nu(E)\rangle^2}.
\ee
As was found by Dyson and Mehta (see \cite{mehta,bohigas}), 
for the GUE the two-level correlation function has the form
\be
\label{e1.40}
R_2(s)=\delta(s)+1-{\sin^2(\pi s)\over (\pi s)^2}\ ,
\ee
where $s=\omega/\Delta$, and $\Delta$ is the mean level spacing,
$\Delta= 1/N\langle\nu(E)\rangle$. 
Note that we assume that $\omega$ is much
smaller than the energy band width; in particular, we neglect the
change of the mean level density on the scale of $\omega$. This is
justified by the fact that, according to (\ref{e1.40}), 
the scale for correlations is set by the mean level spacing,
$\Delta\propto 1/N$. The delta function  in the r.h.s of
Eq.~(\ref{e1.40}) corresponds to a ``self-correlation'' of an energy
level, the second term (unity) is a disconnected part of $R_2$
(corresponding to the absence of correlations), and the last term is
the non-trivial part of $R_2$ describing the correlations of different
levels. We show below how Eq.~(\ref{e1.40}) is obtained in the
framework of the supersymmetric $\sigma$-model method.

We follow essentially the same strategy as was outlined in
Sec.~\ref{s2.1}, \ref{s2.3} for the average DOS. The level corelation
function is expressed in terms of the Green's functions as follows
\be
\label{e1.41}
R_2(\omega)={1\over 2(\pi\nu N)^2}{\rm Re}\,\left\langle 
{\rm Tr}G_{\rm R}^{E+\omega/2} 
{\rm Tr}(G_{\rm R}^{E-\omega/2}-G_{\rm A}^{E-\omega/2})
\right\rangle\ . 
\ee
The product of  two retarded Green's functions here is trivial;
it can be calculated in the same way as the average DOS above,
yielding 
\be
\label{e1.42}
N^{-2}\langle 
{\rm Tr}G_{\rm R}^{E+\omega/2} {\rm Tr}G_{\rm R}^{E-\omega/2}\rangle
\simeq N^{-2}\langle {\rm Tr}G_{\rm R}^E\rangle^2=
\left(E/2-i\sqrt{1-E^2/4}\right)^2\ .
\ee
In other words, the average of the product of  retarded Green's
functions decouples into the product of the averages (the same is valid
for the product of the advanced Green's functions, of course).
All the non-trivial information about correlations is contained
therefore in the product of the type $\langle G_{\rm R} G_{\rm
A}\rangle$, 
\be
\label{e1.43}
T_2(E_1,E_2)={1\over N^2}\langle {\rm Tr}(E_1-\hat{H})^{-1}
{\rm Tr}(E_2-\hat{H})^{-1} \rangle\ ,
\ee
where
\be
\label{e1.44}
E_{1,2}=E\pm\left({\omega\over 2}+i\eta\right) \equiv E \pm r/2N\ ; \ \ 
{\rm Im}\, r>0\ .
\ee

To represent $T_2(E_1,E_2)$ as a superintegral, we introduce
supervectors of a double size,
\be
\label{e1.45}
\Phi_i=\left\lgroup\begin{array}{cc} 
S_1(i) \\ \chi_1(i)\\S_2(i) \\ \chi_2(i)
\end{array}\right\rgroup\ ,
\ee
with the first two components corresponding to the advanced and the
last two to the retarded sector. As before, the $S$-components are
commuting, while the $\chi$-components anticommuting; the index $i$
running from $1$ to $N$ corresponds to the vector space in which the
matrix $\hat{H}$ acts. We have then 
\be
\label{e1.46}
T_2(E_1,E_2)={(-)^N\over N^2}\int[{\rm d}\Phi^*{\rm d}\Phi]\sum_{i,j}
S_1^*(i)S_1(i)S_2^*(j)S_2(j)\exp\{-S[\Phi]\}\ ,
\ee
where the action $S[\Phi]$ is given by (to make notations more compact, 
we combine all $S_1(i)$ into an $N$-component vector $S_1$, and the
same for $S_2$, $\chi_1$, $\chi_2$)
\bea
S[\Phi]&=&-iS_1^\dagger(E_1-\hat{H})S_1
        -i\chi_1^\dagger(E_1-\hat{H})\chi_1
        +iS_2^\dagger(E_2-\hat{H})S_2
        -i\chi_2^\dagger(E_2-\hat{H})\chi_2  \nonumber\\
 &\equiv& -i\Phi^\dagger L\left(E+{r\over
2N}\Lambda-\hat{H}\right)\Phi\ .
\label{e1.47}
\eea
Here the matrices $\Lambda$ and $L$ are defined as $\Lambda={\rm
diag}(1,1,-1,-1)$, $L={\rm diag}(1,1,-1,1)$, with ordering of
components according to (\ref{e1.45}), i.e. $({\rm Rb,Rf,Ab,Af})$, where
${\rm R,A}$ correspond to the 
retarded-advanced and ${\rm b,f}$ to the boson-fermion decomposition. 
After averaging over the ensemble of matrices $\hat{H}$, the action
acquires the form
\be
\label{e1.48}
S[\Phi]=-i\Phi^\dagger L\left(E+{r\over 2N}\Lambda\right)\Phi
+{1\over 2N}\sum_{ij}(\Phi_i^\dagger L\Phi_j)(\Phi_j^\dagger L\Phi_i)\ .
\ee
Decoupling of the quartic
term via the Hubbard-Stratonovich transformation requires the
introduction of a $4\times 4$ supermatrix variable $R$,
\bea
T_2(E,r)&=&{(-)^N\over N^2}\int[{\rm d}\Phi^*{\rm d}\Phi]\sum_{i,j}
S_1^*(i)S_1(i)S_2^*(j)S_2(j)\int {\rm d}R \nonumber\\
&\times&\exp\left\{-i\Phi^\dagger L^{1/2}
\left(E-R+{r\over 2N}\Lambda\right)L^{1/2}\Phi
-{N\over 2}{\rm Str}R^2\right\}\ .
\label{e1.49}
\eea

The next step of our program is to interchange the integrations $\int
{\rm d}R$ and $\int[{\rm d}\Phi^*{\rm d}\Phi]$. This leads, however,
to a  set of requirements of convergence of both the
$\Phi$-integral (when it is taken first) and the $R$-integral.
It turns out that in order to satisfy these requirements, one has to
choose a non-trivial manifold of integration over the
matrix $R$. Indeed, a naive attempt to generalize  Eq.~(\ref{e1.26})
straightforwardly by choosing 
\be
\label{e1.50}
R=\left\lgroup \begin{array}{cc} R_{\rm b}  &  \bar{\chi} \\
                            \chi &   iR_{\rm f}
          \end{array} 
   \right\rgroup \mbox{in boson-fermion decomposition}
\ee
with Hermitian matrices $R_{\rm b}$ and $R_{\rm f}$ fails, since then
the $\Phi$-dependent part of the action,
$i\Phi^\dagger L^{1/2}(E-R)L^{1/2}\Phi$, contains the term 
$$
S_1^\dagger R_{{\rm b};12}S_2+S_2^\dagger R_{{\rm b};21}S_1=2{\rm Re}\,
S_1^\dagger R_{{\rm b};12}S_2\ ,
$$
which is real and has an arbitrary sign, thus leading to a divergent
integral over $S_1$, $S_2$. The natural idea to cure this problem by
multiplying the components $R_{{\rm b};12}$ and $R_{{\rm b};21}$ by
$i$ is immediately seen to fail as well, since it leads to a divergent
$R$-integral. The way out was found by Sch\"afer and Wegner
\cite{schaefweg} for the case of the bosonic replica model. Since the
problem is pertinent to the bosonic sector of the supersymmetric
model, the 
idea of Ref.~\cite{schaefweg} can be straightforwardly generalized
to the supersymmetric case. For a thorough
discussion of this issue, the reader is referred to the review paper
by Verbaarschot, Weidenm\"uller, and Zirnbauer \cite{vwz}. 

The solution of the problem, which we are going to formulate now, is
suggested by the invariance of the quartic term in (\ref{e1.48}) with
respect to the rotations $\Phi\to V\Phi$, where the matrices $V$
satisfy $V^\dagger LV=L$, thus forming a pseudounitary 
supergroup\footnote{Let us remind that $L={\rm diag}(1,1,-1,1)$; 
in the notation $U(1,1|2)$  the content of the brackets 
to the left of the vertical bar refers to the
$(+,-)$ metric in the bosonic sector, while that to the right
corresponds to the $(+,+)$ metric in the fermionic sector. In other
words, the group $U(1,1|2)$ represents a  product
$U(1,1)|_{\rm bb}\times U(2)|_{\rm ff}$, ``dressed'' by Grassmannian
generators transforming the commuting components into anticommuting
and vice versa.} $U(1,1|2)$. The reason for the above problem with
convergence of the integral over the commuting components of $\Phi$
lies in 
the fact that the set of hermitian matrices $R_{\rm b}$ is {\it not}
invariant with respect to the pseudounitary group $U(1,1)$. The proper
integration manifold is to be chosen as follows
\cite{efetov83,ef-book,vwz} (note that we return to the ordering of
components according to Eq.~(\ref{e1.45}), so that the external block
structure shown explicitly in (\ref{e1.51}) corresponds to the
retarded-advanced decomposition; each block being a $2\times 2$
supermatrix): 
\be
\label{e1.51}
R=T\left\lgroup\begin{array}{cc}
P_1-i\delta_0   &           \\
                 & P_2+i\delta_0
\end{array}
\right\rgroup T^{-1}\ ,
\ee
where $\delta_0>0$ is a constant (which is to be chosen below in such
a way that the integration manifold passes through the saddle points)
and $P_1=P_1^\dagger$, $P_2=P_2^\dagger$ are hermitian supermatrices.
Further, the matrix $T$ satisfies $T^\dagger L T=L$
and belongs to the coset space\footnote{For a group $G$ and its
subgroup $K$ the space $G/K$ is formed by the set of (left) cosets
$gK$ with $g\in G$.}
$U(1,1|2)/U(1|1)\times U(1|1)$. This coset space is obtained from the
group $U(1,1|2)$, if the elements of this group which commute with
$\Lambda$ (they form the subgroup $U(1|1)\times U(1|1)$) are
identified with unity. The matrices of this coset space can
be parametrized in the following way
\be
\label{e1.52}
T=\left\lgroup\begin{array}{cc}
(1+t_{12}t_{21})^{1/2}  &  t_{12}\\
t_{21}                  &  (1+t_{21}t_{12})^{1/2}
\end{array}\right\rgroup\ ;
\ee
\be
\label{e1.53}
t_{12}=\left\lgroup\begin{array}{cc}
a         &   -i\eta_1 \\
\eta_2^*  &   ib^*
\end{array}\right\rgroup\ ;\ \ 
t_{21}=\left\lgroup\begin{array}{cc}
a^*         &   -\eta_2 \\
i\eta_1^*  &   ib
\end{array}\right\rgroup\ ;\ \ |b|^2\le 1\ .
\ee
The matrix $R$ has 16 real degrees of freedom (the same number as a
$4\times4$ hermitian matrix); in the above parametrization 8 degrees of
freedom are contained in matrices $P_{1,2}$ and the remaining 8 in
$t_{12}, t_{21}$. The measure ${\rm d}R$ takes in this parametrization
the following form (see \cite{vwz} for details)
\be
\label{e1.54}
{\rm d}R={\cal F}(P_1,P_2) {\rm d}P_1 {\rm d}P_2 {\rm d}\mu(T)\ ,
\ee
where ${\rm d}P_i$ are conventional matrix measures (product of differentials
of all elements) and ${\cal F}(P_1,P_2)$ is a function depending on
eigenvalues of $P_1-i\delta_0$ and $P_2+i\delta_0$ and equal to unity at the
saddle point manifold, for which $P_{1,2}$ are proportional to the unit
matrix. Further, ${\rm d}\mu(T)$ is the invariant measure on the coset
space; in the parametrization (\ref{e1.52}), (\ref{e1.53}) of $T$ it
is given explicitly by
\be
\label{e1.55}
{\rm d}\mu(T)={\rm d}t_{12}{\rm d}t_{21}\ .
\ee

Having specified the manifold of matrices $R$ over which the integration in
(\ref{e1.49}) goes, we can interchange the order of integrals and
evaluate the integral over $\Phi$ first. The result reads, similarly
to (\ref{e1.28}),
\bea
\label{e1.56}
&&T_2(E,r)=\int {\rm d}R\left(E-R+{r\over 2N}\Lambda\right)^{-1}_{{\rm
bb},11} 
\left(E-R+{r\over 2N}\Lambda\right)^{-1}_{{\rm bb},22}
e^{-NS[R]}\ ;\\
&&S[R]={1\over 2}{\rm Str}R^2 + {\rm Str}\ln
\left(E-R+{r\over 2N}\Lambda\right)\ .
\label{e1.57}
\eea
As in the case of the average DOS calculation, 
the large factor $N\gg 1$ in the exponent allows us to use the
saddle-point approximation. The saddle-point equation that is
obtained by varying the action (\ref{e1.57})\footnote{We neglect the
term proportional to $r/N$ when deriving the saddle-point equation,
since it gives a correction $\sim O(1)$ to $NS[R]$. We will take this
term into account below when integrating over the saddle-point
manifold.}  looks identical to 
Eq.~(\ref{e1.30}) [the difference being that $R$ is now a $4\times 4$
supermatrix, while it was $2\times 2$ in Sec.~\ref{s2.3}]. 
Similarly to Eq.~(\ref{e1.31}), diagonal solutions
of this equation have the form
\be
\label{e1.58}
R={E\over 2}-i\sqrt{1-E^2/4}\left\lgroup\begin{array}{cccc} 
               s_1 &      &      &        \\  
                   & s_2  &      &        \\
                   &      & s_3  &        \\
                   &      &      &  s_4    
\end{array}\right\rgroup\ ; \qquad s_1,\ldots,s_4=\pm 1\ .
\ee
When fixing the signs of $s_i$, the same reasons as for the average
DOS calculation are to be taken into account:
\begin{itemize}
\item bosonic sector: shift of the integration contour should not
cross singularities. This requires $s_1=-s_3=-1$;
\item fermionic sector: the leading contribution is found to be given
by the saddle points with $s_2=-s_4=-1$ and $s_2=-s_4=1$;
contributions of the two other saddle-points are suppressed by $1/N$.
\end{itemize}

Therefore, only 2 saddle-points out of 16 survive. In fact, it is
sufficient to consider only one of them [say, with the signs
$(-,-,+,+)$]; the second one [with $(-,+,+,-)$]
will belong to the manifold obtained by
rotating the first one by the matrices $T$. 
Indeed, if $R_0=E/2-i\sqrt{1-E^2/4}\Lambda$
is a solution of Eq.~(\ref{e1.30}), then all matrices of the form
\be
\label{e1.59}
R=TR_0T^{-1}={E\over 2}-i\sqrt{1-E^2/4}\,T\Lambda T^{-1}\equiv
{E\over 2} -i\pi\nu Q
\ee
are solutions [and thus saddle points of the action
(\ref{e1.57})] as well. Here $\nu\equiv\langle \nu(E)\rangle$ [see
(\ref{e1.38})] and matrices $Q$ defined as
\be
\label{e1.60}
Q=T\Lambda T^{-1}
\ee
(and obviously satisfying $Q^2=1$) form a manifold of the supermatrix
$\sigma$-model. 

Comparing (\ref{e1.59}) with (\ref{e1.51}), we see that $\delta_0$
should be chosen as $\delta_0=\pi\nu$ and that the saddle-point
manifold corresponds to $P_1=P_2=E/2$ (and $T$ running over the coset
space). The 
Gaussian integral over $P_1$ and $P_2$ around this saddle-point value 
can be easily calculated, yielding unity (due to compensation of
the bosonic and fermionic massive modes, {\it cf.} Eq.~(\ref{e1.36})). 
We are thus left with an integral over the manifold
(\ref{e1.60}). Expanding the action (\ref{e1.57}) up to the first
order in $r/N\ll 1$, we reduce it to the form\footnote{since in
Eq.~(\ref{e1.61}) and below we
use $Q$ as a variable on the coset space, we denote the corresponding
invariant measure as ${\rm d}\mu(Q)$; it is identical to ${\rm d}\mu(T)$
introduced earlier.}
\be
\label{e1.61}
T_2(E,r)=\int {\rm d}\mu(Q)\left({E\over 2}-i\pi\nu Q_{{\rm
bb},11}\right) 
\left({E\over 2}-i\pi\nu Q_{{\rm bb},22}\right)
\exp\left\{{i\pi\nu r\over 2}{\rm Str}Q\Lambda\right\}\ .
\ee
We have therefore reduced calculation of the level correlation
function for an ensemble of matrices $\hat{H}$ (with $\sim
N^2\to\infty$ 
degrees of freedom) to evaluation of an integral over a supermatrix
$Q$ parametrized by a finite number (eight) variables and belonging to
certain non-linear space. The obtained problem is known as
zero-dimensional (0D) $\sigma$-model. The term ``zero-dimensional''
distinguishes integrals of the type (\ref{e1.61}) over a single matrix
$Q$ from a
field-theoretical $\sigma$-model studied in Sec.~\ref{s3} with the
matrix $Q$ depending on spatial coordinates.

For an explicit evaluation of integrals of the type (\ref{e1.61}) over
the 
coset space the parametrization (\ref{e1.52}), (\ref{e1.53}) is
inappropriate; a much more convenient parametrization was found by
Efetov \cite{efetov83}:
\be
\label{e1.62}
Q=\left\lgroup\begin{array}{cc}
   U_1   &       \\
         &  U_2            \end{array}\right\rgroup
\left\lgroup\begin{array}{cccc}
  \lambda_1   &     0      &      i\mu_1   &     0     \\
     0        &  \lambda_2 &       0      & \mu_2^*   \\
   i\mu_1^*   &     0      &  -\lambda_1  &     0      \\
     0        &   \mu_2    &       0      & -\lambda_2
\end{array}\right\rgroup
\left\lgroup\begin{array}{cc}
   U_1^{-1}   &       \\
         &  U_2^{-1}            \end{array}\right\rgroup\ ;
\ee
\bea
\label{e1.63}
&& U_1=\exp\left\lgroup\begin{array}{cc}
   0   & -\alpha^*      \\
   \alpha      &   0          \end{array}\right\rgroup\ ;\qquad
U_2=\exp i\left\lgroup\begin{array}{cc}
   0   & -\beta^*      \\
   \beta      &   0          \end{array}\right\rgroup\ ,\\
&& 1\le\lambda_1<\infty,\ \ -1\le\lambda_2\le1,\ \
   |\mu_1|^2=\lambda_1^2-1,\ \ |\mu_2|^2=1-\lambda_2^2\ .
\label{e1.64}
\eea
Alternatively, instead of the variables $\lambda_{1,2}$, $\mu_{1,2}$
defined in Eq.~(\ref{e1.64}), one can introduce the set of variables
$\theta_1$, $\theta_2$, $\phi_1$, $\phi_2$ via
\bea
\label{e1.65}
&&\lambda_1=\cosh\theta_1\ ,\ \ \mu_1=\sinh\theta_1e^{i\phi_1}\ ,\ \ 
\lambda_2=\cos\theta_2\ ,\ \ \mu_2=\sin\theta_2e^{i\phi_2}\ ,
\nonumber \\
&& 0\le\theta_1<\infty\ ,\ \ 0\le\theta_2\le\pi\ ,\ \ 
0\le\phi_{1,2}<2\pi\ .
\eea
Note that $\lambda_{1,2}$ are eigenvalues of the boson-boson (and
fermion-fermion) block of the matrix $Q$; we will
call them simply  ``eigenvalues'' for brevity.
The integral (\ref{e1.61}) takes in this parametrization the form
\bea
\label{e1.66}
T_2(E,r)&=&-\int {\rm d}\mu(Q)\left[{E\over
2}-i\pi\nu(\lambda_1-\alpha^*\alpha(\lambda_1-\lambda_2)
)\right]\nonumber \\
&\times & \left[{E\over
2}+i\pi\nu(\lambda_1+\beta^*\beta(\lambda_1-\lambda_2) )\right]
e^{i\pi\nu r(\lambda_1-\lambda_2)}\ ,
\eea
with the measure ${\rm d}\mu(Q)$ given by
\be
\label{e1.67}
{\rm d}\mu(Q)=-{{\rm d}\lambda_1{\rm d}\lambda_2
\over(\lambda_1-\lambda_2)^2}
{\rm d}\phi_1 {\rm d}\phi_2 {\rm d}\alpha {\rm d}\alpha^*{\rm d}\beta
{\rm d}\beta^*\ . 
\ee

The crucial advantage of the above parametrization is that the
exponent in (\ref{e1.61}) acquires a very simple form (see
(\ref{e1.66})) dependent on $\lambda_1$ and $\lambda_2$ only.
This allows to integrate out straightforwardly the Grassmannian
variables.  
The integration rules (\ref{e1.8}) imply that only the highest order
term ($\sim\alpha^*\alpha\beta^*\beta$) in the expansion of the
integrand in a polynomial over Grassmannians gives a non-zero
contribution after integration over 
${\rm d}\alpha^*{\rm d}\alpha {\rm d}\beta^*{\rm d}\beta$. The result
is easily seen to be 
\be
\label{e1.68}
(\pi\nu)^2\int_1^\infty {\rm d}\lambda_1\int_{-1}^1{\rm d}\lambda_2\, e^{i\pi\nu
r(\lambda_1-\lambda_2)}={2i\over r^2}\sin(\pi\nu r)e^{i\pi\nu r}\ .
\ee
There exists, however, one more contribution to the integral
(\ref{e1.66}), namely that of the term of the zeroth order in
Grassmannians. Though naively it gives zero after the Grassmannian
integration, the corresponding integral over $\lambda_{1,2}$ has a
singularity $\int {\rm d}\lambda_1 {\rm
d}\lambda_2/(\lambda_1-\lambda_2)^2$ and 
thus diverges in the vicinity of $\lambda_1=\lambda_2=1$. The reason
for this ambiguity is in the singular character of Efetov's
parametrization at $Q=\Lambda$, i.e  at $\lambda_1=\lambda_2=1$.
The problem arises only for the term which does not contain
Grassmannians; all higher order terms contain additional powers of
$(\lambda_1-\lambda_2)$  [see (\ref{e1.66})], which make the integral
over $\lambda$'s convergent. The value of the integral of the zeroth
order term is determined by the following formula
\be
\label{e1.69}
\int {\rm d}\mu(Q)F(\lambda_1,\lambda_2) = F(1,1) \equiv F(Q=\Lambda)\ ,
\ee
where $F$ is an arbitrary function (depending on eigenvalues
$\lambda_1,\,\lambda_2$ only) which vanishes at $\lambda_1\to\infty$. 
Equation (\ref{e1.69}) is a particular case of a theorem (often called 
Parisi-Sourlas-Efetov-Wegner theorem in the physical literature),
which states that for a broad 
class of supersymmetric models an integral of an invariant function
(in our case the fact that $F(Q)$ depends only on $\lambda$'s  means
that it is invariant with respect to rotations $Q\to VQV^{-1}$ with
$V\in U(1|1)\times U(1|1)$) is given by its value at the origin (in
our case $Q=\Lambda$). For a discussion and a proof, see
Refs.~\cite{zirn86,congr}. A general mathematical treatment of such
anomalous contributions to integrals over supermanifolds can be found
in \cite{roth87,zirn96}.

Applying (\ref{e1.69}) to the zeroth order (in Grassmannians) term of
(\ref{e1.66}) and adding the result to the contribution (\ref{e1.68})
of the highest order term, we finally get
\be
\label{e1.70}
T_2(E,r)=\left({E\over 2}-i\pi\nu\right)\left({E\over
2}+i\pi\nu\right) + {2i\over r^2}\sin(\pi\nu r)e^{i\pi\nu r}\ .
\ee
Substituting this in (\ref{e1.41}), including the
$\langle G_{\rm R}G_{\rm R}\rangle$ contribution (\ref{e1.42}),
and taking into account that $\nu r=\nu\omega N=\omega/\Delta=s$, we
find the two-level correlation function
\bea
\label{e1.71}
R_2(\omega)&=&{1\over 2\pi^2\nu^2}{\rm Re}\,\left[T_2(E,r)-\left({E\over
2}-i\pi\nu\right)^2\right] \nonumber \\
&=& 1 + \delta(s) -{\sin^2(\pi s)\over (\pi
s)^2}\ ,
\eea
in agreement with (\ref{e1.40}).

\subsection{Comments and generalizations}
\label{s2.5}

\subsubsection{Structure of the saddle-point manifold.}
\label{s2.5.1a}

We first note that the topology of the $\sigma$-model manifold was
crucially important for the above calculation. Indeed, 
if we keep only the ordinary part (which does not contain Grassmannians)
of each element $Q$ of this supermanifold\footnote{the discarded part
containing the Grassmannians is called nilpotent}, we get a conventional
(not super-) manifold\footnote{called the base of the
supermanifold}, which is a product of the hyperboloid
$U(1,1)/U(1)\times U(1)$ (parametrized by $\lambda_1=\cosh\theta_1$
and $\phi_1$) and the sphere $U(2)/U(1)\times U(1)$ (parametrized by
$\lambda_2=\cos\theta_2$ and $\phi_2$). This combination of the
compact ($\lambda_2$) and non-compact ($\lambda_1$) degrees of freedom
is clearly reflected in the result, see Eq.~(\ref{e1.68}).

It is appropriate here to add a comment concerning a somewhat subtle
point of the calculation. The non-compact (hyperbolic) symmetry of the
bb-sector originates from the opposite signs of the first and the
third terms in Eq.~(\ref{e1.47}). This sign choice was unambiguously
dictated by the requirement of convergence of the integral over the
bosonic components $S_{1,2}$ of the supervector $\Phi$. On the other
hand, the situation seems to be different 
for the two terms of (\ref{e1.47})
containing the Grassmannians $\chi_{1,2}$, since the integral over the
anticommuting variables is always defined and no condition of
convergence arises. Therefore, the choice of signs of the second and
the fourth terms in (\ref{e1.47}), which has eventually led us to the
compact symmetry of the ff-sector, seems to be arbitrary. A thorough
analysis \cite{vwz} shows, however, that this apparent freedom is
spurious, and the symmetry of the ff-block is uniquely fixed to be
compact on a later stage of the calculation by the requirement of
convergence of the integral over the coset space ({\rm i.e.} over the
matrix $T$).

\subsubsection{Gaussian ensembles of different symmetry.}
\label{s2.5.1b}

Let us recall that for the sake of technical simplicity we considered
the case of GUE (with the matrix $\hat{H}$ being hermitian
without any further restrictions) while calculating the level
statistics. An analogous (though technically more involved)
calculation can be performed for the two
other Gaussian ensembles of Wigner and Dyson, the Gaussian Orthogonal
Ensemble (GOE) of real symmetric matrices and the Gaussian Symplectic
Ensemble (GSE) of real quaternionic matrices \cite{mehta,bohigas}.
If the matrix $\hat{H}$ is  considered as a Hamiltonian, the GUE
describes systems without the (antiunitary) time reversal symmetry,
while GOE and GSE correspond to systems with  the time reversal invariance
realized by an antiunitary operator ${\cal T}$ such that ${\cal
T}^2=1$ (GOE) or ${\cal T}^2=-1$ (GSE). 

Calculation of the two-level correlation function for GOE within the
supersymmetry approach 
is performed essentially in the same way as for GUE, see
\cite{efetov83,ef-book}. However, due to the presence of an additional
symmetry in the problem, one has to double the size of the
supervectors $\Phi_i$, combining $\Phi_i$ with its time reversal
$\Phi_i^*$. Correspondingly, one ends up with a 0D $\sigma$-model of
$8\times 8$ supermatrices $Q$  parametrized by 16 independent
variables. Now, instead of two for GUE, there are three eigenvalues,
$1\le\lambda_1,\lambda_2<\infty$ and $-1\le\lambda\le 1$. A similar
structure is obtained for GSE with $8\times 8$ $Q$-matrix as well, but
with two ``compact'' and one ``non-compact'' eigenvalues, 
$-1\le\lambda_1\le 1$, $0\le\lambda_2\le 1$, 
$1\le \lambda<\infty$. The evaluation of
the corresponding integrals \cite{efetov83,ef-book} reproduces again
the result for the two-level correlation function obtained by Mehta
and Dyson within the orthogonal polynomial method,\footnote{Note that
in the symplectic case all the levels are double 
degenerate (Kramers degeneracy). This degeneracy is disregarded in
(\ref{e1.73}), which thus represents the correlation function of
distinct levels only, normalized to the corresponding level spacing.}
\be
 R_{2}(s)=1+\delta(s)-\frac{\sin^2(\pi s)}{(\pi s)^2}-
\left[{\pi\over 2}{\rm sign}(s)-\mbox{Si}(\pi s)\right]
\left[{\cos\pi s\over\pi
s}-{\sin\pi s\over (\pi s)^2}\right] \qquad ({\rm GOE}), \label{e1.72} 
\ee
\begin{eqnarray}
 && \hspace*{-1cm} R_{2}(s)=1+\delta(s)-\frac{\sin^2(2\pi s)}{(2\pi s)^2}+
\mbox{Si}(2\pi s)\left[{\cos 2\pi s\over 2\pi
s}-{\sin 2\pi s\over (2\pi s)^2}\right] \qquad ({\rm GSE}), \label{e1.73}\\
&& \hspace*{-1cm}
\mbox{Si}(x)=\int_0^x{\sin y\over y} dy\ .  \nonumber
\end{eqnarray}
In the remaining part of the lecture course, we will continue
considering explicitly the unitary class only;  
corresponding results for the other two classes will be sometimes
quoted in the end of the calculation. 

On top of the three ``standard'' Wigner-Dyson ensembles, seven other
ensembles have been introduced more recently. They become relevant, in
various physical situations, near a special point of the spectrum
where the symmetry of the problem gets enlarged. Specifically, this
happens (i) at the center of the band for a particle moving on a
bipartite (AB) lattice with only off-diagonal (A$\to$B or B$\to$A)
hopping allowed \cite{gade}; (ii) in the chiral random matrix
ensembles \cite{verb}, which model the massless Dirac operator in the
quantum 
chromodynamics, near the zero point of the spectrum; (iii) near the
Fermi energy in models of a
mesoscopic metallic grain in contact with a superconductor
\cite{alzirn96}. We will not consider such ensembles in this lecture
course. An interested reader is referred to Ref.~\cite{zirn96} for a
review of related algebraic structures.

A natural question that may be asked at this point is why is there a
need in the supersymmetry formalism if the results discussed above
have been obtained 
much earlier by the methods of ``classical'' RMT. In fact,
Sec.~\ref{s3}-\ref{s5} will give an answer to this question, since the
supersymmetry approach will be used there to study statistical
properties for a much broader class of problems (disordered and
chaotic systems), essentially different
in their formulation from the RMT ensembles.  However, already here we
want to give two examples of the random matrix ensembles which, while
being direct generalizations of the Gaussian ensembles, are not
accessible within the classical RMT methods.

\subsubsection{Ensembles with  non-Gaussian distributions of matrix
elements. } 

\label{s2.5.1}

Let us consider an ensemble of large ($N\to\infty$)
$N\times N$ hermitian matrices
with all matrix elements being independent and equally distributed,
but (in contrast to GUE) with some non-Gaussian distribution
function $f(z,z^*)\equiv f(|z|^2)$. In other words, the overall
probability density for the matrix $\hat{H}$ is assumed to be
\be
\label{e1.74}
{\cal P}(\hat{H})=\prod_{i\le j}f(|H_{ij}|^2)\ .
\ee
For the Gaussian distribution, $f(|z^2|)=\exp(-N|z|^2)$,
this is reduced\footnote{The distribution of the diagonal elements
$H_{ii}$ is not important in the leading order in $1/N$.} 
to the Gaussian ensemble (\ref{e1.1}). However, for
any other distribution function $f$ the probability density
${\cal P}(\hat{H})$ cannot be written in the form $\exp{\rm
Tr}\,F(\hat{H})$ any more. In other words, the ensemble (\ref{e1.74})
is no more invariant with respect to the unitary rotations,
$\hat{H}\to \hat{U}\hat{H}\hat{U}^{-1}$ and the probability
distributions of eigenvalues and eigenvectors of $\hat{H}$ do not
decouple any more. This precludes the application of the orthogonal
polynomial method, which uses an explicit form of the distribution of
the eigenvalues.

On the other hand, the supersymmetry method can still be successfully
used \cite{mf91a}. After averaging over $\hat{H}$ one gets,
instead of the quartic term in the action (\ref{e1.48}),
\bea
\label{e1.75}
&&\hspace*{-1cm}
\left\langle \exp\{-\sum_{ij}
\Phi_i^\dagger H_{ij}L\Phi_j\}\right\rangle = \prod_{i<j}
\int {\rm d}z {\rm d}z^*
\exp\{i\Phi_i^\dagger L\Phi_j z + i\Phi_j^\dagger L\Phi_i
z^*\} f(|z|^2) \nonumber \\
&& \hspace*{-1cm} = \prod_{i<j}\left\{1-
(\Phi_i^\dagger L\Phi_j)(\Phi_j^\dagger L\Phi_i)\langle |z^2|\rangle_f
+{1\over 4}
(\Phi_i^\dagger L\Phi_j)^2(\Phi_j^\dagger L\Phi_i)^2
\langle|z^4|\rangle_f + \dots\right\}\ ,
\eea
where $\langle\ldots\rangle_f$ denotes averaging over $f(|z|^2)$. 
As in the case of GUE, we will assume the normalization which keeps
the band of eigenvalues of $\hat{H}$ finite in the limit
$N\to\infty$. This implies $\langle|H_{ij}|^2\rangle\sim 1/N$, i.e.
$\langle |z^2|\rangle_f\sim 1/N$. If $f$ is a smooth, non-singular
function, then $\langle |z^4|\rangle_f\sim 1/N^2$ etc. For this
reason, one can drop all terms beyond the quartic one in the second
line of (\ref{e1.75}). Indeed,
the integral over $\Phi$ (taken after the Hubbard-Stratonovich
decoupling) will be dominated by $\Phi\sim N^0$, see
Eq.~(\ref{e1.49}). Therefore, the second line 
of (\ref{e1.75}) is in fact an expansion in
$1/N$. Dropping all higher order terms and reexponentiating the
quartic term (which is $\sim 1/N$), we get
\be 
\label{e1.76}
\left\langle \exp\{-\sum_{ij}
\Phi_i^\dagger H_{ij}L\Phi_j\}\right\rangle \simeq \prod_{i,j}
\exp\left\{-{1\over 2}\langle |z^2|\rangle_f
(\Phi_i^\dagger L\Phi_j)(\Phi_j^\dagger L\Phi_i)\right\}\ ,
\ee
which is the same result as for the Gaussian distribution 
$f(z,z^*)=\exp\{-|z|^2/\langle |z^2|\rangle_f\}$. Therefore, the spectral
statistics for the ensemble (\ref{e1.74}) has the same Wigner-Dyson
form as for the GUE. 

\paragraph{Sparse random matrix ensemble.}
A notable exception from this statement is formed by an ensemble of
sparse random matrices \cite{mf91a,fm91a} for which the distribution
function $f(|z|^2)$ is singular at $z=0$. More specifically, 
\be
\label{e1.77}
f(|z|^2)= \left(1-{p\over N}\right)\delta(|z|^2)+{p\over N}h(|z|^2)\ ,
\ee
where $p>0$ is a constant of order unity 
(i.e. independent of $N$) and $h$ is a smooth (in
particular, non-singular at $z=0$) distribution function with
$\langle|z^2|\rangle_h\sim 1$. The distribution (\ref{e1.77}) implies
that almost all matrix elements of $\hat{H}$ are zero; only $p$ (in
average) out of $N$ elements in each row are non-zero (and distributed
according to $h(|z|^2)$). Because of the singular character of
$f(|z|^2)$ truncation of the series in (\ref{e1.75}) is no more
legitimate. Indeed, we have now  $\langle |z^2|\rangle_f\sim 1/N$,
$\langle |z^4|\rangle_f\sim 1/N$, etc., so that all the terms have to
be taken into account. To decouple all of them via a kind of the
Hubbard-Stratonovich transformation, one has to introduce,
in addition to the usual matrix $Q_{\alpha\beta}$, also higher order
tensors  $Q^{(4)}_{\alpha\beta\gamma\delta}$, 
$Q^{(6)}_{\alpha\beta\gamma\delta\mu\nu},\ldots$. It turns out
\cite{mf91a,fm91a} that 
they all can be combined in a function $Y(\Phi,\Phi^\dagger)$, so that
the Hubbard-Stratonovich transformation acquires a functional
form. The analysis show that the model exhibits an Anderson
localization transition if one of the parameters ({\it e.g.} $p$) is
changed. At $p<p_c$ all eigenvectors are localized and the level
statistics is Poisonnian (uncorrelated levels) in the limit
$N\to\infty$; at $p>p_c$ the states are delocalized and the
Wigner-Dyson statistics applies\footnote{To be precise, one should
speak about the states belonging to the infinite cluster 
only, see \cite{mf91a,fm91a}.};
finally $p=p_c$ is the critical point of the Anderson transition.
We mention also that this model has effectively infinite-dimensional
character and is closely related to the problem of Anderson
localization on the tree-like Bethe lattice, see \cite{mf91b}.

\subsubsection{Random banded matrices.}

\label{s2.5.2}

The random banded matrix (RBM) ensemble is defined in the following
way (see {\it e.g.} \cite{izrailev90,casati90}). As in the Gaussian
ensemble, all 
the matrix elements are supposed to be independent and have a Gaussian
distribution. The difference is that the variance of this distribution
is now not the same for all the elements but rather depends on the
distance from the main diagonal,
\be
\label{e1.78}
\langle |H_{ij}|^2\rangle = F(|i-j|)\ .
\ee
The function $F(r)$ is supposed to be roughly constant for
$r\lesssim b$ and negligibly small for $r\gg b$. Therefore, all the
elements of the matrix which are essentially non-zero are located
within a band of width $\sim b$ around the main diagonal (hence the
term ``banded matrices''). Again, the RBM ensemble is not
rotationally-invariant and thus cannot be treated by the classical
RMT methods. In contrast, the supersymmetry approach can be applied to
this ensemble \cite{fm91}. In particular, one finds that if $b/N$ is
kept constant while the limit $N\to \infty$ is considered, the level
statistics acquire the universal Wigner-Dyson form. More generally,
this ensemble describes a quasi-one-dimensional system, with
statistical properties depending on a value of the ratio $X=N/b^2$. If
$X\ll 1$ (as in the above limiting procedure), the level (and
eigenfunction) statistics are essentially  of the GUE form;
in the opposite limit $X\gg 1$ the system is in the strong
localization regime. This will be discussed in more detail in
Sec.~\ref{s3}, \ref{s4}. 

\subsubsection{Parametric level statistics.}
\label{s2.5.3}

We have just demonstrated that the supersymmetry approach is extremely
useful when the matrix ensemble different from the Gaussian ones are
considered. However, also for the Gaussian ensembles the
supersymmetry method has produced a bulk of new  results for
quantities more complicated than the conventional two-level
correlation function (\ref{e1.39}). As an important example, we
mention here the parametric level statistics which is defined in the
following way. Let us perturb the set of matrices $\hat{H}_0$ forming the
GUE by adding some given traceless matrix $\hat{V}$ multiplied by a
parameter $\alpha$,
\be
\label{e1.79}
\hat{H}(\alpha)=H_0+\alpha\hat{V}\ .
\ee
Consider now the following correlation function
\be
\label{e1.80}
R_2(\omega,\delta\alpha)=\langle{\rm Tr}\,\delta(E-\hat{H}(\alpha))\:{\rm Tr}\,
\delta(E+\omega-\hat{H}(\alpha+\delta\alpha))\rangle/\langle\nu\rangle^2\ .
\ee
For $\delta\alpha=0$ this is just the usual two-level
correlation function (\ref{e1.39}). The parametric DOS-DOS correlation
function was calculated via the supersymmetry method by Simons and
Altshuler \cite{simons93}. As for the simple level correlation
function (\ref{e1.39}), the result has a universal form if the
perturbation parameter $\delta\alpha$ is properly rescaled,
\be
\label{e1.81}
R_2(\omega,\delta\alpha)=1+{1\over 2}{\rm Re}\int_1^\infty
{\rm d}\lambda_1\int_{-1}^1 {\rm d}\lambda_2\exp\left[i\pi
s(\lambda_1-\lambda_2)-{\pi^2\over
2}x^2(\lambda_1^2-\lambda_2^2)\right]\ ,
\ee
where $x=\delta\alpha\langle(\partial_\alpha E_i)^2\rangle^{1/2}/\Delta$
and, as before, $s=\omega/\Delta$. The corresponding formulas for the
other two ensembles can be found in
\cite{simons93,altshuler94,ef-book}.

\section{Level statistics in a disordered sample: Diffusive
$\sigma$-model.} 
\label{s3}
\setcounter{equation}{0}

\subsection{Derivation of the diffusive $\sigma$-model.}
\label{s3.1}

Having introduced the main ideas and ingredients of the supersymmetry
formalism for the Gaussian Ensemble, we are prepared to proceed with
consideration of a much richer problem of a particle moving in a
random potential. The Hamiltonian has now the form\footnote{We set
$\hbar=1$.}
\begin{equation}
\hat{H}={\hat{\bf p}^2\over 2m}+ U({\bf r})\ ,\qquad
\hat{\bf p}=-i\nabla
\label{e3.1}
\end{equation}
where $U({\bf r})$ is a disorder potential. For simplicity we choose
it to be of the white-noise type (this is not essential for the
derivation of the $\sigma$-model)
\begin{equation}
\langle U({\bf r})U({\bf r'})\rangle={1\over
2\pi\nu\tau}\delta({\bf r}-{\bf r'})\ , 
\label{e3.2}
\end{equation}
where $\tau$ is the mean free time. We will assume that the time
reversal invariance is broken (say, by weak homogeneous or random
magnetic field) and restrict our consideration to the (technically
simpler) case of the unitary symmetry.\footnote{We will use a somewhat
sloppy (but convenient) terminology and refer to systems having the
symmetry of GOE, GUE, and GSE as to systems of orthogonal, unitary, and
symplectic symmetry, respectively. For a disordered electronic
problem, the orthogonal symmetry corresponds to purely potential
scattering in the absence of magnetic fields (both time reversal and
spin-rotation invariance preserved), the symplectic symmetry to the
spin-orbit scattering (spin-rotation invariance broken, but time
reversal preserved), while the unitary symmetry class is realized in
the presence of homogeneous or random magnetic field breaking the time
reversal invariance \cite{gorel,efetov83,ef-book}.}
The DOS and the Green's functions 
are given by a direct generalization of
Eqs.~(\ref{e1.3})--(\ref{e1.5}),  
\be
\label{e3.3}
\nu(E)={1\over 2\pi i V}\int {\rm d}^d{\bf r}\left[G_{\rm A}^E({\bf
r},{\bf r})- G_{\rm R}^E({\bf r},{\bf r})\right]\ ,
\ee
\begin{equation}
G^E_{\rm R,A}({\bf r}_1,{\bf r}_2)=\langle
{\bf r}_1|(E-\hat{H}\pm i\eta)^{-1}|{\bf r}_2\rangle\ ,\qquad \eta\to
+0\ .
\label{e3.4}
\end{equation}
In the two-level correlation function (\ref{e1.39}) the 
$\langle G_{\rm R}G_{\rm R}\rangle$ and
$\langle G_{\rm A}G_{\rm A}\rangle$ terms are again trivial (decouple
in the product of averages), and all the non-trivial information is
contained in the $\langle G_{\rm R}G_{\rm A}\rangle$
terms. Introducing the supervector field
\be
\label{e3.5}
\Phi({\bf r})=\left\lgroup\begin{array}{c} 
S_1({\bf r}) \\ \chi_1({\bf r})\\S_2({\bf r}) \\ \chi_2({\bf r})
\end{array}\right\rgroup\ ,
\ee
we write, in analogy with (\ref{e1.46}), (\ref{e1.47}), the
corresponding product of the Green's functions as a functional
integral,
\begin{eqnarray}
&&G_{\rm R}^{E+\omega/2}({\bf r}_1,{\bf r}_1) 
G_{\rm A}^{E-\omega/2}({\bf r}_2,{\bf r}_2 )
= \int {\rm D}\Phi\,
{\rm D}\Phi^\dagger S_1({\bf r}_1) S_1^*({\bf r}_1) 
S_2({\bf r}_2)S_2^*({\bf r}_2)
\nonumber\\&&\ \ \ \times 
\exp\left\{i\int {\rm d}^d{\bf r}\, \Phi^\dagger({\bf r}) L
[E+(\omega/2+i\eta)\Lambda-\hat{H}] \Phi({\bf r})\right\}\ .
\label{e3.6}
\end{eqnarray}
Averaging over the disorder with the correlator (\ref{e3.2}) produces
a quartic term,
\be
\label{e3.7}
\left\langle\exp\{i\int {\rm d}^d{\bf r}\,\Phi^\dagger({\bf r}) 
LU({\bf r})\Phi({\bf r})\}\right\rangle\ 
=\exp\left\{-{1\over 4\pi\nu\tau}\int {\rm d}^d{\bf
r}\,[\Phi^\dagger({\bf r}) 
L\Phi({\bf r})]^2\right\} ,
\ee
the Hubbard-Stratonovich decoupling of which requires the introduction
of an ${\bf r}$-dependent supermatrix field $R({\bf r})$,
\bea
\label{e3.8}
\exp\left\{-{1\over 4\pi\nu\tau}\int {\rm d}^d{\bf
r}\,[\Phi^\dagger({\bf r}) 
L\Phi({\bf r})]^2\right\} &=& \nonumber\nonumber
\int {\rm D}R\,  \exp \{-i\int {\rm d}^d{\bf r}\, 
\Phi^\dagger({\bf r})L^{1/2}R({\bf r})L^{1/2}\Phi({\bf r}) \nonumber\\ 
&-&\pi\nu\tau\int {\rm d}^d{\bf r}\,{\rm Str}R^2({\bf r})\}\ .
\eea
For any given ${\bf r}$ the matrix $R({\bf r})$ has the structure
specified by Eqs.~(\ref{e1.51})--(\ref{e1.53}). 
Substituting (\ref{e3.8}), (\ref{e3.7}) in (\ref{e3.6}) and performing
the $\Phi$-integral, we obtain, similarly to (\ref{e1.57}), an
integral over the $R$-field with the action
\be
\label{e3.9}
S[R]=\pi\nu\tau\int {\rm d}^d{\bf r}\,{\rm Str} R^2 + {\rm
Str}\ln\left(E+{\omega\over 2}\Lambda-{{\hat{\bf p}}^2\over
2m}-R\right)\ .
\ee
The corresponding saddle-point equation reads (for $\omega\to 0$)
\be
\label{e3.10}
R({\bf r})={1\over 2\pi\nu\tau}g({\bf r},{\bf r})\ ;\qquad
g=\left(E-{{\hat{\bf p}}^2\over 2m}-R\right)^{-1}\ .
\ee
We look first for a diagonal, ${\bf r}$-independent solution of
(\ref{e3.10}), $R={\rm diag}(q_1,q_2,q_3,q_4)$,
which has, in the weak-disorder regime $E\tau\gg 1$,
the form
\be
\label{e3.11}
q_j={1\over 2\pi\nu\tau}{\rm Re}
\langle G_{\rm R}({\bf r},{\bf r})\rangle\pm{i\over 2\tau}\ , 
\ee
with 
\be
\label{e3.12}
\langle G_{\rm R}({\bf r},{\bf r})\rangle=\int
{{\rm d}^d{\bf p}\over(2\pi)^d}(E-{\bf p}^2/2m+i/2\tau)^{-1}.
\ee
 The first term in
(\ref{e3.11}) gives a non-interesting constant real
contribution to $R$,
which can be absorbed\footnote{As it stands, Eq.~(\ref{e3.12}) has an
ultraviolet divergence at $p\to\infty$ in $d\ge 2$. This is related to
the white-noise (zero correlation length) character of the disorder
potential. Assuming a finite correlation length of the disorder would
introduce an UV-cutoff and make the integral (\ref{e3.12}) finite. In
a realistic model of a disordered
metal the correlation length is set by the screening
length, which is of order of the Fermi wave length. Introducing a
cutoff $\sim p_F=(2mE)^{1/2}$ in (\ref{e3.12}), we find that 
the first term in (\ref{e3.12}) is much smaller than $E$ in the considered
weak-disorder regime.} into the energy $E$.

The choice of signs in the second term of (\ref{e3.11}) is dictated by
the same considerations as in the GUE case (see the text below
Eq.~(\ref{e1.58})), leaving us with 
\be
\label{e3.13}
R_0={1\over 2\pi\nu\tau}{\rm Re}
\langle G_{\rm R}({\bf r},{\bf r})\rangle
-{i\over 2\tau}\Lambda \equiv \sigma -{i\over 2\tau}\Lambda.
\ee
The manifold of saddle-points is generated from $R_0$ by rotations
with matrices $T$ defined in (\ref{e1.52}), (\ref{e1.53}),
\be
\label{e3.14}
R=\sigma -{i\over 2\tau}T\Lambda T^{-1} \equiv 
\sigma-{i\over 2\tau}Q\ ,
\ee
with $Q$ from Eq.~(\ref{e1.60}). The set of matrices (\ref{e3.14})
constitutes a manifold of degenerate (at $\omega\to 0$) saddle points
of the action (\ref{e3.9}). 

We allow now the matrix $T$ (and consequently $Q$) to vary slowly in
space,
\be
\label{e3.15}
R({\bf r})=\sigma-{i\over 2\tau}Q({\bf r})\ ,\qquad
Q({\bf r})=T({\bf r})\Lambda T^{-1}({\bf r})\ .
\ee
While (\ref{e3.15}) with an ${\bf r}$-dependent $Q$ is not an exact
saddle-point any more, the fields of this form constitute the
low-lying excitations for the functional integral 
$\int {\rm D}R \ldots e^{-S[R]}$ with the action
(\ref{e3.9}). Performing the gradient expansion of the second term in
Eq.~(\ref{e3.9}) for the configurations defined by
Eq.~(\ref{e3.15})(and also expanding in $\omega$ up to the linear
term), we find the action for these low-lying modes
\be
\label{e3.16}
S[Q]={\pi\nu\over 4}\int {\rm d}^d{\bf r}\,{\rm Str}
[-D(\nabla Q)^2-2i\omega\Lambda Q]\ ,
\ee
where $D=v_F^2\tau/d$ is the diffusion constant (here
$v_F=(2E/m)^{1/2}$ is the particle velocity and $d$ the spatial
dimensionality). 
The two-level correlation function is thus reduced to a functional
integral over the $Q$-matrix field slowly varying on the coset space,
\be
\label{e3.17}
R_2(\omega)=\left({1\over 4V}\right)^2 \mbox{Re} \int {\rm D}Q({\bf r}) 
\left[\int {\rm d}^d{\bf r}\,\mbox{Str} Q\Lambda k\right]^2 e^{-S[Q]}\ ,
\ee
where $k={\rm diag}(1,-1,1,-1)$, i.e $k$ is equal to 1 $(-1)$ in the
boson-boson (resp. fermion-fermion) block. 

The field theory characterized by the action (\ref{e3.16})  
is called a ($d$-dimensional) non-linear $\sigma$-model. It
characterizes the low-frequency long-wavelength physics of the
original electronic problem. The fact that such a matrix
$\sigma$-model in the replica limit $n\to 0$ is the effective field
theory for the problem of a particle in a random potential was first
realized by Wegner \cite{wegner79}; a derivation (in the replica
formulation) was given by Sch\"afer and Wegner \cite{schaefer80}
and by Efetov, Larkin, and Khmelnitskii \cite{elk}. 
The supersymmetric version of the model was presented by Efetov
\cite{efetov83}. To elucidate the physical content of the $\sigma$-model
(\ref{e3.16}), it is instructive to draw a parallel with a model of
classical magnetic moments with a ferromagnetic interaction in an
external magnetic field,
\be
\label{e3.18}
H[{\bf S}]=-\sum_{\bf rr'}J_{\bf rr'}{\bf S}({\bf r}){\bf S}({\bf r'})
-\sum_{\bf r}{\bf B}{\bf S}({\bf r})\ ,
\ee
where ${\bf S}({\bf r})$ are $n$-component vectors with ${\bf
S}^2({\bf r})=1$. At low temperatures, only the low-energy sector is
important, which corresponds to a slowly varying  vector ${\bf S}({\bf
r})$. The Hamiltonian (\ref{e3.18}) is then reduced to a vector
$\sigma$-model,
\be
\label{e3.19}
H[{\bf S}]=\int {\rm d}^d {\bf r}\,\left[{\kappa\over 2}
(\nabla {\bf S}({\bf r}))^2 - {\bf B}{\bf S}({\bf r})\right]\ ,
\ee
where $\kappa$ is the spin stiffness. Let us now demonstrate the
analogies between (\ref{e3.16}) and (\ref{e3.19}). The $n$-component
unit vector ${\bf S}$ sweeps the $(n-1)$-dimensional sphere $S^{n-1}$
isomorphic to the coset space $O(n)/O(n-1)$. If the external magnetic
field ${\bf B}$ is directed, say, along ${\bf e}_1=(1,0,\ldots,0)$,
then ${\bf e}_1$ plays in (\ref{e3.19}) the same role as
$\Lambda$ in (\ref{e3.16}). Indeed, $O(n-1)$ is the subgroup of $O(n)$
which does not rotate ${\bf e}_1$; for this reason the
space of matrices from $O(n)$ which generate different vectors when
acting  on ${\bf e}_1$ is the coset space
$O(n)/O(n-1)$. Fully analogously, $U(1|1)\times U(1|1)$ is the
subgroup of those matrices from  $U(1,1|2)$ which do not rotate (i.e
commute with) $\Lambda$, so that the manifold of matrices $Q$ is
generated by rotations $T$ belonging to the coset space 
$U(1,1|2)/U(1|1)\times U(1|1)$. Furthermore, while the first term of
(\ref{e3.19}) is invariant with respect to a global rotation of ${\bf
S}({\bf r})$, the second term breaks this $O(n)/O(n-1)$
symmetry. In the ferromagnetic phase the symmetry is broken
spontaneously  at $B\to 0$; the corresponding Goldstone modes are the
spin waves. To write their Hamiltonian explicitly, one should
represent $H$ in terms of independent degrees of freedom, {\it i.e.}
in terms of  
the transverse (with respect to ${\bf e}_1$) part ${\bf S}_\bot$ (the
longitudinal component being $S_1=(1-{\bf S}_\bot^2)^{1/2}$),
\be
\label{e3.20}
H[{\bf S}_\bot]=  {1\over 2}\int {\rm d}^d{\bf r}\,
\left[\kappa[\nabla{\bf S}_\bot({\bf r})]^2 + B{\bf S}_\bot^2({\bf r})
+O({\bf S}^4_\bot({\bf r}))\right]\ ,
\ee
with the last term describing interaction of the spin waves. 
Equation (\ref{e3.20}) implies that the
correlation function of the transverse magnetization has the
Goldstone-type low-momentum behavior,
\be
\label{e3.21}
\int {\rm d}^d{\bf r}\,e^{-i{\bf qr}}\langle {\bf S}_\bot(0)
{\bf S}_\bot({\bf r})\rangle\propto {1\over \kappa {\bf q}^2 + B}\ .
\ee
Comparing (\ref{e3.19}) with
(\ref{e3.16}), we see that the frequency $\omega$ plays, in the case of
a disordered electronic system, the same role of a symmetry breaking
field as the magnetic field ${\bf B}$ for a ferromagnet. The soft
modes for (\ref{e3.16}) are the diffusion modes.
To write their action explicitly, one should choose some
parametrization of $Q$ in terms of independent degrees of freedom. Two
parametrizations, most often used for perturbative calculations,
are
\bea
\label{e3.22}
Q & = & (1-W/2)\Lambda(1-W/2)^{-1} = \Lambda(1+W/2)(1-W/2)^{-1}
\nonumber\\
& = & \Lambda\left(1+W+{W^2\over 2}+{W^3\over 4}
+{W^4\over 8}+\ldots\right) 
\eea
and 
\be
\label{e3.23}
Q=\Lambda\left(W+\sqrt{1+W^2/2}\right)
=\Lambda\left(1+W+{W^2\over 2}-{W^4\over 8}+\ldots\right)\ ,
\ee
with $W$ having the off-diagonal block structure in the RA space,
\be
\label{e3.24}
W=\left\lgroup\begin{array}{cc} 
0     &      W_{12}    \\
W_{21} &     0
\end{array}\right\rgroup\ ;\qquad W_{12}^\dagger = kW_{21}\ .
\ee
In either case the action (\ref{e3.16}) takes the form
\be
\label{e3.25}
S[W]={\pi\nu\over 4}\int {\rm d}^d{\bf r}\,{\rm Str}[D(\nabla W)^2-i\omega W^2
+ O(W^3)]\ ,
\ee
with the last term on the r.h.s. describing the interaction of the
diffusion modes. The analog of (\ref{e3.21}) is the diffusion
propagator (see below),
\be
\label{e3.26}
\int {\rm d}^d{\bf r}\,e^{-i{\bf qr}}\langle {\rm Str} k W_{12}(0) k
W_{21}({\bf r})\rangle\sim {1\over \pi\nu(D{\bf q}^2-i\omega)}\ .
\ee
After this digression we return to the analysis of the two-level
correlation function (\ref{e3.17}).

\subsection{Reduction to the 0D $\sigma$-model: Universal limit.}
\label{s3.2}

According to (\ref{e3.25}), the ``energies'' (eigenvalues of the
quadratic form of the action) of the diffusion modes are
given by
\be
\label{e3.27}
D{\bf q}_\mu^2-i\omega \equiv \epsilon_\mu-i\omega\ ,
\ee
with ${\bf q}_\mu$ being the allowed values of wave vectors. One can
show that in an isolated sample the boundary condition for the
$Q$-field is\footnote{To understand the physical meaning of
(\ref{e3.28}), one should remember that $Q$ was introduced as a field
conjugate to the product $\Phi\Phi^\dagger$, i.e. it is a kind of a
density field. Equation (\ref{e3.28}) is thus the condition of the
absence of current through the sample boundary, which is precisely
what should be imposed on a boundary with vacuum or with an
insulator.}  
\be
\label{e3.28}
\nabla_{\bf n}Q|_{\rm boundary}=0\ ,
\ee
where $\nabla_{\bf n}$ is the normal derivative. If we consider a
rectangular sample of a size 
$L_1\times \ldots \times L_d$, the wave vectors
${\bf q}_\mu$ are quantized according to 
\be
\label{e3.29}
{\bf q}_\mu=\pi\left({n_1\over L_1},\ldots,{n_d\over L_d}\right)\ ,\qquad 
n_i=0, 1, 2,\ldots.
\ee
For periodic boundary conditions, which are often used by
theoreticians, we have instead
\be
\label{e3.30}
{\bf q}_\mu=2\pi\left({n_1\over L_1},\ldots,{n_d\over L_d}\right)\ ,\qquad 
n_i=0,\pm 1,\pm 2,\ldots.
\ee
In either case, one can distinguish the zero mode with ${\bf q}_0=0$
(and thus $\epsilon_0=0$) from all other modes with $\epsilon_\mu\ge
D(\pi/L)^2$, where $L={\rm max}\{L_1,\ldots,L_d\}$. The energy
$E_c=D/L^2$ is called the Thouless energy; it is of the order of the
inverse time of diffusion through the sample. Equation (\ref{e3.27})
suggests that for $\omega\ll E_c$ one can neglect all non-zero modes
when calculating the level correlation function. This approximation
used by Efetov \cite{efetov83} is known as the zero-mode
approximation. As a result, the functional integral (\ref{e3.17})
reduces to an integral over a single supermatrix $Q$, acquiring
the form of the correlation function of the 0D
$\sigma$-model, Eq.~(\ref{e1.61}). Therefore, in the zero-mode
approximation the level statistics is described by the 0D
$\sigma$-model and thus has (according to Sec.~\ref{s2.4}) the RMT
form (\ref{e1.71}). 

For systems with preserved time-reversal invariance the same
consideration leads \cite{efetov83,ef-book} to the $\sigma$-models of
the orthogonal or symplectic symmetry (depending on the presence of
spin-dependent scattering). In the zero-mode approximation they reduce
to corresponding 0D $\sigma$-models yielding the level statistics
(\ref{e1.72}), (\ref{e1.73}) of Gaussian ensembles of the
corresponding symmetries.

In order for the zero-mode approximation to make sense, the Thouless
energy $E_c$ should be much larger than the mean level spacing
$\Delta$. It is easy to see that the ratio $g=2\pi E_c/\Delta$ is the
dimensionless (measured in units of $e^2/h$) conductance of the sample
(for a current flowing in the direction of $L$). Therefore, the above
condition is equivalent to $g\gg 1$; we will term the corresponding
situation ``metallic regime'' (or ``weak localization regime''). 
The opposite case, $g\lesssim 1$,
corresponds to the strong localization of a particle and is
realized in 2D or 3D if the disorder is not weak\footnote{Strictly
speaking, in 2D even for a weak disorder all states are localized, but
the corresponding localization length is exponentially large.},
$E\tau\sim 1$, or in sufficiently long systems of quasi-1D geometry
(see below). The notion of the dimensionless conductance plays a
central role in the scaling theory of localization, see \cite{leeram}
for review.

\subsection{Deviations from universality}
\label{s3.3}

\subsubsection{Perturbation theory.}
\label{s3.3.1}

Deviations from the universal (RMT) behavior are due to the non-zero
diffusion modes. The first calculation of the level
statistics in a diffusive grain taking into account the non-zero modes
was performed by Altshuler and Shklovskii \cite{ashk}. They used
the 
perturbation theory, which amounts to keeping only the terms of the
leading order in $W$ in the expansion of the preexponential factor and
of the action (\ref{e3.25}) in Eq.~(\ref{e3.17}). This yields
(for the unitary symmetry)
\bea
\label{e3.31}
R_{2,{\rm AS}}(\omega)-1 & \simeq &
{1\over (4V)^2} {\rm Re}\int {\rm D}W({\bf r})
\left[\int {\rm d}^d{\bf r}\,{\rm Str}Wk\right]^2 \nonumber \\
&& \qquad \times
\exp\left\{{\pi\nu\over 4}\int {\rm d}^d{\bf r}\,
{\rm Str}W[D\nabla^2+i\omega]W\right\} \nonumber \\
& = & {\Delta^2\over 2\pi^2}{\rm Re}
\sum_\mu{1\over (\epsilon_\mu-i\omega)^2}.
\eea
The same result is obtained for the other symmetry classes, with an
additional factor $2/\beta$, where $\beta$ is the commonly used
parameter equal to 1, 2, and 4 for the orthogonal, unitary, and
symplectic symmetry, respectively.

We analyze first the perturbative expression (\ref{e3.31}) in the
limit $\omega\ll E_c$ (or, equivalently, $s\equiv\omega\Delta\ll g$),
when the zero-mode approximation applies. Keeping only the term with
$\epsilon_0=0$ in (\ref{e3.31}) yields
\be
\label{e3.32}
R_{2,{\rm AS}}(\omega)-1 =-{1\over 2\pi^2 s^2}\ ,\qquad s\ll g\ .
\ee
To compare this perturbative formula with the exact result
(\ref{e1.71}), we note that the connected part of the RMT correlator
(\ref{e1.71}) can be decomposed into the smooth and the oscillatory
contributions,
\be
\label{e3.33}
-{\sin^2(\pi s)\over (\pi s)^2}= - {1\over 2(\pi s)^2}
+  {\cos(2\pi s)\over 2(\pi s)^2}\ .
\ee
We see that the Altshuler-Shklovskii calculation reproduces the smooth
term, but fails to give the oscillatory contribution. This is because
their method is perturbative in $1/s$, while the oscillations
$\propto\cos(2\pi s)$ are of non-perturbative (in $1/s$)
character. Furthermore, the Altshuler-Shklovskii result gives no
information about the actual small-$s$ behavior of $R_2(\omega)$. 

We consider now the opposite regime, $\omega\gg E_c$ ($s\gg g$), in
which the summation in (\ref{e3.31}) can be replaced by integration,
yielding\footnote{We assume the spatial dimensionality $d<4$
(otherwise, the sum in (\ref{e3.31}) has an ultraviolet
divergence). We also note that in 2D the Altshuler-Shklovskii
calculation gives the coefficient $c_2=0$; a more careful consideration
\cite{kraler95} taking into account higher order terms of the
perturbative expansion yields $c_2\sim 1/g^2$ for the unitary and
$c_2\sim 1/g$ for the orthogonal and symplectic symmetry.}
\be
\label{e3.34}
R_{2,{\rm AS}}(\omega)-1 \simeq c_d g^{-d/2}s^{d/2-2}\ , \qquad s\gg
g\ ,
\ee
with a numerical coefficient $c_d$.
Therefore, in the domain $s\gg g$ the level statistics differs
completely from its universal (RMT) form. The slower (compared to RMT)
decay of DOS correlations at $s\gg g$ corresponds to the diffusive
motion of the particle at times $t\ll t_D\sim E_c^{-1}$. In contrast,
at long times, $t\gg t_D$, the trajectory covers ergodically the whole
sample volume, and the correlations acquire their universal form.

It should be mentioned here that Altshuler and Shklovskii used in fact
not the $\sigma$-model formalism but the impurity diagram technique,
with the ladder diagrams representing the diffusion modes.
This diffuson-cooperon diagrammatics (completely equivalent to the
perturbative expansion of the $\sigma$-model) is by now a standard 
tool, which has allowed to discover and to study a number of remarkable
effects in mesoscopic physics, in particular the weak
localization, the Altshuler-Aronov effect of interplay of the Coulomb
interaction and disorder, and the universal conductance
fluctuations. Several excellent reviews of these subjects are
available, see Refs.~\cite{leeram,alar,lsf,simons93}. However, as the
above discussion shows, the perturbative methods are not sufficient
for studying the statistics of energy levels (and also of
eigenfunctions, see Sec.~\ref{s4}), and the non-perturbative
$\sigma$-model approach has to be used.

\subsubsection{Deviations from universality at $\omega\ll E_c$.}
\label{s3.3.2}

We return now to the region $s\ll g$ and consider deviations from the
universal (RMT) behavior. The method that allows us to 
calculate such deviations from the universality 
was developed in \cite{km} and can be outlined as follows. 
We first decompose $Q$ (in a proper way taking into account the
non-linear constraint $Q^2=1$) into the
constant part $Q_0$ and the contribution $\tilde{Q}$ of higher modes
with non--zero momenta. Then we 
integrate out all non-zero modes. This can be done perturbatively
provided the dimensionless 
conductance $g\gg 1$. As a result,
we get an integral over the matrix $Q_0$ only, which has to be
calculated non-perturbatively. 

We begin by presenting  the correlator $R_2(\omega)$ in the form
\begin{eqnarray}
&& R_2(\omega)={1\over (2\pi i)^2} {\partial^2\over \partial u^2}\int
{\rm D}Q \exp\{-S[Q]\}|_{u=0}\ ,  \nonumber\\
&& S[Q]=-{\pi\nu D\over 4}\int \mbox{Str}(\nabla Q)^2
+\tilde{s}\int \mbox{Str}\Lambda Q +\tilde{u}\int\mbox{Str} Q\Lambda
k\ ,
\label{e2.6}
\end{eqnarray}
where $\tilde{s}=\pi s/2iV$, $\tilde{u}=\pi u/2iV$. Now we
decompose $Q$ in the following way:
\begin{equation}
Q({\bf r})=T_0\tilde{Q}({\bf r})T_0^{-1}\ ,
\label{e2.7}
\end{equation}
where $T_0$ is a spatially uniform matrix from the coset space
and $\tilde{Q}$ describes all modes with non-zero momenta. 
When $\omega\ll E_c$, the matrix $\tilde{Q}$ fluctuates only weakly
near the 
origin $\Lambda$ of the coset space. In the leading order, 
$\tilde{Q}=\Lambda$, thus reducing (\ref{e2.6}) to the 0D 
$\sigma$--model, which leads to the Wigner--Dyson result (\ref{e1.71}). 
To find the corrections, we should expand the matrix $\tilde{Q}$ around
the origin $\Lambda$. We use the parametrization (\ref{e3.22}),
\be
\tilde{Q}=\Lambda(1+W/2)(1-W/2)^{-1} 
=\Lambda\left(1+W+{W^2 \over 2}+{W^3\over 4}+\ldots\right)\ .
\label{e2.8}
\ee
To keep the number of degrees of freedom unchanged, we should exclude
the zero mode from $\tilde{Q}$, which is achieved by the constraint 
$\int {\rm d}^d{\bf r}\,W=0$. 
Substituting the expansion (\ref{e2.8}) into Eq.~(\ref{e2.6}), we get 
\begin{eqnarray}
&& S = S_0[Q_0]+ S_W[W]+S_1[Q_0,W]\ ;\nonumber\\
&& S _0=\int\mbox{Str}\left[\tilde{s}Q_0\Lambda+ 
\tilde{u}Q_0\Lambda k\right]\ ,               \nonumber\\
&& S_W={\pi\nu D\over 4}\int \mbox{Str}(\nabla
W)^2\ ,  \nonumber\\
&& S _1={1\over 2}\int\mbox{Str}[\tilde{s}U_0\Lambda W^2 +
\tilde{u}U_{0k}\Lambda W^2]+O(W^3)\ ,
\label{e2.10}
\end{eqnarray}
where $Q_0=T_0\Lambda T_0^{-1}$,
$U_0=T_0^{-1}\Lambda T_0$, $U_{0k}=T_0^{-1}\Lambda k T_0$.
Let us define $ S _{\rm eff}[Q_0]$ as a result of elimination of the 
fast (non-zero) modes:
\begin{equation}
e^{-{S}_{\rm eff}[Q_0]}=e^{-{S}_0[Q_0]}\langle e^{-
S_1[Q_0,W]+\ln J[W]} \rangle_W\ ,
\label{e2.11}
\end{equation}
where $\langle\ldots\rangle_W$ denote the integration over $W$ with
the Gaussian weight $\exp\{-S_W[W]\}$, and 
$J[W]$ is the Jacobian of the transformation (\ref{e2.7}),
(\ref{e2.8}) from the variable $Q$ to $\{Q_0,W\}$ (the Jacobian does
not contribute to the leading order correction calculated here, but is
important for higher-order calculations \cite{fm95a,kravtsov97b}). 
Expanding
up to the order $W^4$, we get
\begin{equation} 
{S}_{\rm eff}={S}_0+\langle {S}_1\rangle -
{1\over 2}\langle{S}_1^2\rangle + {1\over 2} \langle{S}_1\rangle^2
+\ldots\ .
\label{e2.12}
\end{equation}
The integral over the fast modes $W$ can be calculated now using the
Wick theorem with the contraction rules induced by the action $S_W$,
\begin{eqnarray}
&&\langle\mbox{Str} W({\bf r})PW({\bf r'})R\rangle=\Pi({\bf r},{\bf r'})
(\mbox{Str}P\mbox{Str}R-\mbox{Str}P\Lambda\mbox{Str}R\Lambda)\ ;\nonumber\\
&&\langle\mbox{Str}[W({\bf r})P]\mbox{Str}[W({\bf r'})R]\rangle=
\Pi({\bf r},{\bf r'})
\mbox{Str}(PR-P\Lambda R\Lambda),
\label{e2.13}
\end{eqnarray}
where $P$ and $R$ are arbitrary supermatrices.
The diffusion propagator $\Pi$ is the solution of the diffusion
equation
\begin{equation} \label{corrrev_diff} 
- D \nabla^2\Pi({\bf  r}_1,{\bf  r}_2) = (\pi \nu)^{-1}
[\delta({\bf  r}_1 - {\bf  r}_2)-V^{-1}]   
\end{equation}
with the Neumann boundary condition (normal derivative equal to
zero at the sample boundary) and can be presented in the form
\begin{equation}
\Pi({\bf r}, {\bf r'})={1\over \pi\nu}\sum_{\mu;\ \epsilon_\mu\ne 0}
{1\over \epsilon_\mu} \phi_{\mu}({\bf r})\phi_{\mu}({\bf r'})\ ,
\label{e2.13a}
\ee
where $\phi_{\mu}({\bf r})$ 
are the eigenfunctions of the diffusion operator $-D\nabla^2$
corresponding to the eigenvalues $\epsilon_\mu$ (equal to $D{\bf q}^2$
for a rectangular geometry). As a result, we find
\begin{eqnarray}
&&\langle{S}_1\rangle=0\ ,\nonumber\\
&&\langle{S}_1^2\rangle=
{1\over 2}\int {\rm d}^d{\bf r}{\rm d}^d{\bf r'} 
\Pi^2({\bf r},{\bf r'})
(\tilde{s}\mbox{Str}Q_0\Lambda+\tilde{u}\mbox{Str}Q_0\Lambda k)^2\ .
\label{e2.14}
\end{eqnarray}
Substitution of Eq.~(\ref{e2.14}) into Eq.~(\ref{e2.12}) yields
\begin{eqnarray}
&&\hspace{-1cm}
{S}_{\rm eff} [Q_0]={\pi \over 2i}s\mbox{Str}Q_0\Lambda+{\pi\over 2i} u
\mbox{Str} Q_0\Lambda k + A {\pi^2\over 16} (s\mbox{Str} Q_0\Lambda +
u\mbox{Str}Q_0\Lambda k)^2  \ ; \nonumber\\
&& \hspace{-1cm}
A= {1\over V^2}\int {\rm d}^d{\bf r} {\rm d}^d{\bf r'}
\Pi^2({\bf r}, {\bf r'}) = {1\over \pi^2}\sum_{\mu\ne
0}\left({\Delta\over\epsilon_\mu}\right)^2= {4a_d\over g^2}\ ,
\eea
where $a_d$ is a numerical coefficient depending on the spatial
dimensionality $d$ and on the sample geometry. 
Assuming a cubic sample with hard-wall boundary conditions, we find 
\be
a_d={1\over \pi^4}
\sum_{\begin{array}{c}n_1,\ldots,n_d=0\\n_1^2+\ldots +n_d^2>0 
\end{array}}^{\infty}{1\over (n_1^2+\ldots+n_d^2)^2}\ ,
\label{e2.15}
\ee
yielding for $d=1,2,3$ the values $a_1=1/90\simeq 0.0111$, 
$a_2\simeq 0.0266$, and $a_3\simeq 0.0527$
respectively.\footnote{While speaking about $d=1$, we mean a sample of
a quasi-1D geometry, i.e. either a 2D strip $b\times L$ with $b\ll L$
or a 3D wire $b_1\times b_2\times L$ with $b_1,b_2\ll L$. For
a strictly 1D sample (a chain) the diffusive $\sigma$-model is not
applicable.}
In the case of a cubic sample with periodic boundary conditions
we get instead 
\begin{equation}
a_d={1\over (2\pi)^4}
\sum_{\begin{array}{c}n_1,\ldots,n_d=-\infty\\n_1^2+\ldots +n_d^2>0 
\end{array}}^{\infty}{1\over (n_1^2+\ldots+n_d^2)^2}\ ,
\label{e2.15a}
\ee
so that $a_1=1/720\simeq 0.00139$, $a_2\simeq 0.00387$, and $a_3\simeq
0.0106$. Note that for $d<4$ the sum in Eqs.~(\ref{e2.15}),
(\ref{e2.15a}) converges, so that no ultraviolet cutoff is needed. 

Using now Eq.(\ref{e2.6}) and calculating the remaining integral over
the supermatrix $Q_0$, we finally get 
the following expression for the
correlator to the $1/g^2$ order:
\begin{equation}
R_2(s)=1+\delta(s)
-\frac{\sin^2(\pi s)}{(\pi s)^2}+A\sin^2(\pi s)\ ; 
\qquad A=\frac{4a_d}{ g^2}\ .
\label{e2.16}
\end{equation}
The last term in Eq.(\ref{e2.16}) represents the sought correction 
of  order $1/g^2$ to the Wigner--Dyson distribution.

The important feature of Eq.(\ref{e2.16}) is that it relates
corrections to the smooth and oscillatory parts of the level
correlation function,
\be
\label{e2.16a}
\delta R_2(s)= A\sin^2\pi s={A\over 2}(1-\cos 2\pi s)\ .
\ee
While appearing naturally in the framework of the supersymmetric
$\sigma$-model, this fact is highly non-trivial from the point of view
of semiclassical theory \cite{berry1}, which represents the
level structure factor $K(\tau)$ (Fourier transform of $R_2(s)$) in
terms of a sum over periodic orbits. The smooth part of $R_2(s)$
corresponds then to the small-$\tau$ behavior of $K(\tau)$, which is
related to the properties of short periodic orbits. On the other hand,
the oscillatory part of $R_2(s)$ is related to the behavior of $K(\tau)$
in the vicinity of the Heisenberg time $\tau=2\pi$ ($t=2\pi/\Delta$
in dimensionful units), and thus to the properties of long periodic
orbits. We will return to the non-universal corrections to $K(\tau)$
below. 

The calculation presented above can be straightforwardly generalized
to the 
other symmetry classes. The result can be presented in a form valid
for all the three cases:
\begin{equation}
R_2^{(\beta)}(s)=\left(1+\frac{A}{2\beta}
\frac{d^2}{ds^2}s^2\right) R^{(\beta)}_{2,RMT}(s)\ ,
\qquad A=\frac{4a_d}{ g^2}\ ,
\label{e2.19}
\end{equation}
where $\beta=1 (2, 4)$ for the orthogonal (unitary, symplectic)
symmetry; $R^{(\beta)}_{\rm 2,RMT}$ denotes the corresponding 
RMT correlation function (\ref{e1.71})--(\ref{e1.73}).  

For $s\to 0$ the RMT correlation functions have the following
behavior: 
\begin{eqnarray}
&& R^{(\beta)}_{\rm 2,RMT}\simeq C_\beta s^{\beta}\ ,\qquad  s\to 0\ ; 
\nonumber\\
&& C_1={\pi^2\over 6}\ ,\ \ \  
C_2={\pi^2\over 3}\ ,\ \ \  C_4={(2\pi)^4\over 135}\ .
\label{e2.20}
\end{eqnarray}
As is clear from Eq.(\ref{e2.19}), the found correction does not change
the exponent $\beta$, but renormalizes the prefactor $C_\beta$:
\begin{equation}
R_2^{(\beta)}(s)=\left(1+\frac{(\beta+2)(\beta+1)}{2\beta} A
\right) C_\beta s^\beta\ ;\quad s\to 0
\label{e2.21}
\end{equation} 
The correction to $C_\beta$ is positive, which means that 
the level repulsion becomes weaker. This is related to a tendency of
eigenfunctions to localization with decreasing $g$.

\subsubsection{Stationary-point method.}
\label{s3.3.3}

Let us return now to the 
behavior of the level correlation function in its
high-frequency tail $s\gg g$. We have already discussed
the non-oscillatory part of $R_2(s)$ in
this region, see Eq.~(\ref{e3.34}). 
What is the fate of the oscillations in $R_2(s)$
in this regime? The answer to this question was given by Andreev and
Altshuler \cite{aa} who calculated $R_2(s)$ using the stationary-point
method for the $\sigma$-model integral (\ref{e3.17}) and treating the
zero mode and the higher modes on equal footing. Their crucial
observation was that on top of the trivial stationary point
$Q=\Lambda$ (expansion around which is just the usual perturbation
theory), there exists another one, $Q=k\Lambda$, whose vicinity
generates the oscillatory part of $R_2(s)$. (In the case of symplectic
symmetry there exists an additional family of stationary points, see
\cite{aa}). The saddle-point approximation\footnote{To avoid possible
confusion, we remind that the matrices of the manifold (\ref{e3.14})
are exact saddle-points of the action (\ref{e3.9}) ({\it i.e.} they
all have exactly the same action) for $\omega=0$
only. At finite $\omega$ this becomes a manifold of
quasi-saddle-points, with the action difference determined by the term
$(-i\pi s/2){\rm Str}Q\Lambda$, see Eqs.~(\ref{e1.61}) and
(\ref{e3.16}). The corresponding soft ($\sigma$-model) modes should be
contrasted to massive modes (describing fluctuations in $P_{1,2}$, see
Eq.~(\ref{e1.51})), which have been integrated out in the course of
derivation of the $\sigma$-model. Now, on this manifold of
almost-saddle-points $Q$ there are two true (even at non-zero
$\omega$) stationary points, namely, $Q=\Lambda$ and $Q=k\Lambda$. In
fact, we have already mentioned the existence of the second diagonal
saddle point below Eq.~(\ref{e1.58}). It is easy to see that the
choice of signs $s_2=-s_4=1$ produces there precisely the matrix
$Q=k\Lambda$.}  
of Andreev and Altshuler is valid for $s\gg 1$; at $1\ll s\ll g$ it
reproduces the above results of Ref.~\cite{km} (we remind that the
method of \cite{km}
works for all $s\ll g$). The result of \cite{aa} has the following
form:
\begin{eqnarray}
R_{\rm 2, osc}^{\rm U}(s)&=&{\cos 2\pi s\over 2\pi^2}D(s)\ ,
\label{e2.23} \\
R_{\rm 2,osc}^{\rm O}(s)&=&{\cos 2\pi s\over 2\pi^4}D^2(s)\ ,
\label{e2.23o} \\
R_{\rm 2,osc}^{\rm Sp}(s)&=&{\cos 2\pi s\over 4 }D^{1/2}(s)+
{\cos 4\pi s\over 32\pi^4}D^2(s)\ ,
\label{e2.23sp}
\end{eqnarray}
where $D(s)$ is the spectral determinant
\begin{equation}
D(s)={1\over s^2}\prod_{\mu\ne 0}
\left(1+{s^2\Delta^2\over\epsilon_\mu^2}\right)^{-1}.
\label{e2.24}
\end{equation}
The product in Eq.(\ref{e2.24}) goes over the non-zero eigenvalues
$\epsilon_\mu$ of the diffusion operator. This demonstrates again the 
relation between $R_{\rm 2,osc}(s)$ and the perturbative part
(\ref{e3.31}), which can be also expressed through $D(s)$,
\begin{equation}
R^{(\beta)}_{\rm 2,AS}(s)-1
=-{1\over 2\beta\pi^2}{\partial^2\ln D(s)\over \partial s^2}\ .
\label{e2.25}
\end{equation}
In the high-frequency region $s\gg g$ the spectral determinant is
found to have the following behavior: 
\begin{equation}
D(s)\sim\exp\left\{-{\pi\over \Gamma(d/2)d\sin(\pi
d/4)}\left({2s\over g}\right)^{d/2}\right\}\ ,
\label{e2.26}
\end{equation}
so that the amplitude of the oscillations vanishes exponentially with
$s$ in this region. 

Taken together, the results of \cite{km} and \cite{aa}
provide a complete description of the deviations of the
level correlation function from universality in the metallic regime
$g\gg 1$. They show that in the whole region of frequencies these
deviations are controlled by the classical (diffusion) operator
governing the dynamics in the corresponding classical
system. 

\subsection{Spectral characteristics related to $R_2(s)$.}
\label{s3.4}

\subsubsection{Spectral formfactor.}
\label{s3.4.1}

The spectral formfactor is defined as the Fourier transform of the
connected part $R_2^{\rm (c)}(s)=R_2(s)-1$ of the two-level correlation
function, 
\be
\label{e3.101}
K(\tau)=\int_{-\infty}^\infty R_2^{\rm (c)}(s)e^{is\tau}{\rm d}s\ .
\ee
By definition, $K(\tau)$ (as well as $R_2(s)$) is an even function, so
that it is sufficient to discuss it at $\tau>0$. In GUE it has the
form
\be
\label{e3.102}
K(\tau)=\left\{\begin{array}{ll}
\tau/2\pi\ , &\qquad 0\le\tau/2\pi\le 1 \\
1\ ,         &           \tau/2\pi\ge 1\ .
\end{array}
\right.
\ee
Let us analyze, what kind of corrections to $K_2(\tau)$ is implied by
the deviations of $R_2(s)$ from universality studied in
Sec.~\ref{s3.3}. For this purpose, let us use Eq.~(\ref{e2.16a}),
forgetting for a moment about the condition of its validity ($s\ll
g$). The Fourier transformation of (\ref{e2.16a}) then yields
\be
\label{e3.103}
\delta K(\tau)={A\pi\over 2}[2\delta(\tau)-\delta(\tau-2\pi)
-\delta(\tau+2\pi)]\ .
\ee
Taking now into account the existence of the cutoff $s\sim g$ for
(\ref{e2.16a}) leads to smearing of the $\delta$-functions in
(\ref{e3.103}) over an interval $\sim 1/g$ around $\tau=0$ and
$\tau=\pm 2\pi$, respectively. Thus, we conclude that
\be
\label{e3.104}
\delta K(\tau)=\delta K_0(\tau)+\delta K_{2\pi}(\tau)\ ,\qquad \tau>
0\ ,
\ee
where $\delta K_0(\tau)$ is located in the interval between $\tau=0$
and $\tau\sim 1/g$, while $\delta K_{2\pi}(\tau)$ is concentrated in
the interval of a width $\sim 1/g$ around $\tau=2\pi$. Since $\delta
R_2(0)=0$, the integrals of $\delta K_0(\tau)$ and  $\delta
K_{2\pi}(\tau)$ are equal up to a sign,
\be
\label{e3.105}
\int {\rm d}\tau\, \delta K_0(\tau) =-\int {\rm d}\tau\, 
\delta K_{2\pi}(\tau)\ . 
\ee
Furthermore, using the identity
$$
{{\rm d}^2\over {\rm d}s^2}\delta R_2(s)|_{s=0}=
-\int {\rm d}\tau\, \tau^2\delta K(\tau)\ ,
$$
we find
\be
\label{e3.106}
\int {\rm d}\tau\, \delta K_{2\pi}(\tau)\simeq -{A\over 4}\ ,
\ee
so that the correction around the Heisenberg time $\tau=2\pi$ is
negative. Since the decay of $R_{2,{\rm osc}}(s)$ at $s > g$ is
exponential, $\delta K_{2\pi}(\tau)$ is a smooth function. Using the
fact that it is essentially located in an interval of width $\sim
1/g$ and that $A\sim 1/g^2$, we conclude 
that the magnitude of $\delta K_{2\pi}(\tau)$ in the above interval is
$\sim 1/g$. The correction $K_{2\pi}(\tau)$ has thus both the
magnitude and the width of the order of $1/g$ and leads to a rounding
of the 
singularity in the spectral form-factor at the Heisenberg time
$\tau=2\pi$, as was first realized by Andreev and Altshuler
\cite{aa}. In the quasi-1D case, the spectral determinant $D(s)$,
Eq.~(\ref{e2.24}), and thus the correction $\delta K_{2\pi}(\tau)$ can
be calculated analytically, see \cite{aa}.

As to $\delta K_0(\tau)$, its small-$\tau$ behavior depends on spatial
dimensionality, in view of the
the Altshuler-Shklovskii ``tail'' (\ref{e3.34}). Taking the Fourier
transform, we find\footnote{In 3D the spectral formfactor
(\ref{e3.107}) seems to diverge as $\tau\to 0$. It should be taken
into account, however, that the above considerations are valid only
for frequencies $\omega$ corresponding to diffusive motion, {\it i.e.}
$\omega\ll \tau_{\rm e}^{-1}$, where $\tau_{\rm e}$ is the elastic
mean free path (denoted by $\tau$ in
Sec.~\ref{s3.1}). Correspondingly, the applicability of (\ref{e3.107})
is restricted by the condition $\tau\gtrsim \tau_e\Delta$.} 
\be
\label{e3.107}
\delta K_0(\tau)\sim {1\over g}(g\tau)^{1-d/2}\ ,\qquad \tau\lesssim
1/g\ .
\ee
For semiclassical treatment of the spectral form-factor
applicable for $\tau/2\pi\ll 1$ [where it  reproduces the formula
(\ref{e3.107})] see Ref.~\cite{argaman93}.

\subsubsection{Level number variance.}
\label{s3.4.2}

Consider the variance $\Sigma_2(s)$ of the number of energy levels in
a spectral window of a width $\delta E=s\Delta$ (the average
number of levels in this interval being equal to $s$). It is easy to see
that $\Sigma_2(s)$ is related to the two-level correlation function as
follows:
\be
\label{e3.108}
\Sigma_2(s)=\int_{-s}^s {\rm d}\tilde{s}\,(s-|\tilde{s}|)
R_2^{\rm (c)}(\tilde{s})\ .
\ee
The relation (\ref{e3.108}) can be also presented in the following way
\be
\label{e3.109}
{d\over ds}\Sigma_2(s)=\int_{-s}^s d\tilde{s}\,R_2^{\rm (c)}(\tilde{s})\ .
\ee
The level number variance is commonly used to characterize the
long-range behavior of spectral correlations.
The $1/s^2$ decay of the smooth part\footnote{The oscillatory part of
$R_2(s)$ is not important for the behavior of $\Sigma_2(s)$ at $s\gg
1$, because it gives a negligible contribution after integration
(\ref{e3.108}).} of the two-level correlation function at $s\gg 1$ in
RMT implies the $\ln s$ behavior of $\Sigma_2(s)$. In
particular, for GUE the large-$s$ asymptotics reads
\cite{mehta,bohigas} 
\be
\label{e3.110}
\Sigma_{2,{\rm RMT}}^{\rm U}(s)={1\over\pi^2}[1+\gamma+\ln(2\pi s)]\
,\qquad s\gg 1\ .
\ee
According to (\ref{e2.16a}), the correction to the RMT form of
$\Sigma_2(s)$ in a diffusive sample is small (and positive) at $s\ll
g$ and has the form
\be
\label{e3.111}
\delta\Sigma_2(s)={A\over 2}s^2=2a_d\left({s\over g}\right)^2\ , 
\qquad s\ll g\ .
\ee
On the other hand, at $s\gg g$ the level number variance is determined
by the Altshuler-Shklovskii behavior of $R_2(s)$ and is totally different
from its RMT form:
\be
\label{e3.112}
\Sigma_2(s)\sim \left({s\over g}\right)^{d/2}\gg\Sigma_{2,{\rm
RMT}}(s)\ ,\qquad s\gg g\ .
\ee

\section{Eigenfunction statistics.}
\label{s4}
\setcounter{equation}{0}

Not only the energy levels statistics but also the
statistical properties of wave functions are of considerable
interest. In the case of nuclear spectra (the statistical description
of which was the original motivation for the development of the RMT),
they determine fluctuations 
of widths and heights of the resonances \cite{porter,brody}.
A more recent growth of interest to statistical properties
of eigenfunctions in disordered and
chaotic systems has been motivated, on the experimental side, by the
possibility of fabrication of small systems (quantum dots)
with well resolved electron energy levels
\cite{sivan94,tarucha96,dotrev,ralph95}. Fluctuations in the tunneling
conductance of such a dot measured in recent experiments
\cite{chang96,folk96} are related to statistical properties of
wavefunction amplitudes
\cite{jsa,prigefii,alhassid96,bruus96}. Furthermore,  the
eigenfunction fluctuations determine the statistics of matrix
elements of the Coulomb interaction, which is  important for
understanding the properties of excitation and addition spectra of the
dot \cite{blanter,bmm1,agkl,agam97}. It is also worth mentioning that
the microwave cavity
technique allows one to observe experimentally spatial fluctuations of
the wave amplitude in chaotic and disordered cavities
\cite{stoeckmann,kudrolli,alt} (though in this case one considers the
intensity of a classical wave rather than of a quantum particle, all
the results are equally applicable).

\subsection{Random matrix theory}
\label{s4.1}

Within the RMT, the eigenfunction statistics has a very simple form in
the limit $N\gg 1$. The components of an eigenvector $\psi^{(i)}$ of a
matrix $\hat{H}$ from the Gaussian ensemble become then uncorrelated
Gaussian random numbers (real, complex, or quaternionic for $\beta=1$,
2, and 4, respectively) with the distributions
\be
\label{wv.1}
{\cal P}\left(\psi^{(i)}_k\right)\propto\exp\left\{-{N\beta\over 2}
|\psi^{(i)}_k|^2\right\}\ .
\ee
If we introduce the ``intensity''
\be
\label{wv.2}
y^{(i)}_k=N|\psi^{(i)}_k|^2
\ee
(we have chosen the normalization $\langle y\rangle=1$), its
distribution will then have the form of the $\chi^2$-distribution with
$\beta$ degrees of freedom. In particular, for GUE and GOE we
have\footnote{For GSE this consideration gives ${\cal
P}(y)=4ye^{-2y}$. Note, however, that in terms of the electronic wave
function this corresponds to defining  $y$ as the total (summed over
the spin projections) 
intensity, $y=V(|\psi_\uparrow|^2+|\psi_\downarrow|^2)$. For the
distribution of the spin-projected intensity $y=2V|\psi_\uparrow|^2$
the RMT result would have the same form (\ref{er3.2}) as for
GUE. We do not consider the symplectic symmetry below;
deviations from universality have in the symplectic symmetry
class a form similar to the results for the systems of the unitary
and the orthogonal symmetry.} 
\cite{porter,brody} 
\begin{eqnarray}
&&{\cal P}^{\rm U}(y)=e^{-y}\ , \label{er3.1} \\
&&{\cal P}^{\rm O}(y)={e^{-y/2}\over \sqrt{2\pi y}}\ . \label{er3.2}
\end{eqnarray}
Equation (\ref{er3.2}) is known as the Porter-Thomas distribution; it
was originally introduced to describe
fluctuations of widths and heights of resonances in nuclear spectra
\cite{porter}. 

\subsection{Eigenfunction statistics in terms of the supersymmetric
$\sigma$-model} 
\label{s4.2}

Theoretical study of the eigenfunction statistics in a $d$-dimensional
disordered system is again possible with making use of the
supersymmetry method \cite{mf93a,fm94a,fm95a,fe2}. We consider the
local intensity of an eigenfunction at some point ${\bf r}_0$ of the
sample, $y_i({\bf r}_0)=V|\psi_i^2({\bf 
r}_0)|$ (we have again normalized it to $\langle y\rangle =1$). In
order to calculate the distribution function ${\cal P}(y)$, we
introduce its moments,
\be
\label{wv.3}
I_q({\bf r}_0)=\Delta\langle\sum_i
|\psi_i({\bf r}_0)|^{2q}\delta(E-E_i)\rangle
\equiv \langle|\psi({\bf r}_0)|^{2q}\rangle_E\ .
\ee
The next step is to observe that $I_q({\bf r}_0)$ is related for
$q=2,3,\dots$ to the 
following product of the Green's functions\footnote{For $q=0$ and
$q=1$ the moments are 
trivial, $I_0({\bf r}_0)=1$, $I_1({\bf r}_0)=1/V$.}
\bea
\label{wv.4}
K_{l,m}({\bf r}_0,\eta)&=&
\langle {\bf r}_0|(E-\hat{H}+ i\eta)^{-1}|{\bf r}_0\rangle^l
\langle {\bf r}_0|(E-\hat{H}- i\eta)^{-1}|{\bf r}_0\rangle^m
\nonumber \\
& \equiv & 
[G^E_{\rm R}({\bf r}_0,{\bf r}_0)]^l
[G^E_{\rm A}({\bf r}_0,{\bf r}_0)]^m\ ,
\eea
where $l,m\ge 1$ and $l+m=q$.
Indeed, in the limit $\eta\to 0$ $\langle K_{l,m}\rangle$ possesses a
singularity  $\propto \eta^{1-l-m}$ determined by the probability
$\propto \eta$ to find an eigenvalue of $\hat{H}$ within a distance
$\sim\eta$ from $E$, in which case $K_{l,m}\propto \eta^{-l-m}$.
Extracting this singularity, we get
\be
\label{wv.5}
I_q({\bf r}_0)={\Delta\over 2\pi}i^{l-m}{(l-1)!(m-1)!\over (l+m-2)!}
\lim_{\eta\to +0}(2\eta)^{l+m-1}
\langle K_{l,m}({\bf r}_0,\eta) \rangle \ .
\ee
Note that at this stage we have for $q>2$ the freedom in choosing
$l$ and $m$ (constrained only by $l+m=q$). The final result will
not, of course, depend on this choice.

We express now the product of Green's functions as an integral over the
supervector field
\begin{eqnarray}
&&K_{l,m}({\bf r}_0,\eta)
= {i^{m-l}\over l!m!}\int D\Phi\,
D\Phi^\dagger [S_1({\bf r}_0) S_1^*({\bf r}_0)]^{l} 
[S_2({\bf r}_0)S_2^*({\bf r}_0)]^m
\nonumber\\&&\ \ \ \times 
\exp\left\{i\int {\rm d}^d{\bf r}\, \Phi^\dagger({\bf r}) L^{1/2}
(E+i\eta\Lambda-\hat{H})L^{1/2} \Phi({\bf r})\right\}\ .
\label{anom17}
\end{eqnarray}
Proceeding in the same way as in the case of the level correlation
function (Sec.~\ref{s2.1}), we represent the r.h.s. of
Eq.~(\ref{anom17}) in terms of a $\sigma$-model correlation function.
In the case of the unitary symmetry the results reads
\bea
\label{wv.6}
K_{l,m}({\bf r}_0,\eta) &=& (-i\pi\nu)^{l+m}\int {\rm D}Q\sum_j
\left(\begin{array}{c} l \\ j\end{array}\right)
\left(\begin{array}{c} m \\ j\end{array}\right) \nonumber\\
& \times &
Q_{\rm 11,bb}^{l-j}({\bf r}_0)Q_{\rm 22,bb}^{m-j}({\bf r}_0)
Q_{\rm 12,bb}^j({\bf r}_0)Q_{\rm 21,bb}^j({\bf r}_0)e^{-S[Q]}\ ,
\eea
where $S[Q]$ is the $\sigma$-model action,
\begin{equation}
S[Q]=-
\int {\rm d}^d{\bf r}\,\mbox{Str}\left[{\pi\nu D\over 4}(\nabla
Q)^2-\pi\nu\eta\Lambda Q\right]\ .
\label{anom_2}
\end{equation}
We remind that the two pairs of indices of the $Q$-matrices refer to
the retarded-advanced (1, 2) and the boson-fermion (b, f)
decomposition, respectively. 

Since the preexponential factor in (\ref{wv.6}) depends on the
$Q$-field at the point ${\bf r}_0$ only, it is convenient to
introduce the function $Y(Q_0)$ as the result of integrating out all
other degrees of freedom,
\begin{equation}
Y(Q_0)=\int_{Q({\bf r}_0)=Q_0} {\rm D}Q({\bf r})
\exp\{-S[Q]\}\ .
\label{anom_1}
\end{equation}
With this definition, Eq.(\ref{e3.4}) takes the form of an integral
over the single matrix $Q_0$,
\bea
\label{wv.7}
K_{l,m}({\bf r}_0,\eta) &=& (-i\pi\nu)^{l+m}\int {\rm d}\mu(Q_0)\sum_j 
\left(\begin{array}{c} l \\ j\end{array}\right)
\left(\begin{array}{c} m \\ j\end{array}\right) \nonumber \\
& \times &
Q_{\rm 0;11,bb}^{l-j}Q_{\rm 0;22,bb}^{m-j}
Q_{\rm 0;12,bb}^jQ_{\rm 0;21,bb}^j Y(Q_0)\ .
\eea
For invariance reasons, the function $Y(Q_0)$ turns out to be
dependent in the unitary symmetry case on the two eigenvalues $1\le
\lambda_{1}<\infty $ and $ -1\le \lambda_{2} \le 1$ only,
when the parametrization   (\ref{e1.62}), (\ref{e1.63}) for the matrix
$Q_0$ is used. 
Moreover, in the limit $\eta \to 0$ (at a fixed  value of the
system volume, and thus of the level spacing $\Delta$) only the
dependence on $\lambda_1$ persists, 
\begin{equation}
Y(Q_0)\equiv Y(\lambda_1,\lambda_2)\to 
Y_{\rm a}(2\pi{\eta\over\Delta}\lambda_1)\ ,
\label{anom_3}
\end{equation}
with relevant values of $\lambda_1$ being $\lambda_1\sim
\Delta/\eta\gg 1$. One can further show that in this asymptotic domain
$$
Q_{\rm 11,bb}Q_{\rm 22,bb}\simeq Q_{\rm 12,bb}Q_{\rm 21,bb}\ ,
$$
so that all terms in $\sum_j$ in Eq.~(\ref{wv.7}) are equal.
Evaluating the integral over all coordinates but $\lambda_1$  in the
parametrization (\ref{e1.62}), (\ref{e1.63}) of $Q_0$, we get 
\begin{equation}
I_q({\bf r}_0)={1\over V^q}q(q-1)\int {\rm d}z \,z^{q-2}
Y_{\rm a}(z).
\label{wv.8}
\end{equation}
Consequently, the distribution function of the eigenfunction intensity
is given by \cite{mf93a}
\begin{equation}
{\cal P}^{\rm U}(y)={{\rm d}^2 \over {\rm d}y^2} Y_{\rm a}(y)\ . 
\label{anom_4}
\end{equation}

In the case of the orthogonal symmetry, $Y(Q_0)\equiv
Y(\lambda_1,\lambda_2,\lambda)$, where 
$1\le \lambda_{1},\lambda_{2} <\infty $ and $ -1\le \lambda \le 1$.
In the limit $\eta\to 0$, the relevant region of values is 
$\lambda_1\gg\lambda_2,\lambda$, where
\begin{equation}
Y(Q_0)\to Y_{\rm a}(\pi{\eta\over\Delta}\lambda_1)\ .
\label{anom_3o}
\end{equation}
The distribution of eigenfunction intensities is expressed
in this case through the function $Y_{\rm a}$ as follows \cite{mf93a}:
\begin{eqnarray}
{\cal P}^{\rm O}(y)&=& {1\over\pi y^{1/2}}\int_{y/2}^{\infty}
{\rm d}z  (2z-y)^{-1/2}{{\rm d}^2\over {\rm d}z ^2} Y_{\rm a}(z)
\nonumber\\ 
&=& {2\sqrt{2}\over\pi y^{1/2}}{{\rm d}^2\over {\rm
d}y^2}\int_0^\infty {{\rm d}z \over 
z^{1/2}}  Y_{\rm a}(z+y/2) \ . 
\label{anom_4o}
\end{eqnarray}

In a metallic sample, typical configurations of the $Q$--field are
nearly constant in space, so that the zero-mode approximation is
expected to be a good starting point (see Sec.~\ref{s3.2}). It amounts
to approximating the 
functional integral (\ref{wv.6}), (\ref{anom_1}) by an integral
over a single supermatrix $Q$, yielding
\begin{equation}
Y_{\rm a}(z)\simeq e^{-z} \qquad ({\rm O,U})\ ,
\label{anom1}
\end{equation}
which reproduces [after substitution in (\ref{anom_4}),
(\ref{anom_4o})] the RMT results (\ref{er3.1}) and (\ref{er3.2}). 
Therefore, like for the level correlations, the zero mode
approximation yields the RMT results for the distribution of the
eigenfunction amplitudes. To calculate deviations from RMT, one has to
go beyond the zero-mode approximation and to evaluate the function
$Y_{\rm a}(z)$ determined by Eqs.~(\ref{anom_1}), (\ref{anom_3}) for a
$d$-dimensional diffusive system. In the case of a quasi-1D geometry
this can be done exactly via the transfer-matrix method
\cite{mf93a,fm94a}. The result depends crucially on the ratio $L/\xi$,
where $L$ is the system size and $\xi=2\pi\nu AD$ the localization
length ($A$ being the transverse cross-section of the sample). The two
limiting cases correspond to the metallic regime $L\ll \xi$ (with the
dimensionless conductance $g=\xi/L\gg 1$) and to the strong
localization regime $L\gg\xi$. For higher $d$, the exact solution is not
available any more, 
and one should rely on approximate methods. Corrections to the ``main
body'' of the distribution can be found by treating the non-zero modes
perturbatively \cite{fm95a}, while the asymptotic ``tail'' can be
found \cite{fe2} via an instanton method \cite{mk1}. For the quasi-1D
geometry 
these approximate methods reproduced the results obtained earlier
\cite{mf93a}  from the exact solution. 

%We find it more convenient here
%to present first the results for the metallic regime (for arbitrary
%$d$) and then to return to the quasi-1D geometry and to consider,
%in particular, the regime of strong localization.

\subsection{Quasi-one-dimensional geometry}
\label{s4.3}

\subsubsection{Exact solution of the $\sigma$-model}
\label{s4.3.1}

In the case of quasi-1D geometry an exact solution of the
$\sigma$-model is possible due to the
transfer-matrix method. The idea of the method, quite general for 
one-dimensional problems, is in reducing the functional integral
of the type (\ref{anom_1})  to solution of a differential
equation. This is completely analogous to  constructing  the
Schr\"odinger equation from the quantum-mechanical Feynman path
integral. In the present case, the role of the time is played by the
coordinate along the wire, while the role of the particle coordinate
is played by the supermatrix $Q$. In general, at a finite value of the
frequency $\eta$ in Eq.~(\ref{anom_2}) (more precisely, $\eta$ plays
the role of imaginary frequency), the corresponding differential
equation is too complicated and cannot be solved analytically.
However, a simplification appearing in the limit $\eta\to 0$, when
only the non-compact variable $\lambda_1$ survives, allows to find an
analytical solution \cite{fm91,mf93a,fm94a} 
of the 1D $\sigma$-model\footnote{Let
us stress that we consider a sample with the hard-wall (not
periodic) boundary conditions in the longitudinal direction, i.e a wire
with two ends (not a ring).}.

There are several different microscopic models which can be
mapped onto the 1D supermatrix $\sigma$-model. First of all, this is a
model of a particle in a random potential (discussed above) in the
case of a quasi-1D sample geometry. Then one can neglect the transverse
variation of the $Q$-field in the $\sigma$-model action, thus reducing
it to the 1D form \cite{eflar,efetov83}. Secondly, the random
banded matrix (RBM) model  (already mentioned
in Sec.~\ref{s2.5.2}) has been mapped onto the 1D $\sigma$-model
\cite{fm91,mf93a,fm94a}. The RBM model is relevant
to various problems in the field of quantum chaos. In particular, the
evolution operator of a kicked rotor (paradigmatic model of a
periodically driven quantum system) has a structure of a quasi-random
banded matrix, which makes this system  belong to the ``quasi-1D
universality class'' (see Sec.~\ref{s5.4} for a more detailed
discussion). Finally, the 
Iida-Weidenm\"uller-Zuk random matrix model \cite{iwz} of the
transport in a disordered wire can be also reduced to the 1D
$\sigma$-model, which allows to study analytically the wire
conductance and its fluctuations \cite{iwz,mmz}. 

The result for the function $Y_{\rm a}(y)$ determining the
distribution of the eigenfunction intensity reads (for the unitary
symmetry) 
\begin{equation}
Y_{\rm a}(y)=
W^{(1)}(y\xi/L,\tau_+)W^{(1)}(y\xi/L,\tau_-)\ .
\label{anom11}
\end{equation}
Here  $\xi=2\pi\nu DA$ is the localization
length (where $A$ is the wire cross-section), and
$\tau_-=x/\xi$, $\tau_+=(L-x)/\xi$,  with $0<x<L$ being
the coordinate of the observation point ${\bf r}_0$ along the
sample. For the orthogonal symmetry $\xi$ is replaced by $\xi/2$. 
The function $W^{(1)}(z,\tau)$ satisfies the equation
\begin{equation}
{\partial W^{(1)}(z,\tau)\over \partial\tau}=
\left( z^2 {\partial^2\over \partial z^2}-z\right) 
W^{(1)}(z,\tau)
\label{anom12}
\end{equation}
and the boundary condition
\begin{equation}
W^{(1)}(z,0)=1\ .
\label{anom13}
\end{equation}
The solution to Eqs.(\ref{anom12}), (\ref{anom13}) can be found in
terms of the expansion in eigenfunctions of the operator
$z^2 {\partial^2\over \partial z^2}-z$. The functions
$2z^{1/2}K_{i\mu}(2z^{1/2})$, with $K_\nu(x)$ being the modified Bessel
function (Macdonald function),  form the
proper basis for such an expansion \cite{marichev}, which is known as
the Lebedev--Kontorovich expansion; the corresponding
eigenvalues are $-(1+\mu^2)/4$.  The result is
\begin{equation}
W^{(1)}(z,\tau)=2z^{1/2}\left\{K_{1}(2z^{1/2})+\frac{2}{\pi}\int_
{0}^{\infty}{\rm d}\mu\frac{\mu}{1+\mu^{2}}
\sinh{\frac{\pi\mu}{2}}K_{i\mu}(2z^{1/2})
e^{-\frac{1+\mu^{2}}{4}\tau}\right\}\ .
\label{anom14}
\end{equation}
The formulas (\ref{anom_4}), (\ref{anom_4o}), (\ref{anom11}),
(\ref{anom14}) give therefore the exact solution for the 
eigenfunction statistics for arbitrary value of the parameter
$X=L/\xi$. The form of the distribution function ${\cal
P}(y)$ is essentially different in the metallic regime $X\ll 1$ (in
this case $X=1/g$) and in the insulating one, $X\gg 1$. We  discuss
these two limiting cases below.

\subsubsection{Short wire}
\label{s4.3.2}

In the case of a short wire, $X=1/g\ll 1$, Eq.~(\ref{anom_4}),
(\ref{anom_4o}), (\ref{anom11}), (\ref{anom14}) yield
\cite{mf93a,fm94a,m97} 
\begin{eqnarray}
& {\cal P}^{\rm (U)}(y)= e^{-y}\left[1+{\alpha X\over 6}(2-4y+y^2)+
\ldots\right]\
; &\ \    y\lesssim X^{-1/2} 
\label{anom_adm15a}\\
& {\cal P}^{\rm (O)}(y)=\frac{e^{-y/2}}{\sqrt{2\pi y}}
\left[1+{\alpha X\over 6}\left({3\over 2}-3y+{y^2\over 2}\right)
+\ldots\right]\ ;&\ \    y\lesssim X^{-1/2} 
\label{anom_adm8a} \\
& {\cal P}^{\rm (U)}(y)= \exp\left\{ 
-y+{\alpha \over 6}y^2X+\ldots\right\}\
;& \ \    X^{-1/2}\lesssim y\lesssim X^{-1} 
\label{anom_adm15b}\\
& {\cal P}^{\rm (O)}(y)={1\over\sqrt{2\pi y}} \exp\left\{ {1\over 2}
\left [-y+{\alpha \over 6}y^2X+\ldots\right]\right\}
;&\ \    X^{-1/2}\lesssim y\lesssim X^{-1} 
\label{anom_adm15bo}\\
& {\cal P}(y)\sim\exp\left[-2\beta\sqrt{y/X}\right]\ ;&\ \ 
 y\gtrsim X^{-1}\ . 
\label{anom_adm15c}
\end{eqnarray}
Here the coefficient $\alpha$ is equal to
$\alpha=2[1-3x(L-x)/L^2]$. We see that there exist three different
regimes of the behavior of the distribution function.
For not too
large amplitudes $y$, Eqs.(\ref{anom_adm15a}), (\ref{anom_adm8a})  
are just the RMT results with relatively small corrections.
In the intermediate range (\ref{anom_adm15b}), (\ref{anom_adm15bo})
the correction {\it in the
exponent} is small compared to the leading term but much larger than
unity, so that ${\cal P}(y)\gg {\cal P}_{RMT}(y)$ though
$\ln{\cal P}(y)\simeq \ln{\cal P}_{RMT}(y)$. 
Finally, in the large-amplitude region, 
(\ref{anom_adm15c}), the distribution function ${\cal P}(y)$ differs
completely from the RMT prediction.\footnote{Note that 
Eq.~(\ref{anom_adm15c}) is not valid when
the observation point is located close to the sample boundary, in
which case the exponent of (\ref{anom_adm15c}) becomes smaller by a
factor of 2, see \cite{m97}.}

\subsubsection{Long wire}
\label{s4.3.3}

In the limit of a long sample, $X=L/\xi\gg 1$, Eqs.~(\ref{anom_4}),
(\ref{anom_4o}), (\ref{anom11}), (\ref{anom14}) reduce to
\begin{eqnarray}
&&{\cal P}^{\rm (U)}(u)\simeq {8\xi^2 A\over L}
\left[K_1^2(2\sqrt{uA\xi})+ K_0^2(2\sqrt{uA\xi})\right] \ ,
\label{anom3u}\\
&& {\cal P}^{\rm (O)}(u)\simeq {2\xi^2 A\over L}
{K_1(2\sqrt{uA\xi})\over \sqrt{uA\xi} } \ ,
\label{anom3o}
\end{eqnarray}
with $\xi=2\pi\nu AD$ as before, and $u=|\psi^2({\bf r}_0)|$.
Note that in this case it is not appropriate to use  $y=uV$ as a
variable, since typical intensity of a localized wave function is
$u\sim 1/A\xi$ in contrast to $u\sim 1/V$ for a delocalized one. The
asymptotic behavior of Eqs.~(\ref{anom3u}), (\ref{anom3o}) at $u\gg
1/A\xi$ has precisely the same form, 
\be
{\cal P}(u)\sim\exp(-2\beta\sqrt{uA\xi})\ ,
\label{anom3as}
\ee
 as in the region of very large
amplitude in the metallic sample, Eq.~(\ref{anom_adm15c}). On this
basis, it was conjectured in \cite{fm94a} that the asymptotic
behavior 
(\ref{anom_adm15c}) is controlled by the probability to have a
quasi-localized eigenstate with an effective spatial extent much less
than $\xi$ (``anomalously localized state''). This conjecture was
proven rigorously in \cite{m97} where the shape of the anomalously
localized state (ALS) responsible for the large-$u$ asymptotics was
calculated via the transfer-matrix method. 

The transfer-matrix method allows to study, in the quasi-1D geometry,
not only statistics of the eigenfunction amplitude in a given
point, but also correlation functions of amplitudes in different points.
Relegating a more extensive discussion of this issue to
Sec.~\ref{s4.5}, we 
mention here a distribution function which characterizes fluctuations
in the rate of exponential decay of eigenfunctions (Lyapunov
exponent). Specifically, let us consider the product of the
eigenfunction intensity in the two points close 
to the opposite edges of the sample $x_1\to 0$, $x_2\to L$,
\be
\label{wv.10}
v=(2\pi\nu DA^2)^2 |\psi_\alpha^2({\bf r}_1)\psi_\alpha^2({\bf r}_2)|\ .
\ee
The corresponding distribution function is found to be
\cite{fm93b,fm94a} 
\begin{eqnarray}
\label{ijmpb_123}
&&{\cal P}(-\ln{v})= 
F^{(\beta)}[-(\beta\ln v)/2X]
\left({\beta\over 8\pi X}\right)^{1/2}
\exp\left\{-\frac{\beta}{8X}
\left({2X\over\beta}+\ln{v}\right)^{2}\right\} \nonumber\\
&& F^{\rm U}(u)=u{\Gamma^2[(3-u)/2]\over\Gamma(u)}\ ,\qquad
F^{\rm O}(u)={u\Gamma^2[(1-u)/2]\over\pi\Gamma(u)}
\end{eqnarray}
Therefore, $\ln v$ is asymptotically distributed according to the
Gaussian law with mean value $\langle-\ln
v\rangle=(2/\beta)X=L/\beta\pi\nu AD$ and variance
$\mbox{var}(-\ln v)=2\langle-\ln v\rangle$. The same log-normal
distribution is found for the conductance and for transmission
coefficients of a quasi-1D sample from the
Dorokhov-Mello-Pereyra-Kumar formalism \cite{smmp,been1}.

Note that the formula (\ref{ijmpb_123}) is valid in the region of
$v\ll 1$ (i.e. negative $\ln v$) only, which contains almost all
normalization of the distribution function. In the region of still
higher values of $v$ the log-normal form of ${\cal P}(v)$ changes into
the much faster
 stretched-exponential fall-off $\propto\exp\{-2\sqrt{2}\beta
v^{1/4}\}$, as can be easily found from the exact solution given in
\cite{fm93b,fm94a}. The decay rate of all the moments
$\langle v^k\rangle$, $k\ge 1/2$, is four times less than $\langle-\ln
v\rangle$ and does not depend on $k$: $\langle v^k\rangle\propto
e^{-X/2\beta}$. This is because the moments $\langle v^k\rangle$,
$k\ge 1/2$ are determined by the probability to find an ``anomalously
delocalized state'' with $v\sim 1$.

\subsection{Metallic regime (arbitrary $d$).}
\label{s4.4}

For arbitrary  dimensionality $d$, deviations from the RMT
distribution ${\cal P}(y)$ for not too large $y$ can be calculated
\cite{fm95a} via the method of Ref.~\cite{km} described in
Section~\ref{s3.3.2}. Applying this method to the moments
(\ref{wv.5}), (\ref{wv.6}),  one gets
\be
I_q({\bf r}_0)={q!\over
V^{q}}\left[1+{\kappa\over 2} q(q-1)+\ldots\right]\qquad {\rm (U)}\ ,
\label{er3.10}
\ee
where 
\be
\label{wv.9}
\kappa=\Pi({\bf r}_0, {\bf r}_0)=\sum_{\mu\ne 0}{\phi_\mu^2({\bf
r}_0)\over\pi\nu\epsilon_\mu} \ .
\ee
Correspondingly, the correction
to the distribution function reads
\be
{\cal P}^{\rm (U)}(y)=e^{-y}\left[1+{\kappa\over 2}(2-4y+y^2)+\ldots\right]
\label{er3.12}\ .
\ee
Similar results are obtained for the orthogonal symmetry class,
\be
I_q({\bf r}_0)= {(2q-1)!!\over
V^{q}}[1+\kappa q(q-1)+\ldots]\qquad {\rm (O)}\ , 
\label{er3.11}
\ee
\be
{\cal P}^{\rm (O)}(y)={e^{-y/2}\over\sqrt{2\pi y}}
\left[1+{\kappa\over 2}\left({3\over 2}-3y+{y^2\over 2}\right)
+\ldots\right] \ . 
\label{er3.13}
\ee
%Deviations of the eigenfunction distribution function ${\cal P}(y)$
%from its RMT form are illustrated for the orthogonal symmetry case in
%Fig.~\ref{stat_wave_func}. 
Numerical studies of the statistics of
eigenfunction amplitudes in the weak-localization regime have been
performed in Ref.~\cite{mueller97} for the 2D and in
Ref.~\cite{uski98} for the 3D case. The found deviations from RMT are
well described by the above theoretical results.
Experimentally, statistical properties of the eigenfunction
intensity have been studied for microwaves in a disordered cavity
\cite{kudrolli}. For a weak disorder the found deviations are in
good agreement with (\ref{er3.13}) as well. 

%\begin{figure}
%\centerline{\epsfxsize=120mm\epsfbox{stat_wave_func.eps}}
%\vspace{3mm}
%\caption{Distribution ${\cal P}(y)$ of the normalized eigenfunction
%intensities $y=V|\psi^2({\bf r})|$ in the orthogonal symmetry
%case. The dashed line shows the RMT result, Eq.~(\ref{er3.2}), while the
%full line corresponds to Eq.~(\ref{er3.13}) with $\kappa=0.4$.} 
%\label{stat_wave_func} 
%\end{figure}

As we see from the above formulas, the magnitude of the corrections is
governed by the parameter $\kappa=\Pi({\bf r}_0,{\bf r}_0)$ (the
one-diffuson loop in the diagrammatic language). 
In the quasi-one-dimensional case (with hard wall boundary conditions
in the longitudinal direction), it is equal to
\be
\kappa\equiv\Pi({\bf r}_0,{\bf r}_0)={2\over g}\left[{1\over 3}-{x\over
L}\left(1-{x\over L}\right)\right]\ ,\qquad 0\le x\le L\ ,
\label{er3.14}
\ee
where $x$ is the longitudinal coordinate of the observation point
${\bf r}_0$,
so that Eqs.~(\ref{e3.12}), (\ref{e3.13}) agree with the results
(\ref{anom_adm15a}), (\ref{anom_adm8a}) obtained from the exact
solution. For the periodic boundary conditions in the longitudinal
direction (a ring) we have $\kappa=1/6g$. 
In the case of 2D geometry,
\be
\Pi({\bf r},{\bf r}) = {1\over \pi g}\ln {L\over l}\ ,
\label{er3.15}
\ee
with $g=2\pi\nu D$. Finally, in the 3D case the sum over the momenta 
$\Pi({\bf r},{\bf r}) =(\pi\nu V)^{-1}\sum_{{\bf q}}(D{\bf q}^2)^{-1}$
diverges linearly at large $q$. The diffusion approximation is valid
up to $q\sim l^{-1}$; the corresponding cutoff gives
$\Pi({\bf r},{\bf r}) \sim 1/2\pi\nu Dl=g^{-1}(L/l)$. This divergency
indicates that more accurate evaluation of $\Pi({\bf r},{\bf r})$
requires taking into account also the contribution of the ballistic
region ($q>l^{-1}$) which depends on microscopic details of the
random potential; see \cite{m-rev} for details.

The formulas (\ref{er3.12}), (\ref{er3.13}) are valid in the region of
not too large amplitudes, where the perturbative correction is smaller
than the RMT term, i.e. at $y\ll \kappa^{-1/2}$. In the region of
large amplitudes, $y>\kappa^{-1/2}$ the distribution function was
found by Fal'ko and Efetov \cite{fe2} who applied to
Eqs.~(\ref{anom_4}), 
(\ref{anom_4o}) the saddle-point method suggested by Muzykantskii and
Khmelnitskii \cite{mk1}. We relegate the discussion of the method to
Sec.~\ref{s4.6} and only present the results here:
\bea
&& \hspace{-1.3cm} {\cal P}(y)\simeq\exp\left\{{\beta\over
2}(-y+{\kappa\over 2} y^2+\ldots)\right\}\times\left\{
\begin{array}{ll} 1 & \ (U)\\
                  {1\over\sqrt{2\pi y}} &\  (O)
\end{array}\right.
\ , \  \kappa^{-1/2}\lesssim y\lesssim
\kappa^{-1}, \label{er3.16} \\
&& \hspace{-1.3cm} {\cal P}(y)\sim\exp\left\{-{\beta\over
4\kappa}\ln^d(\kappa 
y)\right\}\ ,\qquad\qquad y\gtrsim\kappa^{-1}. \label{er3.17}
\eea
Again, as in the quasi-one-dimensional case, there is an intermediate
range where a correction in the exponent is large compared to unity,
but small compared to the leading RMT term [Eq.~(\ref{er3.16})] and a
far asymptotic region (\ref{er3.17}), where the decay of ${\cal P}(y)$
is much slower than in RMT. Similarly to the quasi-1D result 
(\ref{anom_adm15c}), the asymptotic behavior (\ref{er3.17}) is
determined by anomalously localized states (see \cite{m97,m-rev} for
a review).

\subsubsection{2D: Weak multifractality of eigenfunctions}
\label{s4.4.1}

Since $d=2$ is the lower critical dimension for the Anderson
localization problem, metallic 2D samples (with $g\gg 1$) share many
common properties with systems at the critical point of the 
metal-insulator transition. Although the localization length $\xi$ in
2D is 
not infinite (as for truly critical systems), it is exponentially
large, and the criticality takes place in the very broad range
of the system size $L\ll\xi$. 

The criticality of
eigenfunctions shows up via their multifractality. The multifractal
structures first introduced by Mandelbrot \cite{mandelbrot74}
are characterized by an infinite set of critical exponents describing
the scaling of the moments of a distribution of some quantity.
Since then, this feature has been observed in various objects, such
as the energy-dissipating set in turbulence
\cite{frisch83,benzi84,frisch95},  
strange attractors in chaotic dynamical systems
\cite{grassberger83,hentschel83,benzi85,halsey86}, and the growth
probability distribution in diffusion-limited aggregation
\cite{halsey86a,j_lee88,blumenfeld89}; see Ref.~\cite{paladin87} for a
review.  

The fact that an eigenfunction at the mobility edge has the
multifractal structure was emphasized in
\cite{castpel} on the basis of renormalization group
calculations done by Wegner several years earlier
\cite{wegner80}.  For this problem, the
probability distribution is just  the eigenfunction intensity
$|\psi^2({\bf r})|$ and the corresponding moments are  
the inverse participation ratios (IPR's),  
\be
\label{multifr1}
P_q=\int {\rm d}^d{\bf r}|\psi^{2q}({\bf r})|\ .
\ee
The  multifractality is characterized by the anomalous
scaling of $P_q$ with the system size $L$,
\be
\label{multifr2}
P_q\propto L^{-D_q(q-1)}\equiv L^{-\tau(q)},
\ee
with $D_q$ different from the spatial dimensionality $d$ and dependent
on $q$. 
Equivalently, the eigenfunctions are characterized by the singularity
spectrum $f(\alpha)$ describing the measure $L^{f(\alpha)}$ of 
the set of those points ${\bf r}$ 
where the eigenfunction takes the value $|\psi^2({\bf
r})|\propto L^{-\alpha}$. The two sets of exponents $\tau(q)$ and
$f(\alpha)$ are  related via the Legendre transformation,
\be
\tau(q)=q\alpha-f(\alpha)\ ;\ \ f'(\alpha)=q\ ;\ \
\tau'(q)=\alpha\ .
\label{multifr3}
\ee
By now, the multifractality of critical wave functions is confirmed by
numerical simulations \cite{huck92,pook,schreiber}; for more
recent reviews see Refs.~\cite{janssen94,huck95,janssen98}.

We turn now to the wave function statistics in 2D.  We  note first
that the 
formulas (\ref{er3.10}), (\ref{er3.11}) for the IPR's with $q\lesssim
\kappa^{-1/2}$ can be rewritten in the 2D case (with (\ref{er3.15})
taken into account) as
\be
\label{multifr4}
{\langle P_q\rangle \over P_q^{\rm RMT}}\simeq\left({L\over
l}\right)^{{1\over\beta\pi g}q(q-1)}\ ,
\ee
where $P_q^{\rm RMT}$ is the RMT value of $P_q$ equal to
$q!L^{-2(q-1)}$ for GUE and $(2q-1)!!L^{-2(q-1)}$ for GOE. We see that
(\ref{multifr4}) has precisely the form (\ref{multifr2}) with 
\be
D_q=2-{q\over\beta\pi g}
\label{multifr5}
\ee
As was found in \cite{fe2}, the eigenfunction amplitude
distribution (\ref{er3.16}), (\ref{er3.17}) leads to the same result
(\ref{multifr5}) for all $q\ll 2\beta\pi g$. Since the deviation of $D_q$
from the normal dimension 2 is proportional to the small parameter
$1/\pi g$, it can be termed ``weak multifractality'' (in analogy with
weak localization). The result (\ref{multifr5}) was in fact obtained
for the first time by Wegner \cite{wegner80} via the renormalization
group calculations. 

The limits of validity of Eq.~(\ref{multifr5}) are not unambiguous and
should be commented here. The singularity spectrum $f(\alpha)$
corresponding to (\ref{multifr5}) has the form
\be
\label{multifr6}
f(\alpha)=2-{\beta\pi g\over 4}\left(2+{1\over\beta\pi
g}-\alpha\right)^2\ ,
\ee
so that $f(\alpha_\pm=0)$ for
\be
\label{multifr7}
\alpha_\pm=2\left[1\pm{1\over(2\beta\pi g)^{1/2}}\right]^2\ .
\ee
If $\alpha$ lies outside the interval $(\alpha_-,\,\alpha_+)$, the
corresponding $f(\alpha)<0$, which means that the most likely the
singularity $\alpha$ will not be found for a given
eigenfunction. However, if one considers the {\it average} $\langle
P_q\rangle$ over a sufficiently large ensemble of
eigenfunctions,
a negative value of $f(\alpha)$ makes sense (see a related discussion
in \cite{mandelbrot90}). This is the definition which
was assumed  in \cite{fe2} where
Eq.~(\ref{multifr5}) was obtained for all positive $q \ll 2\beta\pi g$. 

In contrast, if one studies
a {\it typical} value of $P_q$, the regions $\alpha>\alpha_+$ and
$\alpha<\alpha_-$ will not contribute. In this case, Eq.~(\ref{multifr5})
is valid only within the interval $q_-\le q\le q_+$ with
$q_\pm=\pm(2\beta\pi g)^{1/2}$; outside this region one finds
\cite{chamon96,castillo97} 
\be
\tau(q)\equiv D_q(q-1)=\left\{  \begin{array}{ll}
q\alpha_-\ , &\ \ q>q_+\\
q\alpha_+\ , &\ \ q<q_-\ .
\end{array}
\right.
\label{multifr8}
\ee
Therefore, within this definition the multifractal dimensions $D_q$
saturate at the values $\alpha_+$ and $\alpha_-$ for $q\to+\infty$ and
$q\to-\infty$ respectively. This is in agreement with results of
numerical simulations \cite{huck92,pook,schreiber,janssen94,huck95}.

\subsection{Spatial correlations of eigenfunction amplitudes.}
\label{s4.5}

Correlations of amplitudes of an eigenfunction in different spatial
points are characterized by a set of correlation functions (we
consider, as usual, the unitary symmetry for definiteness)
\be
\label{wv.11}
\Delta\langle\sum_i\psi_i^*({\bf r}_1)\psi_i({\bf r}_1')\ldots
\psi_i^*({\bf r}_q)\psi_i({\bf r}_q')\delta(E-E_i)\rangle\equiv
\langle \psi^*({\bf r}_1)\psi({\bf r}_1')\ldots
\psi^*({\bf r}_q)\psi({\bf r}_q')\rangle\ .
\ee
Using the supersymmetry formalism and performing the same
transformations that have led us to Eq.~(\ref{wv.6}), we get
\bea
\label{wv.12}
&&\!\!\!\langle \psi^*({\bf r}_1)\psi({\bf r}_1')\ldots
\psi^*({\bf r}_q)\psi({\bf r}_q')\rangle  \nonumber \\
&&\!\!\!
=-{1\over 2V(q-1)!}\lim_{\eta\to 0}(2\pi\nu\eta)^{q-1}\int
{\rm D}Q\sum_\sigma\prod_{i=1}^q {1\over\pi\nu}
g_{p_i p_{\sigma(i)},{\rm bb}}({\bf r}_i, {\bf r}'_{\sigma(i)})
e^{-S[Q]}\ ,
\eea
where the summation goes over all transpositions $\sigma$ of the set
$\{1,2,\dots,q\}$, $p_i$ is equal to $1$ for $i=1,\ldots,q-1$ and to 2
for $i=q$, and $g$ is the Green's function in the field $Q({\bf r})$, 
\be
\label{wv.13}
 g=\left(E-{\hat{\bf p}^2\over 2m}+i{Q\over 2\tau}\right)^{-1}\ .
\ee
Taking into account that the field $Q({\bf r})$ varies only weakly on
the scale of the mean free path $l$ yields\footnote{Since $Q$ is a
slowly varying field, the argument of $Q$ in the second term of
(\ref{wv.14}) can be chosen to be either ${\bf r}_1$ or ${\bf r}_2$,
or $({\bf r}_1+{\bf r}_2)/2$.}
\bea
&& g({\bf r}_1,{\bf r}_2)\simeq {\rm Re}\, 
G_{\rm A}({\bf r}_1-{\bf r}_2) - i\,{\rm Im}\,
G_{\rm A}({\bf r}_1-{\bf r}_2) Q({\bf r}_1)\ , 
\label{wv.14}\\
&& G_{\rm A}({\bf r})=\int{{\rm d}^d{\bf p}\over (2\pi)^d}{e^{i{\bf
pr}}\over E-{\bf p}^2/2m-i/2\tau}\ .
\label{wv.15}
\eea
For $|{\bf r}_1-{\bf r}_2|\gg l$ the Green's function 
$g({\bf r}_1,{\bf r}_2)$ vanishes exponentially, in view of $G_{\rm
A}({\bf r})\propto e^{-r/2l}$. Since  in the limit $\eta\to 0$ the
characteristic magnitude  of $Q$ is $\propto\Delta/\eta\gg 1$ [see the
text around Eq.~(\ref{anom_3})], the real part  
${\rm Re}\, G_{\rm A}({\bf r})$ can be neglected in all the Green's
function factors in (\ref{wv.12}), and only the products of the
imaginary parts survive. The imaginary part ${\rm Im} G_{\rm A}({\bf
r})$ is given explicitly by
\be
\label{wv.16}
{{\rm Im} G_{\rm A}({\bf r})\over \pi\nu}\equiv f_{\rm F}(r)\simeq
e^{-r/2l}\times \left\{\begin{array}{ll}
J_0(p_{\rm F}r)\ ,                     & \qquad {\rm 2D} \\
{\sin(p_{\rm F}r)\over p_{\rm F}r}\ ,  & \qquad {\rm 3D}
\end{array} \right.
\ee
Let us note that $f_{\rm F}(r)$
is determined by microscopic (short-scale) dimensionality of the
sample rather than by its global geometry. In particular, a quasi-1D
sample may be microscopically of 2D (strip) or 3D (wire) nature.

\subsubsection{Zero-mode approximation.}
\label{s4.5.1}

In the zero-mode approximation, Eq.~(\ref{wv.12}) reduces to
\be
\label{wv.17}
V^q\langle \psi^*({\bf r}_1)\psi({\bf r}_1')\ldots
\psi^*({\bf r}_q)\psi({\bf r}_q')\rangle
=\sum_\sigma\prod_{i=1}^qf_{\rm F}({\bf r}_i,{\bf r}'_{\sigma(i)})\ .
\ee
Equation (\ref{wv.17}) implies
that the wave function $\psi({\bf r})$ 
has a global Gaussian distribution,
\be
\label{wv.18}
{\cal P}\{\psi({\bf r})\}\propto\exp\left\{-{\beta\over 2}\int
{\rm d}^d{\bf r}{\rm d}^d{\bf r'}\psi^*({\bf r})K({\bf r},{\bf r'})\psi({\bf r'})
\right\}
\ee
determined by the correlation function
\be
\label{wv.19}
V\langle\psi^*({\bf r})\psi({\bf r'})\rangle=
f_{\rm F}(|{\bf r}-{\bf r'}|)\ ,
\ee 
the kernel $K({\bf r},{\bf r'})$ being the operator inverse of 
$V^{-1}f_{\rm F}(|{\bf r}-{\bf r'}|)$. An analogous consideration for
the orthogonal symmetry class (when $\psi({\bf r})$ is real) leads, in
the zero-mode approximation, to the same conclusion. 
The result (\ref{wv.18}), (\ref{wv.19}) was first obtained by Berry
\cite{berry3} from the conjecture that a wave function in a
classically chaotic system is given by a random superposition of plane
waves.\footnote{Historically, the equivalence of the Berry's
conjecture and the zero-mode supersymmetry calculation has been
established in a somewhat convoluted and less general way. Prigodin
{\it et al} \cite{prigodin95}  calculated, in the zero-mode
approximation for the 
$\sigma$-model, the joint distribution function ${\cal P}(u,v)$  
of the wave function amplitudes in two different spatial points,
$u=|\psi^2({\bf r})|$, $v=|\psi^2({\bf r'})|$. Srednicki then
demonstrated \cite{srednicki96} that the same results for ${\cal
P}(u,v)$ are obtained from the Berry's conjecture of Gaussian
statistics of $\psi({\bf r})$. We have demonstrated this equivalence
above in a more transparent and general form.}

As has been explained above, quite generally there are corrections to
the zero-mode approximation induced by the diffusion modes. They
change the eigenfunction correlations qualitatively by inducing
correlations on long scales $r\gg l$ (which are exponentially small in 
the zero-mode approximation). Similarly to our discussion of  
the wave function fluctuations
(Sec.~\ref{s4.3}, \ref{s4.4}), we proceed by first presenting results 
of the exact solution in the quasi-1D case and then consider the
metallic regime for an arbitrary dimensionality.

\subsubsection{Quasi-1D geometry.}
\label{s4.5.2}

In the case of the quasi-1D geometry of the sample the method
described in Sec.~\ref{s4.3} allows to calculate analytically all the
multipoint correlation functions (\ref{wv.11}); the results are
presented in terms of multiple integrals of the type (\ref{anom14}),
see reviews \cite{fm94a,m-rev}. Without going into technical details,
we quote here some important conclusions concerning the global
statistics of eigenfunctions. A wave function $\psi({\bf r})$ can be
represented as a product
\be
\psi({\bf r})=\Phi({\bf r})\Psi(x)\ ,
\label{glob9}
\ee
where $x$ is the coordinate of the point ${\bf r}$ along the
sample. Here $\Phi({\bf r})$ is a quickly fluctuating in space
function, which has the Gaussian statistics (\ref{wv.18}),
(\ref{wv.19}) obtained above from the 0D $\sigma$-model. The (real)
function $\Psi(x)$ determines, in contrast, a smooth envelope of
the wave function. Its fluctuations are long-range correlated and are
described by the probability density
\be
\label{glob10}
{\cal P}\{\theta(x)\}\propto e^{-L/2\beta\xi}
e^{-[\theta(0)+\theta(L)]/ 2}
\exp\left\{-{\beta\over 8}\xi\int_0^L {\rm d} x\,
\left({{\rm d}\theta\over {\rm d} x}\right)^2\right\}
\delta\left(L^{-1}\int {\rm d}x\,e^\theta-1\right)\ ,
\ee
where $\theta(x)=\ln \Psi^2(x)$. 

The physics of these results is as follows. The short-range
fluctuations of the wave function (described by the function
$\Phi({\bf r})$) originate, as explained above, from the 
superposition of plane waves with random amplitudes and phases
leading to the Gaussian fluctuations of eigenfunctions with the
correlation function (\ref{wv.16}).  
The second factor $\Psi(x)$ in the decomposition
(\ref{glob9}) describes the smooth envelope of the eigenfunction
(changing on a scale $\gg l$), whose statistics given by
(\ref{glob10}) is determined by diffusion and localization
effects. This factor is responsible for the long-range
correlations ({\it i.e.} those on scales $\gg l$) of the wave
amplitude. 
In the metallic regime, $\xi/L=g\gg 1$, $\Psi(x)$ fluctuates
relatively weakly around unity and the above long-range correlations
of $\psi({\bf r})$ are parametrically small (see
Sec.~\ref{s4.5.3}). In the opposite regime of a long wire, $L/\xi\gg
1$, the strong localization manifests itself in extremely strong
spatial correlations of eigenfunctions. 

Finally, we compare the eigenfunction statistics in the quasi-1D case
with that in a strictly 1D disordered system. In the latter case, the
eigenfunction can be written as
\be
\label{glob13}
\psi_{\rm 1D}(x)=\sqrt{{2\over L}}\cos(kx+\delta)\Psi(x)\ ,
\ee
where $\Psi(x)$ is again a smooth envelope function. 
The local statistics of 
$\psi_{\rm 1D}(x)$ (i.e. the moments (\ref{wv.3})) was studied in
\cite{alprig89}, while the global statistics (the correlation
functions of the type (\ref{wv.11})) in \cite{kolokol93}. Comparison
of the 
results for the quasi-1D and 1D systems shows  that the
statistics of the smooth envelopes $\Psi$ is exactly the same in the
two cases, for a given value of the ratio of the system length $L$ to
the localization length 
(equal to $\beta\pi\nu A D$ in quasi-1D
and to the mean free path $l$ in 1D). The
equivalence of the statistics of the eigenfunction envelopes implies,
in particular, that the distribution of the inverse participation
ratios $P_q$ [see Eq.~(\ref{multifr1})] 
is identical in the 1D and quasi-1D cases. This is confirmed by
explicit calculations of the distribution function of 
$P_2$, see Refs.~\cite{bergor80,kolokol94}
(1D) and Refs.~\cite{fm93a,fm94a} (quasi-1D).

\subsubsection{Metallic regime (arbitrary $d$).}
\label{s4.5.3}

Correlations of eigenfunction amplitudes in the regime of a good
conductor can be again studied via
the method of Ref.~\cite{km} described in Sec.~\ref{s3.3.2}; see
Refs.~\cite{fm95a,bm97}.  The result has the form of the expansion
in powers of the diffusion propagator $\Pi$.
In particular, for the simplest correlation function
showing long-range correlations we find (up to the linear-in-$\Pi$
terms) 
\begin{equation} 
\label{corrrev_fin1}
 V^2\langle | \psi({\bf  r}_1)|^2 |\psi({\bf  r}_2)|^2
\rangle = 1+  {2\over \beta}
   f_{\rm F}^2(|{\bf  r}_1-{\bf  r}_2|)\left 
 [1 + {2\over \beta}\Pi({\bf  r}_1,{\bf  r}_1)\right] 
+{2\over \beta} \Pi({\bf  r}_1,{\bf  r}_2)\ .
\end{equation}
The last term on the r.h.s. of (\ref{corrrev_fin1}) describes
the long-range correlations between $|\psi({\bf  r}_1)|^2$ and 
$|\psi({\bf  r}_2)|^2$ induced by diffusion (or, in other words, by
classical dynamics). 

In a similar way one can calculate also higher order
correlation functions of the eigenfunction amplitudes. In particular, the
correlation function $\langle|\psi^4({\bf r}_1)||\psi^4({\bf
r}_2)|\rangle$ determines fluctuations of the inverse participation
ratio $P_2$,  the result
for the relative variance of $\delta(P_2)=\mbox{var}(P_2)/\langle
P_2\rangle^2$ being \cite{fm95a}
\begin{equation}
\delta(P_2)={8\over\beta^2}\int 
{{\rm d}^d{\bf r}{\rm d}^d{\bf r'}\over V^2}
\Pi^2({\bf r}, {\bf r'})={32 a_d\over\beta^2 g^2},
\label{corrrev_varipr}
\end{equation}
with the numerical coefficient $a_d$ defined in Sec.~\ref{s3.3.2}
(see Eqs.~(\ref{e2.15}), (\ref{e2.15a})). 
The fluctuations (\ref{corrrev_varipr}) have the same relative
magnitude ($\sim 1/g$)
as the famous universal conductance fluctuations. Note also
that extrapolating Eq.(\ref{corrrev_varipr}) to the Anderson
transition point, 
where $g\sim 1$, we find $\delta(P_2)\sim 1$, so that the magnitude of
IPR fluctuations is of the order of its mean value [which is, in turn,
much larger than in the metallic regime; see Eq.~(\ref{multifr2})].

Equation (\ref{corrrev_varipr}) can be generalized onto higher IPR's
$P_q$ with $q>2$,
\be
{\mbox{var}(P_q)\over\langle P_q\rangle^2}\simeq {2\over
\beta^2}q^2(q-1)^2 
\int {{\rm d}^d{\bf r}{\rm d}^d{\bf r'}\over V^2}
\Pi^2({\bf r}, {\bf r'})={8 q^2(q-1)^2 a_d\over\beta^2 g^2},
\label{ipr41}
\end{equation}
so that the relative magnitude of the fluctuations of $P_q$ is $\sim
q(q-1)/g$. Furthermore, the higher irreducible moments (cumulants)
$\langle\langle P_q^n \rangle\rangle$, $n=2,3,\ldots$, have the form
\begin{eqnarray}
{\langle\langle P_q^n \rangle\rangle\over \langle P_q\rangle^n}&=&
{(n-1)!\over 2} \left[{2\over\beta}q(q-1)\right]^n\int
{{\rm d}^d{\bf r}_1\ldots {\rm d}^d{\bf r}_n\over V^n}\Pi({\bf
r}_1,{\bf r}_2)\ldots 
\Pi({\bf r}_n,{\bf r}_1) \nonumber \\
&=&{(n-1)!\over 2} \mbox{Tr}  \left[{2\over\beta}q(q-1)\Pi\right]^n
\nonumber\\
&=&{(n-1)!\over 2}\left[{2\over\beta}q(q-1)\right]^n
\sum_{\mu\ne 0}\left({\Delta\over\pi\epsilon_\mu}\right)^n\ ,
\label{ipr42}
\end{eqnarray}
where $\Pi$ is the integral operator with the kernel $\Pi({\bf r},{\bf
r'})/V$. This is valid provided $q^2 n\ll 2\beta\pi
g$. Prigodin and Altshuler  \cite{prigal98} obtained
Eq.~(\ref{ipr42}) starting from the assumption that the eigenfunction
statistics is described by the Liouville theory. 
According to (\ref{ipr42}), the ``central body'' of the distribution
function ${\cal P}(P_q)$ of the IPR $P_q$ (with $q^2/\beta\pi g\ll 1$)
is determined by the spectrum of eigenvalues $\epsilon_\mu$ of the
diffusion operator $-D\nabla^2$. For a more detailed study of this 
distribution function see \cite{prigal98,m-rev}.

The perturbative calculations show that the cumulants of the IPR's are
correctly reproduced 
(in the leading order in $1/g$) if one assumes  \cite{prigal98}
that the statistics of the eigenfunction envelopes 
$|\psi^2({\bf r})|_{\rm smooth}=e^{\theta({\bf r})}$ is governed by
the Liouville theory  (see {\it e.g.} \cite{zz,i_kogan96} and
references therein)  defined by the functional integral 
\be
\label{ipr44}
\int {\rm D}\theta\,\delta\left(\int {{\rm d}^d {\bf r}\over V} 
e^\theta-1\right)\exp\left\{
-{\beta\pi\nu D\over 4}\int {\rm d}^d{\bf r}({\bf
\nabla}\theta)^2\right\}\ldots
\ee
The ``tails'' of the IPR distribution function
governed by rare realizations of disorder
are described by saddle-point solutions
which can also be obtained from the Liouville theory description
(\ref{ipr44}), see \cite{m-rev}.
The multifractal dimensions (\ref{multifr5}) in 2D can 
be reproduced by starting from the Liouville theory as well
\cite{zz,i_kogan96}. It should 
be stressed, however, that this agreement between the supermatrix
$\sigma$-model governing the eigenfunctions statistics and the
Liouville theory is not exact, but only holds in the leading order in
$1/g$. In particular, the Liouville theory does not describe the wave
function localization by weak disorder in 2D. 

Up to now, we have considered correlations between amplitudes of one
and the same eigenfunction at different spatial points. One can also
study correlations of different eigenfunctions, see \cite{bm97}. 
Understanding of both types of correlations is important for
evaluation of fluctuations of matrix elements 
({\it e.g.} those of Coulomb interaction) computed on
eigenfunctions $\psi_k$ of the one-particle Hamiltonian in a random
potential. Such a problem naturally arises when one investigates
the effect of interaction onto statistical properties of excitations in
a mesoscopic sample (see Refs.~\cite{blanter,agkl,bmm1}).

\subsection{Anomalously localized states and long-time relaxation.}
\label{s4.6}

In this subsection we discuss one more method that can be used within
the supermatrix $\sigma$-model formalism to investigate statistical
properties of eigenfunctions. This is the instanton method introduced 
by Muzykantskii and Khmelnitskii \cite{mk1} in order to
calculate the long-time dispersion of the average conductance $G(t)$. 
Soon after the paper \cite{mk1} appeared, 
it was realized that the method allows one to study
the asymptotic behavior of distribution functions of different
quantities,
including relaxation times, eigenfunction intensities,
local density of states, inverse participation ratio, and level
curvatures. These asymptotics are determined by rare realization of
disorder producing the states which show much stronger localization
features than typical states in the system -- 
anomalously localized states already mentioned in Sec.~\ref{s4.3.3}
and \ref{s4.4}.  
For a review of the obtained results and relevant
references the reader is referred to \cite{m97,m-rev}.  
In fact, we have already quoted the results obtained in this way
by Fal'ko and Efetov \cite{fe2} 
for the ``tail'' of the eigenfunction statistics, see
Eqs.~(\ref{er3.16}), (\ref{er3.17}).
Here we wish to present the ideas of the method by discussing the
original problem considered by Muzykantskii and Khmelnitskii.

Let us consider (following Refs.~\cite{mk1,m95})
the asymptotic (long-time) behavior of
the relaxation processes in an open disordered conductor. One possible
formulation of the problem is to consider the time-dependence of the
average conductance $G(t)$ defined by the non-local (in time)
current-voltage relation
\be
I(t)=\int_{-\infty}^t\,{\rm d}t'G(t-t')V(t')
\label{e4.1}
\ee
Alternatively, one can study the decay law, i.e. the survival
probability $P_{\rm s}(t)$ 
for a particle injected into the sample at $t=0$  to be found 
there after a time $t$. Classically, $P_{\rm s}(t)$  decays
according to the exponential law, $P_{\rm s}(t)\sim e^{-t/t_{\rm D}}$,
where 
$t_{\rm D}^{-1}$ is the lowest eigenvalue of the diffuson operator
$-D\nabla^2$ with the proper boundary conditions. The time $t_{\rm D}$
has 
the meaning of the time of diffusion through the sample, and
$t_{\rm D}^{-1}$ is of the order of the Thouless energy. The same
exponential decay holds for the conductance $G(t)$, where it is
induced by the weak-localization correction. The quantities
of interest can be expressed in the form of the $\sigma$-model
correlation function
\be
G(t),\ P_{\rm s}(t)\sim \int {{\rm d}\omega\over 2\pi}e^{-i\omega t}
\int {\rm D}Q({\bf r})A\{Q\} e^{-S[Q]}\ ,
\label{e4.2}
\ee
where $S[Q]$ is given by Eq.~(\ref{e3.16}).
The preexponential factor $A\{Q\}$ depends on the specific
formulation of the problem but is not important for the leading
exponential behavior studied here.

The main idea of the method is that the asymptotic behavior is
determined by a non-homogeneous ({\it i.e.} ${\bf r}$-dependent)
stationary point of the action $S[Q]$ (instanton). Such a
stationary point is found by varying the exponent in
Eq.~(\ref{e4.2}) with respect to $Q$ and $\omega$, which yields
the equations \cite{mk1}
\bea
&& 2D\nabla(Q\nabla Q)+i\omega[\Lambda,Q]=0 
\label{e4.3}\\
&& {\pi\nu\over 2} \int {\rm d}^d{\bf r}\,\mbox{Str}(\Lambda Q)=t 
\label{e4.4}
\eea
[We assume unitary symmetry; in the orthogonal symmetry
case the calculation is applicable with minor modifications.]
It remains 
\begin{enumerate}
\item[i)] to find a solution $Q_\omega$ of Eq.(\ref{e4.3}) (which will
depend on $\omega$);
\item[ii)] to substitute it into the self-consistency equation
(\ref{e4.4}) and thus to fix $\omega$ as a function of $t$;
\item[iii)] to substitute the found solution $Q_t$ into
Eq.(\ref{e4.2}), 
\end{enumerate}
\be
P_{\rm s}(t)\sim\exp\left\{ {\pi\nu D\over 4}\mbox{Str}
\int (\nabla Q_t)^2\right\}\ .
\label{e4.5}
\ee
Note that Eq.~(\ref{e4.3}) is to be supplemented by the boundary
conditions $Q=\Lambda$ at the open part of the boundary ({\it i.e.}
boundary with an ideal metal)  and (\ref{e3.28})
at the insulating part of the boundary (if it exists).

It is not difficult to show \cite{mk1} that the solution of
Eq.~(\ref{e4.3}) has in the standard parametrization (\ref{e1.62}),
(\ref{e1.63}) the only non-trivial variable -- bosonic eigenvalue
$\lambda_1=\cosh\theta_1$, all 
other coordinates being equal to zero. As a result, Eq.~(\ref{e4.3}) 
reduces to an equation for $\theta_1({\bf r})$ (we drop the subscript
``1'' below):
\be
\nabla^2\theta+{i\omega\over D}\sinh\theta=0\ ,
\label{e4.8}
\ee
the self-consistency condition (\ref{e4.4}) takes the form
\be
\pi\nu\int {\rm d}^d{\bf r}(\cosh\theta-1)=t\ ,
\label{e4.9}
\ee
and Eq.~(\ref{e4.5}) can be rewritten as
\be
\ln P_{\rm s}(t)=-{\pi \nu D\over 2}\int {\rm d}^d{\bf r}(\nabla\theta)^2\ .
\label{e4.10}
\ee
For sufficiently small times, $\theta$ is small according to
(\ref{e4.9}), so that Eqs.~(\ref{e4.8}), (\ref{e4.9}) can be linearized:
\bea
&&\nabla^2\theta+2\gamma\theta=0\ ;\qquad 2\gamma=i\omega/D\ ;
\label{e4.11} \\
&&{\pi\nu\over 2}\int {\rm d}^d{\bf r} \theta^2=t\ .
\label{e4.12}
\eea
This yields
\be
\theta({\bf r})=\left({2t\over\pi\nu}\right)^{1/2}\phi_1({\bf r})\ ,
\label{e4.13}
\ee
where $\phi_1$ is the  eigenfunction of the diffusion operator
corresponding to the lowest eigenvalue 
$\epsilon_1=t_{\rm D}^{-1}$. The survival probability (\ref{e4.10})
reduces thus to 
\be
\ln P_{\rm s}(t) = {\pi\nu D\over 2}
\int {\rm d}^d{\bf r}\,\theta\nabla^2\theta=-{\pi\nu
\epsilon_1\over 2}\int {\rm d}^d{\bf r}\,\theta^2=-t/t_{\rm D}\ ,
\label{e4.14}
\ee
as expected. Eq.~(\ref{e4.14}) is valid (up to relatively small
corrections) as long as $\theta\ll 1$, i.e. for $t\Delta\ll 1$.
To find the behavior at $t\gtrsim\Delta^{-1}$, as well as the
corrections at $t<\Delta^{-1}$, one 
should consider the exact (non-linear) equation (\ref{e4.8}), 
the solution of which depends on the sample geometry.

\subsubsection{Quasi-1D geometry}
\label{s4.6.1}

We consider a wire of length $L$ and a cross-section $A$
  with open boundary conditions at
both edges, $\theta(-L/2)=\theta(L/2)=0$. Equations (\ref{e4.8}),
(\ref{e4.9}) take the form
\be
\label{e4.15} 
 \theta''+2\gamma\sinh\theta=0\ ,\qquad 
\int_{-L/2}^{L/2}{\rm d}x(\cosh\theta-1)=t/\pi\nu A\ .
%\label{e4.16}
\ee
The solution of (\ref{e4.15}) yields the log-normal asymptotic
behavior of $P_{\rm s}(t)$ at large times \cite{mk1}:
\be
\ln P_{\rm s}(t)\simeq -{\beta g\over 2}\ln^2
(t\Delta) \ ; \qquad t\Delta\gg 1\ ,
\label{e4.22}
\ee
with $g=2\pi\nu AD/L\gg 1$ being the dimensionless conductance. 

Equation (\ref{e4.22}) has essentially the same form as the asymptotic
formula for $G(t)$ found by Altshuler and Prigodin \cite{alprig88} for a
{\it strictly} 1D sample with a length much {\it exceeding} 
the localization length:
\begin{equation}
G(t)\sim\exp\left\{-{l\over L}\ln^2(t/\tau)\right\}
\label{relax_13}
\end{equation}
If we replace in Eq.~(\ref{relax_13}) the 1D localization length
$l$ by the 
quasi-1D localization length $\beta\pi\nu AD$, we  reproduce 
the asymptotics (\ref{e4.22}). This is one more manifestation of the
equivalence of statistical properties of smooth envelopes of the
wave functions in 1D and quasi-1D samples (see
Sec.~\ref{s4.5.2}). Furthermore, the agreement of the results for the
metallic and the insulating samples demonstrates clearly that the
asymptotic ``tail'' (\ref{e4.22}) in a metallic 
sample is indeed due to anomalously localized eigenstates.

As another manifestation of this fact, Eq.~(\ref{e4.22}) can be
represented 
as a superposition of the simple relaxation processes with mesoscopically
distributed relaxation times \cite{akl}:
\begin{equation}
P_{\rm s}(t)\sim\int {\rm d}t_\phi\, e^{-t/t_\phi} {\cal P}(t_\phi)
\label{anom106a}
\end{equation}
The distribution function ${\cal P}(t_\phi)$ then behaves as follows:
\begin{equation}
{\cal P}(t_\phi)\sim\exp\left\{-{\beta g\over 2}
\ln^2(g\Delta t_\phi)\right\}\ ;\qquad t_\phi\gg
{1\over g\Delta}\equiv t_{\rm D}\ .
\label{anom107}
\end{equation}
Since the Thouless energy $t_{\rm D}^{-1}$ determines
the typical width of a level of an open system, 
the formula (\ref{anom107}) concerns  indeed 
the states with anomalously small energy widths $t_\phi^{-1}$.

The saddle-point method allows us also to find the corrections to
Eq.~(\ref{e4.14}) in the intermediate region $t_{\rm D}\ll t\ll
\Delta^{-1}$  \cite{mirmuz,m-rev},  
\be
-\ln P_{\rm s}(t) = {t\over t_{\rm D}}
\left(1-{1\over \beta\pi^2 g}{t\over t_{\rm D}}+\ldots\right)\ ,
\label{e4.28}
\ee
with $t_{\rm D}=L^2/\pi^2D$.  Equation (\ref{e4.28}) is completely
analogous to the formula (\ref{anom_adm15b}),
(\ref{anom_adm15bo}) for the statistics of
eigenfunction amplitudes. It shows that the correction to the leading
term $-t/t_{\rm D}$ in $\ln P_{\rm s}$ becomes large compared to unity
at $t\gtrsim t_{\rm D}\sqrt{g}$, though it remains small compared to
the leading term up to $t\sim gt_{\rm D}\sim\Delta^{-1}$. 
The result (\ref{e4.28}) was also obtained
by Frahm \cite{frahm97} from rather involved calculations based
on the equivalence between the 1D $\sigma$-model and the Fokker-Planck
approach and employing the approximate solution of the
Do\-ro\-khov-Mello-Pe\-rey\-ra-Kumar 
equations in the metallic limit. The fact that the logarithm of the
quantum decay probability, $\ln P_{\rm s}(t)$, starts to deviate
strongly (compared to unity) from the classical law, 
$\ln P_{\rm s}^{\rm cl}(t)=-t/t_{\rm D}$
at $t\sim t_{\rm D}\sqrt{g}$ was observed in numerical simulations by
Casati, Maspero, and Shepelyansky \cite{casati97}. For related results
in the framework of a random matrix model see Sec.~\ref{s4.6.3}.

\subsubsection{2D geometry}
\label{s4.6.2}

Considering a 2D disk-shaped sample of a radius $R$, one gets 
the following 
long-time asymptotics of $P_{\rm s}(t)$ (or $G(t)$) \cite{mk1,m95}:
\begin{eqnarray}
&  P_{\rm s}(t)\sim
(t\Delta)^{-2\pi\beta g}\ , & \quad 1\ll t\Delta\ll (R/l)^2 
\label{relax_9d}\\
& P_{\rm s}(t)\sim \exp\left\{-{\pi\beta g\over 4} {\ln^2(t/g\tau)\over
\ln(R/l)}\right\}\ ,  &\quad 
t\Delta\gg (R/l)^2
\label{relax_9}
\end{eqnarray}
where $g=2\pi\nu D$ is the dimensionless conductance per square in 2D
and $\tau$ is the mean free time. 
Equivalently, these results can be represented in terms of the 
distribution function ${\cal P}(t_\phi)$ of relaxation times,
\begin{equation}
{\cal P}(t_\phi)\sim\left\{
\begin{array}{ll}
(t_\phi/t_{\rm D})^{-2\pi\beta g}\ , &\ \ 
t_{\rm D}\ll t_\phi\ll t_{\rm D} \left({R\over l}\right)^2 \\
\exp\left\{-{\pi\beta g\over 4} {\ln^2(t_\phi/\tau)\over \ln
(R/l)}\right\} \ ,&\ \ t_\phi\gg t_{\rm D} \left({R\over l}\right)^2\ , 
\end{array}
\right.
\label{relax_11t}
\end{equation}
where $t_{\rm D}\simeq R^2/D$ is the time of diffusion through the sample. 

The far log-normal asymptotics (\ref{relax_9})
agrees with the one obtained earlier by
Altshuler, Kravtsov, and Lerner from the renormalization-group (RG)
treatment. Analogous agreement between the instanton and the RG
calculations was found for the asymptotics of the local DOS
distribution function \cite{m96}. Note that the instanton method is
superior to the RG treatment in several respects: (i) it is not
restricted to 2D or $2+\epsilon$ dimensions; (ii) it is much more
transparent physically, since the stationary-point solution
$\theta({\bf r})$ describes directly the spatial shape of the
anomalously localized state, $|\psi^2({\bf r})|_{\rm smooth}\propto
e^{\theta({\bf r})}$; (iii) in some cases it allows to find
intermediate asymptotics [see {\it e.g.} Eq.~(\ref{relax_9d})] missed
by the RG calculation.

\subsubsection{Random matrix model}
\label{s4.6.3}

Here we mention briefly the results on the quantum decay law obtained
by Savin and Sokolov \cite{savinsok} within the RMT model. 
This will allow us to see the similarities and the differences between
the diffusive systems and the random matrix model.
The model describes 
a Hamiltonian of an open chaotic system by a Gaussian random matrix
coupled to $M$ external (decay) channels. The found decay law has the
form
\be
P_{\rm s}(t)\sim (1+\Gamma t/M)^{-M}\ ,
\label{e4.500}
\ee 
where $\Gamma=MT\Delta/2\pi$ is a typical width of the eigenstate,
with $T$ characterizing the channel coupling ($T=1$ for ideal
coupling). In this case, the product $MT$
plays the role of the dimensionless conductance $g$ (in contrast to the
diffusive case where $g$ is governed by the bulk of the system,
here it is determined by the number of decay channels and the strength
of their coupling). For not too large
$t$ ($t\Delta T\ll 1$), Eq.~(\ref{e4.500}) yields the classical decay
law, $P_{\rm s}(t)\sim e^{-t\Gamma}$, with corrections of the form
\be
\ln P_{\rm s}(t)=-t\Gamma(1-\Gamma t/2M+\ldots)\ ,
\label{e4.501}
\ee
which is similar to the results found for the diffusive systems (see
Eq.~(\ref{e4.28})). 
At large $t\gg (\Delta T)^{-1}$, the decay takes the power-law
asymptotic form \cite{lewenkopf91}
\be
\ln P_{\rm s}(t)\simeq -M\ln(\Gamma t/M)\ , 
\label{e4.502}
\ee
which is to be compared with Eqs.~(\ref{e4.22}) and (\ref{relax_9d}),
(\ref{relax_9}).

\section{Supersymmetry approach to the Quantum Chaos.}
\label{s5}
\setcounter{equation}{0}

The aim of this section is to review recently emerged ideas concerning
application of the supersymmetry formalism to chaotic ballistic systems
(``billiards''). It should be stressed that this field is quite young,
and most of the calculations performed so far are less rigorous than
in the diffusive case; the limits of applicability of the obtained 
results are not well understood yet. In this sense the status of this
section is somewhat different from the preceding part of this lecture
course.

\subsection{Introduction: What have we learned from the diffusive
problem.} 
\label{s5.1}

It is instructive to begin by summarizing what we have learned
concerning the diffusive problem. As has been explained in
Sec.~\ref{s3}, \ref{s4}, the statistical properties of energy levels and
eigenfunctions in a diffusive disordered sample are described by the
diffusive supermatrix $\sigma$-model (\ref{e3.16}). In the metallic
regime, when the dimensionless conductance is large, $g\gg 1$, and the 
eigenfunctions are not localized ({\it i.e.} cover roughly uniformly
the 
whole sample volume), the zero-mode approximation is a good starting
point for treating the $\sigma$-model. It reduces the problem 
to the 0D $\sigma$-model, yielding the RMT results for the level and
eigenfunction statistics. Deviations from RMT (which are typically
small in the ``body'' of the distribution functions but become large in
the ``tail'') are controlled by the diffusion modes. In contrast to
the universal RMT results, these deviations are {\it system-specific},
since they depend on the sample dimensionality, shape and size, as
well as on the disorder strength. More specifically, the deviations
are controlled by the diffusion operator $-D\nabla^2$, which
determines the quadratic part of the action of the diffusion modes,
\be
\label{b.1}
S[W]={\pi\nu\over 2}\int {\rm d}^d{\bf r}\,{\rm Str} 
W_{21}[-D\nabla^2-i\omega]W_{12}\ .
\ee
The magnitude of the deviations from RMT is controlled by the small
parameter 
$g^{-1}\sim\Delta/\epsilon_1$, where $\epsilon_1$ is the lowest
non-zero eigenvalue of the diffusion operator (the Thouless energy).

\subsection{Ballistic $\sigma$-model.}
\label{s5.2}

Let us now turn to the case of  ballistic chaotic systems, {\it i.e.}
to the quantum chaos. We will first guess what the field-theoretical
description of such a problem should look like and then will discuss
how it can be derived.

\subsubsection{Heuristic arguments.}
\label{s5.2.1}

A bulk of numerical simulations data have unambiguously demonstrated
that {\it generically} the statistics of energy levels and
eigenfunctions 
amplitudes in a ballistic system (billiard) whose classical
counterpart is chaotic is well described by RMT (this is known as
Bohigas-Giannoni-Schmit conjecture \cite{bohigas84}) . Since we 
already know 
that the 0D $\sigma$-model is essentially equivalent to RMT, we expect
that the sought field theory of quantum chaos should reduce in the
leading (zero-mode) approximation to the 0D $\sigma$-model described
in Sec.~\ref{s2.4}. Therefore, the field variable of this theory
should be the supermatrix field belonging to the same coset space
as the field $Q({\bf r})$ of the $\sigma$-model. 

Furthermore, in
analogy with the diffusive case, we expect deviations from RMT to be
controlled by the modes of density relaxation in the system. 
Indeed, deviations from ergodicity in a
diffusive sample are physically due to the fact that the process of
a particle spreading over the whole sample volume is not instantaneous
but requires a time $\sim t_{\rm D}$, the relaxation being governed by
the diffusion operator. Since the momentum relaxation in a diffusive
sample takes place on a much shorter time scale (of the order of the
mean free time $\tau$), the diffusion operator (describing the
dynamics on time scales $\gg\tau$) is an operator in ${\bf r}$-space
only. This is why the $Q$-field depends on the coordinate but not on
the velocity in the diffusive case. Such a separation of the fast and
slow dynamics is not applicable to a generic clean sample. Therefore,
in this case the supermatrix field $Q$ should depend both on the
coordinate ${\bf r}$ and the velocity direction ${\bf n}={\bf
v}/v_{\rm F}$, $Q=Q({\bf r},{\bf n})$ (the absolute value of the
velocity $|{\bf v}|=v_{\rm F}$ being fixed by the energy
conservation). The classical dynamics in the phase space is governed
by the Liouville operator
\be
\label{b.2}
{\cal L}=\{\cdot,H\}={\partial H\over\partial {\bf p}}
{\partial \over\partial {\bf r}}-
{\partial H\over\partial {\bf r}}{\partial \over\partial {\bf p}}\ ,
\ee
which reduces for the case of a billiard (no potential energy inside)
to 
\be
\label{b.3}
{\cal L}=v_{\rm F}{\bf n}{\partial \over\partial {\bf r}}
\ee
supplemented by the boundary conditions corresponding to the
particle reflection by the sample boundary. 
It is clear from what has been said above that the Liouville operator
${\cal L}$ is expected to replace the diffusion operator in the
$\sigma$-model description. Therefore, the analog of (\ref{b.1})
should read\footnote{The angular integral $\int {\rm d}{\bf n}\ldots$
is assumed to be normalized to unity, $\int {\rm d}{\bf n}=1$.}
\be
\label{b.4}
S[W({\bf r},{\bf n})]=
{\pi\nu\over 2}\int {\rm d}^d{\bf r}{\rm d}{\bf n}\,{\rm Str} 
W_{21}({\bf r},{\bf n})[{\cal L}-i\omega]W_{12}({\bf r},{\bf n})\ .
\ee
It remains to restore the action in terms of the $Q$-field from the
quadratic form (\ref{b.4}). It turns out, however, that in the
ballistic case the corresponding action cannot be written in a simple
form in terms of the $Q$-field,\footnote{The action can be expressed in
terms of the $Q$-field only at the expense of introduction of an
additional coordinate, with the action taking the Wess-Zumino-Witten
form, see \cite{mk3}.} and it is more convenient to write it in terms
of the $T$-matrix field parametrizing the $\sigma$-model manifold
according to [{\it cf.} (\ref{e1.60}), (\ref{e3.15})]
\be
\label{b.5}
Q({\bf r},{\bf n})=T({\bf r},{\bf n})\Lambda T^{-1}({\bf r},{\bf n})\ . 
\ee
Using that $T=1-W/2+\ldots$, that ${\cal L}$ is the first-order
differential operator and that 
the action should be invariant at $\omega\to 0$ 
with respect to the global rotations $Q({\bf r},{\bf n})\to 
UQ({\bf r},{\bf n})U^{-1}$ ({\it i.e.} with respect to 
$T({\bf r},{\bf n})\to UT({\bf r},{\bf n})$), one concludes that the
only allowed form is
\be
\label{b.6}
S[Q]={\pi\nu\over 2}\int {\rm d}^d{\bf r}{\rm d}{\bf n}\,
{\rm Str}[-2T^{-1}({\bf r},{\bf n}){\cal L}T({\bf r},{\bf n})\Lambda
-i\omega Q({\bf r},{\bf n})\Lambda]\ .
\ee

\subsubsection{Ballistic $\sigma$-model from disorder averaging.}
\label{s5.2.2}

The ballistic $\sigma$-model action was derived for the first time by
Muzykantskii and Khmelnitskii (MK) \cite{mk3}. 
Their starting point was a disordered system, and they
followed the route outlined in Sec.~\ref{s3.1} up to
Eq.~(\ref{e3.9}). Their further aim was to derive a theory
which describes the physics at all momenta $q\ll k_F$ and not only at
$q\ll l^{-1}$, {\it i.e.} which is quasiclassical but is not
restricted to
the diffusion approximation. Clearly, in this case one cannot use the 
gradient expansion leading to the diffusive $\sigma$-model
(\ref{e3.16}). Instead of this, MK employed an analogy with the
Eilenberger quasiclassical approach in the kinetic theory of
disordered superconductors. Along these lines, they defined the field
$Q({\bf r},{\bf n})$ as a result of application of the following two
operations to the Green's function $g({\bf r},{\bf r'})$ [defined in
Eq.~(\ref{e3.10})]: \\
(i) the Wigner transformation
\be
\label{b.7}
g({\bf r},{\bf r'})=\int{{\rm d}^d{\bf p}\over (2\pi)^d}
e^{i{\bf p}({\bf r}-{\bf r'})}
\tilde{g}({{\bf r}+{\bf r'}\over 2},{\bf p})\ ;
\ee
(ii) integration over the kinetic energy $\xi=v_{\rm F}(|{\bf
p}|-p_{\rm F})$,
\be
\label{b.8}
Q({\bf r},{\bf n})={1\over\pi}\int d\xi\,
\tilde{g}({\bf r},{\bf n}{\xi\over v_{\rm F}})\ .
\ee
They were then able to derive the ballistic $\sigma$-model\footnote{As
has already benn mentioned, MK obtained the ballistic action
in a different (Wess-Zumino-Witten) form, which is however equivalent
to (\ref{b.6}).} (\ref{b.6}) with an additional term describing the
scattering by impurities, 
\be
\label{b.9}
S_{\rm imp}[Q]={\pi\nu\over 4}\int 
{\rm d}^d{\bf r}{\rm d}{\bf n}d{\bf n'} 
w({\bf r};{\bf n},{\bf n'}){\rm Str}Q({\bf r},{\bf n})Q({\bf r},{\bf
n'})\ ,
\ee
where $w({\bf r};{\bf n},{\bf n'})$ is the differential cross-section
of the scattering ${\bf n}\to{\bf n'}$ at the point ${\bf r}$ (for the
isotropic scattering  $w({\bf r};{\bf n},{\bf n'})$ is equal to
$1/\tau({\bf r})$). 

MK conjectured further that the derived $\sigma$-model makes sense
also in the limit $\tau\to\infty$, when it describes the clean
system. Let us note, however, that one cannot simply set $\tau^{-1}=0$
({\it i.e.} remove the disorder completely). Indeed, then we would get
a particular clean system with uniquely defined energy levels, so that
the DOS will be a sum of $\delta$-functions. This is certainly not
what we 
want (or what we are able) to calculate in the $\sigma$-model
approach. The correlation or distribution functions that we are
discussing imply necessarily some averaging. The MK
derivation seems to remain meaningful if $\Delta\tau\ll 1$. 
In this case the disorder is strong enough from the quantum point of
view, {\it i.e.} we study not a particular quantum system but rather a
large ensemble of systems. On the other hand, the condition that the
disorder does not influence the classical dynamics is $\tau\gg
L/v_{\rm F}$, where $L$ is the system size. Since
$L/v_F\Delta\sim(k_{\rm F}L)^{d-1}\gg 1$ in the semiclassical limit,
the double inequality $L/v_{\rm F}\ll\tau\ll\Delta^{-1}$ can be
satisfied. It means that the disorder is classically negligible, while
it mixes strongly energy levels of the quantum system. 

In a diffusive sample the density relaxation is governed by the
diffusion operator with eigenvalues $\epsilon_\mu>0$
(and $\epsilon_0=0$ corresponding
to the particle number conservation in a closed sample). The
positiveness of $\epsilon_\mu$ corresponds to the exponential decay of
a density perturbation with time. What are
their counterparts  $\gamma_\mu$ in a  chaotic sample? One might
naively think 
that, since the evolution operator $e^{-{\cal L}t}$ is unitary, the
corresponding eigenvalues $e^{-\gamma_\mu t}$ are of absolute value
unity, so that $\gamma_\mu$ are purely imaginary. This is, however,
incorrect. It is known that an ultraviolet regularization (projection
onto the subspace of smooth functions) shifts all $\gamma_\mu$ (except
$\gamma_0=0$) from the imaginary axis, giving them a positive real
part, ${\rm Re}\gamma_\mu>0$. These eigenvalues are known as Ruelle
resonances \cite{ruelle}. The corresponding regularized evolution
operator (called 
Perron-Frobenius operator) describes irreversible classical dynamics
in a chaotic system. The above ultraviolet regularization may be
physically understood as an infinitesimally weak noise introduced in
the 
system to make the dynamics irreversible. The identification of the
eigenvalues $\gamma_\mu$ determining the non-universal corrections to
the spectral statistics in a chaotic system with the Ruelle resonances
was done in \cite{aaa}. 

\subsubsection{ $\sigma$-model from energy averaging.}
\label{s5.2.3}

In a subsequent paper, Andreev, Agam, Simons, and Altshuler (AASA)
\cite{aasa} proposed another derivation of the ballistic
$\sigma$-model. Instead of averaging over disorder, they started
from a completely clean system and performed the {\it energy}
averaging in 
a certain spectral window. After the Hubbard-Stratonovich
transformation, they arrived at the action of the form
\be
\label{b.10}
S[\hat{Q}]\propto {\rm Str}_{\bf r}
(-2\hat{T}^{-1}i[\hat{H},\hat{T}]\Lambda-i\omega\hat{Q}\Lambda)\ ,
\ee
where $\hat{H}$ is the quantum Hamiltonian, $\hat{T}$ and
$\hat{Q}=\hat{T}\Lambda\hat{T}^{-1}$ are operators in the Hilbert
space (and have on top of this the usual supermatrix structure), and 
${\rm Str}_{\bf r}$ includes the supertrace over the supermatrix
indices and the trace over the Hilbert space. The next (and crucial)
step is the semiclassical expansion. Going to the Wigner
representation and using the fact that in the semiclassical limit the
Wigner transform of a commutator is a Poisson bracket of the
corresponding Wigner transforms, AASA reduced (\ref{b.10}) to the form
(\ref{b.6}). More recently Zirnbauer \cite{zirnbauer99} demonstrated,
however, that an additional averaging is necessary to 
guarantee the condition of a slow variation of the Wigner transform
$T({\bf r},{\bf p})$ required by the semiclassical expansion.

More specifically, Zirnbauer considered the level correlations for a
unitary map (rather than for a Hermitian Hamiltonian). Similarly to
the AASA approach, he averaged over the quasienergy $e^{i\phi}$.
This can be done via the ``color-flavor transformation''
\cite{zirn96},
\be
\label{b.11}
\int {\rm d}\phi\exp i(\Psi_1^\dagger  e^{i\phi}\Psi_1+
\Psi_2^\dagger  e^{-i\phi}\Psi_2)=\int {\rm d}\mu(Z,\bar{Z})\exp(
\Psi_1^\dagger  Z\Psi_2 + \Psi_2^\dagger  \bar{Z}\Psi_1)\ ,
\ee
where $\Psi_1$, $\Psi_2$ are supervectors, 
$T=\left\lgroup\begin{array}{cc} 1 & Z\\ \bar{Z} & 1\end{array}
\right\rgroup$ is the matrix from the coset space and 
${\rm d}\mu(Z,\bar{Z})$ is the corresponding invariant measure.\footnote{$Z$
and $\bar{Z}$ are identical to $-W_{12}/2$ and $-W_{21}/2$
respectively if the parametrization (\ref{e3.22}) is used.}  This
transformation replaces (in the case of unitary maps)
the energy averaging and the Hubbard-Stratonovich
transformation. Moreover, the saddle-point approximation is not needed
in this case, since the r.h.s. of (\ref{b.11}) is already an integral
over the coset space. Zirnbauer showed further that, in order to
justify the 
semiclassical expansion, one has to average over an ensemble of maps,
\be
\label{b.12}
U(\xi)=\exp\left(i\sum_{k=1}^s\xi_k X_k/\hbar\right)U\ ,
\ee
where $\xi_k$ are random variables with the disorder strength scaling
as $\langle\xi_k^2\rangle\propto\hbar^\alpha$, $0<\alpha<1$, in the
limit $\hbar\to 0$. Since $\alpha>0$, all the members of the
introduced ensemble have the same classical limits. On the other hand,
the condition $\alpha<1$ ensures that the disorder is strong from the
quantum point of view. This procedure is thus physically similar to
the derivation of MK with $L/v_{\rm F}\ll\tau\ll\Delta^{-1}$ (see above).

\subsubsection{Non-universal corrections and statistical noise.}
\label{s5.2.4}

Let us discuss now the implications of the ballistic $\sigma$-model
(\ref{b.6}) for the level statistics. Since the system is assumed to
be chaotic, the only zero-mode [{\it i.e.} a field $Q({\bf r})$
yielding zero when substituted in the first (kinetic) term of
(\ref{b.6})] is $Q={\rm const}$, so that in the zero-mode
approximation the 0D $\sigma$-model and thus the RMT are reproduced, as
expected.\footnote{For an integrable system any function $Q$ depending
on integrals of motion only would be a zero mode, thus invalidating
this consideration.} Corrections to the RMT have the same form as
discussed in Sec.~\ref{s3.3}, \ref{s3.4}, with the diffusion
eigenvalues $\epsilon_\mu$ replaced by the Perrron-Frobenius
eigenvalues $\gamma_\mu$. In a strongly chaotic system of a
characteristic size $L$, a typical relaxation time (the counterpart of
the diffusion time $t_{\rm D}$) is of the order of the flight time,
$t_{\rm B}\sim L/v_{\rm F}$. Therefore, the ballistic counterpart of
the dimensionless conductance $g\sim 1/\Delta t_{\rm D}$ can be
estimated as $g\sim 1/\Delta t_{\rm B}\sim (k_F L)^{d-1}\sim
N^{(d-1)/d}$, where $N\sim E/\Delta$ is the level number around which
the statistics is studied. In particular, for the most often
considered case of a 2D billiard $g\sim N^{1/2}$. We will discuss the
deviations in more detail in Sec.~\ref{s5.3}, where important
differences compared to the diffusive case will be demonstrated. 

Prange \cite{prange97} has discussed conditions of observability of
the non-universal corrections to the spectral form-factor $K(\tau)$
around $\tau=2\pi$ predicted by the $\sigma$-model. He pointed out
that $K(\tau)$ is a strongly fluctuating function with the
r.m.s. deviation equal to the mean value. After averaging over a
window of width $\sim 1/g$ around $\tau=2\pi$ the noise amplitude is
reduced to $(g/N)^{1/2}$. Therefore, to detect the deviation $\delta
K_{2\pi}(\tau)$, whose amplitude is $\sim 1/g$, the following
inequality should be satisfied:
\be
\label{b.13}
\left({g\over N}\right)^{1/2}\ll {1\over g} \qquad
\Longrightarrow  \qquad N\gg g^3\ .   
\ee
However, as we have just discussed, $g\sim N^{1/2}$ for a generic 2D
chaotic billiard, so that the condition (\ref{b.13}) is not
fulfilled. Therefore, the non-universal correction to the spectral
form-factor is not observable for an individual quantum system, when
the only averaging available is the energy averaging. This points
again to the necessity of the  additional averaging
over an ensemble of quantum systems having the same classical limit
\cite{prange97}. Let us note that for a diffusive system,
Eq.~(\ref{b.13}) can be satisfied without problems. In this case $g$
and $N$ are two independent parameters, and one can consider the limit
of arbitrarily large $N$ at fixed $g$. Therefore, it is in principle
possible to extract the non-universal correction from a single
disordered sample by using the energy averaging.

\subsubsection{Problem of repetitions.}
\label{s5.2.5}

The following  subtle point concerning the
non-universal correction to $R_2(\omega)$ predicted by the ballistic
$\sigma$-model is worth mentioning here. The smooth
(Altshuler-Shklovskii) part of the 
correction, Eqs.~(\ref{e3.31}), (\ref{e2.25}), can be written as
follows 
\bea
\label{b.14}
R_{2,{\rm AS}}^{\rm (c)}(\omega) & = &
{1\over 2\pi^2}{\rm Re}\sum_\mu{\Delta^2\over(-i\omega+\gamma_\mu)^2}
\nonumber\\ 
& = & {\Delta^2\over 2\pi^2}{\rm Re} \int_0^\infty dt\, t 
e^{i\omega t}\,{\rm Tr}\, e^{-{\cal L}t}\ ,
\eea
which is easily checked by using the fact that the eigenvalues of
the evolution operator $e^{-{\cal L}t}$ are equal to $e^{-\gamma_\mu
t}$. Expressing the trace in terms of the Gutzwiller sum over periodic
orbits, one gets \cite{cvitanovic91,eckhardt93} 
\be
\label{b.15}
{\rm Tr}\, e^{-{\cal L}t}=\sum_p T_p\sum_{r=1}^\infty
{\delta(t-rT_p)\over|{\rm det}(M_p^r-1)|}\ ,
\ee
where the index $p$ labels primitive orbits with periods $T_p$, the
summation over $r$ takes into account repetitions of the primitive
orbits, and $M_p$ is the monodromy matrix characterizing the dynamics
in the vicinity of the orbit $p$. Substitution of (\ref{b.15}) into
(\ref{b.14}) readily yields
\be
\label{b.16}
R_{2,{\rm AS}}^{\rm (c)}(\omega)={\Delta^2\over 2\pi^2}{\rm Re}
\sum_p T_p^2\sum_{r=1}^\infty 
{r\exp(i\omega T_pr)\over|{\rm det}(M_p^r-1)|} \ .
\ee
On the other hand, one can calculate the two-level correlation
function semiclassically, by starting from the Gutzwiller trace
formula for the DOS \cite{gutzwiller,eckhardt93}
\be
\label{b.17}
{\nu(E)\over \langle\nu\rangle}=1+{\rm Re}{\Delta\over
\pi}\sum_pT_p\sum_{r=1}^\infty {\exp\{ir[S_p(E)-\mu_p\pi/2]\}
\over|{\rm det}(M_p^r-1)|^{1/2}}\ ,
\ee
where $S_p$ is the action and $\mu_p$ the Maslov index of the orbit
$p$. The two-level correlation function is thus represented as a
double 
sum $\sum_{pp'}$ over the periodic orbits. Assuming that the terms
with $p\ne p'$ vanish upon energy averaging due to randomly varying
phase factors (``diagonal approximation'' introduced by Berry), one
finds \cite{berry1}
\be
\label{b.18}
R_{2,{\rm diag}}^{\rm (c)}(\omega)={\Delta^2\over 2\pi^2}{\rm Re}
\sum_p T_p^2\sum_{r=1}^\infty 
{\exp(i\omega T_pr)\over|{\rm det}(M_p^r-1)|} \ .
\ee
Comparing equations (\ref{b.16}) and (\ref{b.18}), we see that they
almost agree, 
the only difference being in the extra factor $r$ in
(\ref{b.18}). This discrepancy between the results of the ballistic
$\sigma$-model and the diagonal approximation was emphasized by
Bogomolny and Keating \cite{bogkeat}. The same semiclassical analysis
was performed
earlier by Argaman, Imry, and Smilansky \cite{argaman93} in the
context of diffusive systems. In that case, however, the relevant
periodic orbits determining the non-universal behavior have a length 
$\sim v_{\rm F}t_{\rm D}\gg v_{\rm F}\tau$, {\it i.e.} they are
much longer than the shortest periodic orbits. For this reason and
in view of the
exponential proliferation of the primitive periodic orbits with
length, one can neglect the repetitions, keeping only the $r=1$
term in the trace formulas. Then Eq.~(\ref{b.16}) and (\ref{b.18}) are
in full agreement with each other. On the other hand, in the ballistic 
case the shortest orbits of the length $\sim L$ are relevant, and
there is no parameter which would justify neglecting the
repetitions. There exists thus a 
real discrepancy between the two formulas, which still awaits its
explanation.

\subsection{Billiard with diffuse surface scattering}
\label{s5.3}

According to  Sec.~\ref{s5.2}, the level and eigenfunction
statistics of a clean chaotic system (with an ensemble averaging
discussed above) are described by the formulas of Sec.~\ref{s3},
\ref{s4} with the Perron-Frobenius operator substituted  for the
diffusion operator. However, straightforward application of these
results to a given chaotic billiard is complicated by the fact that the
eigenvalues of the Perron-Frobenius operator are unknown, while its
eigenfunctions are extremely singular.  For this reason the
$\sigma$-model approach has so far failed to provide explicit results
for any particular ballistic system. In this subsection, we consider a
ballistic system with  surface disorder (a rough boundary) leading to
diffusive scattering of  a particle in each  collision with the
boundary \cite{bmm2,DEK}. Since the particle loses memory of its
direction of motion  after a single collision, this model describes a
limit of an ``extremely  chaotic'' ballistic system, with typical
relaxation time being of order of the flight time. (This should be
contrasted with the case of a relatively slight distortion of an
integrable billiard \cite{borgonovi,frahm97a}.) This is a natural
problem to be studied by the ballistic $\sigma$-model approach. While
the assumption of the diffuse surface scattering makes possible an
explicit analytic treatment of the problem
\cite{bmm2,DEK}\footnote{Very recently, the same  approach was used
\cite{samokhin99} to calculate the persistent current in a ring with
diffusive scattering.}, the obtained results seem to reflect generic
features of ballistic systems in the regime of hard chaos.   To
simplify the calculations,  a circular geometry of the billiard is
assumed.  A similar problem was studied numerically in
Ref. \cite{louis96} for a square  geometry. As usual, we
consider the case of unitary symmetry; generalization to the
orthogonal case is  straightforward.

Inside the billiard, the motion is free and the Liouville operator
${\cal L}$ is given by Eq.~(\ref{b.3}). Clearly, it should be
supplemented  by a boundary condition,
which depends on the form of the surface roughness.  As a model
approximation we consider purely diffuse scattering
\cite{fuchs,abrikos} for 
which the distribution function $\varphi({\bf r},{\bf n})$ of the
outgoing particles is constant and is fixed by flux
conservation\footnote{The exact form of the boundary condition depends
on the underlying microscopic model. In particular, the diffuse
scattering can be modelled by surrounding the cavity by a 
disordered layer with a bulk mean free path $l$ and a thickness
$d\gg l$. The corresponding boundary condition \cite{chandra,morfesh} 
differs from Eq.~(\ref{e8.1}) by a parameterless function of order
unity. For a review of the boundary conditions
corresponding to various microscopic realizations of the rough surface
see \cite{okulov79}.}:
\begin{equation}
\label{e8.1}
\varphi({\bf r}, {\bf n}) = \pi \int_{({\bf N}{\bf n'}) > 0}
\left( {\bf N} {\bf n'} \right) \varphi ({\bf r}, {\bf n'})
{\rm d}{\bf n'}, \ \ \ \left ({\bf N} {\bf n} \right) < 0.
\end{equation}
Here the point ${\bf r}$ lies at the surface, and ${\bf N}$ is an
outward normal to the surface. This boundary condition should be
satisfied by the eigenfunctions of ${\cal L}$. The boundary condition
breaks the naive anti-hermiticity of ${\cal L}$, and all its
eigenvalues (except the zero one) acquire a positive real
part, as expected for the Perron-Frobenius operator of a chaotic
system. 

Specifically, 
the eigenvalues $\gamma$ of the operator ${\cal L}$ corresponding to 
the angular momentum $l$ obey the equation
\begin{equation} 
\label{surf_values}
\tilde J_l(\xi) \equiv -1 + \frac{1}{2} \int_0^{\pi} 
{\rm d}\theta \sin\theta
\exp \left[ 2il\theta + 2 \xi \sin\theta \right]  = 0,
\end{equation}
where $\xi \equiv R\gamma/v_F$, and $R$ is the radius of the circle. For
each value of $l=0,\pm1,\pm2,\ldots$ Eq.(\ref{surf_values}) has a set of
solutions $\xi_{lk}$ with $\xi_{lk}=\xi_{-l,k}=\xi^*_{l,-k}$, which
can be labeled with $k=0,\pm1,\pm2,\ldots$ (even $l$) or
$k=\pm1/2,\pm3/2,\ldots$ (odd $l$). For $l=k=0$ we have $\xi_{00}=0$,
corresponding to the zero mode $\varphi ({\bf r},
{\bf n})=\mbox{const}$. All other eigenvalues have a positive real part
$\mbox{Re}\,\xi_{lk} > 0$ and govern the relaxation of the
corresponding classical system to the homogeneous distribution in the
phase space.  
The asymptotic form of the solutions of Eq.~(\ref{surf_values}) for large
$\vert k \vert$ and/or $\vert l \vert$ can be obtained by using the
saddle-point method,
\begin{eqnarray} 
\label{surf_valas}
\xi_{kl} \approx \left\{ \begin{array}{lr} 0.66 l + 0.14 \ln l + 0.55 \pi i
k, & 0 \le k \ll l \\ 
(\ln k)/4 + \pi i (k+1/8), & 0 \le  l \ll k
\end{array} \right. .
\end{eqnarray}
Note that for $k=0$ all eigenvalues are real, while for high values of
$k$ they lie close to the imaginary axis and do not depend on $l$ (see
Fig.~\ref{perron}).

\begin{figure}
\centerline{\epsfxsize=160mm\epsfbox{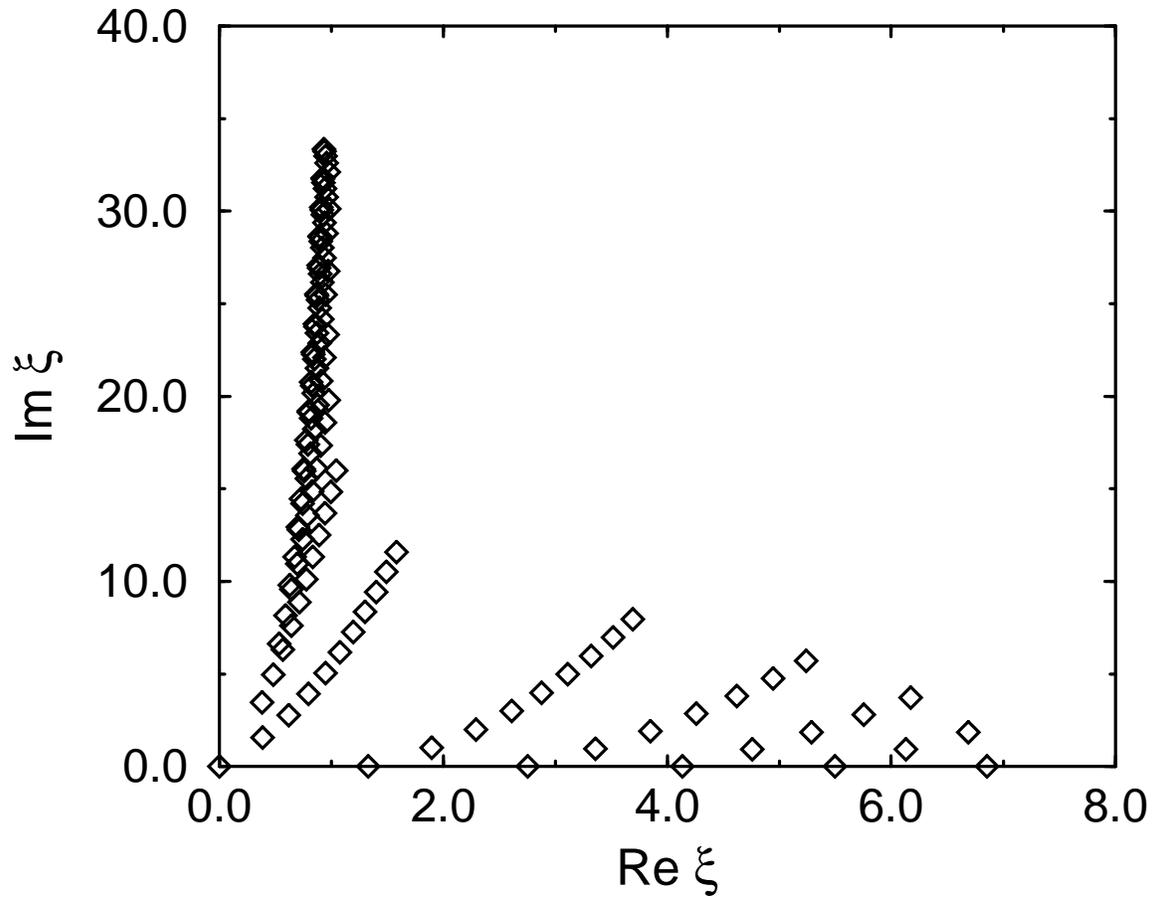}}
\vspace{1cm}
\caption{First $11\times 11$  ($0 \le k,l < 11$)
eigenvalues of the Liouville operator $\cal L$ in
units of $v_F/R$, as given by Eq. (\protect\ref{surf_values}). From
Ref.~\cite{bmm2}.}  
\label{perron}
\end{figure}

\subsubsection{Level statistics.} 
\label{s5.3.1}

The characteristic frequency separating the regions of 
the close-to-RMT and the fully non-universal behavior (analog of the
Thouless energy) for the considered problem is $\omega\sim t_{\rm
B}^{-1}\sim v_{\rm F}/R$. In the low-frequency range, $\omega\ll
v_{\rm F}/R$, the level correlation function is given by
Eq.~(\ref{e2.16}), where we expect $A\sim 1/g^2\sim (R\Delta/v_{\rm
F})^2$. Indeed, the calculation yields \cite{bmm2,DEK}
\begin{equation} 
\label{surf_s0}
A=a\left ({R\Delta\over \pi v_F}\right)^2 \ ,\qquad
  a = -19/27 - 175 \pi^2/1152 + 64/(9\pi^2) \approx -1.48.
\end{equation}
In contrast to the diffusive case, this constant is negative: the level
repulsion is enhanced with respect to the RMT.

The high-frequency behavior of $R_2(\omega)$ is expressed in terms of
the spectral determinant (\ref{e2.24}) by the formulas of
Sec.~\ref{s3.3.3}. Calculating the spectral determinant for the
present problem, we find \cite{bmm2,DEK}
\bea 
\label{spdet1}
D(s)& =& s^{-2}\prod_{kl\ne(00)}  (1- is\Delta/\gamma_{kl} )^{-1} 
(1 + is\Delta/\gamma_{kl} )^{-1} \nonumber\\
&=&\left( \frac{\pi}{2} \right)^6 \frac{1}{N} \prod_l
\frac{1}{\tilde J_l (i s N^{-1/2}) \tilde J_l (-i s N^{-1/2})},
\eea
where $N =(v_{\rm F}/R\Delta)^2 = (p_{\rm F} R/2)^2$ is the number of
electrons 
below the Fermi level. For high frequencies $\omega \gg v_F/R$ this
yields the following expression for the smooth (Altshuler-Shklovskii)
and the oscillating part of the level correlation function: 
\begin{equation} 
\label{surf_highen0}
R_{\rm 2,AS} (\omega) = 
\left( \frac{\Delta R}{v_{\rm F}} \right)^2 \left(
\frac{v_{\rm F}}{2\pi\omega R} \right)^{1/2} \cos \left( 4
\frac{\omega R}{v_{\rm F}} - \frac{\pi}{4} \right),
\end{equation} 
\begin{equation} 
\label{surf_highen2}
R_{\rm 2,osc} (\omega) = \frac{\pi^4}{128} 
\left( \frac{\Delta R}{v_{\rm F}} \right)^2
\cos \left( \frac{2\pi \omega}{\Delta} \right).
\end{equation} 
It is remarkable that the amplitude of the oscillating part 
(\ref{surf_highen2}) does not
depend on frequency. This is in contrast to the diffusive case, where
in the high-frequency
regime ($\omega$ above the Thouless energy) the oscillating
part $R_{\rm osc} (\omega)$ is exponentially small, see Eq.~(\ref{e2.26}).

\subsubsection{The level number variance.} 
\label{s5.3.3}

The smooth part of the level
correlation function can be best illustrated by plotting the level
number variance $\Sigma_2(s)$ [see Sec.~\ref{s3.4.2}]. 
A calculation \cite{bmm2} gives for $ s \ll N^{1/2}$
\begin{equation} 
  \pi^2 \Sigma_2 (s) 
  =  1 + {\bf C} + \ln (2\pi s) +
    a s^2/(2N) 
\label{surf_lnv1}
\end{equation}
and for $s \gg N^{1/2}$
\be
  \pi^2 \Sigma_2 (s) = 1 + {\bf C} + \ln \frac{16 N^{1/2}}{\pi^2}
    - \frac{\pi^2}{16}  \left( \frac{2N^{1/2}}{\pi s}
  \right)^{1/2} \cos \left( \frac{4 s}{N^{1/2}} - \frac{\pi}{4}
  \right). 
\label{surf_lnv2}
\ee
Here ${\bf C} \approx 0.577$ is Euler's constant, and the numerical
constant $a$ is defined by 
Eq.~(\ref{surf_s0}). The first three terms at the r.h.s. of
Eq.~(\ref{surf_lnv1}) 
represent the RMT contribution (curve 1 in Fig.~\ref{surf}).
As  seen from Fig.~\ref{surf}, the two 
asymptotics (\ref{surf_lnv1}) and (\ref{surf_lnv2})
perfectly match in the intermediate regime, $s\sim N^{1/2}$. Taken
together, they provide a complete description of $\Sigma_2(s)$.
According to Eq.(\ref{surf_lnv2}), the level number variance saturates
at the value $\Sigma_2^{(0)} = \pi^{-2} (1 + {\bf C} + \ln
(16N^{1/2}/\pi^2))$, in contrast to the behavior found for diffusive
systems [see Eq.~(\ref{e3.112})] or ballistic systems with weak bulk
disorder \cite{algef93,agfish96}. The saturation occurs at energies $s
\sim N^{1/2}$, or in 
dimensionful units $E \sim v_{\rm F}/R$. This saturation of $\Sigma_2
(s)$, as well as its oscillations on the scale set by short periodic
orbits, is expected for a generic chaotic billiard
\cite{berry1}. It is also in 
good agreement with the results for $\Sigma_2 (s)$ found numerically
for a tight-binding model with moderately strong disorder on boundary
sites \cite{louis96}. 

\begin{figure}
{\epsfxsize=140mm\centerline{\epsfbox{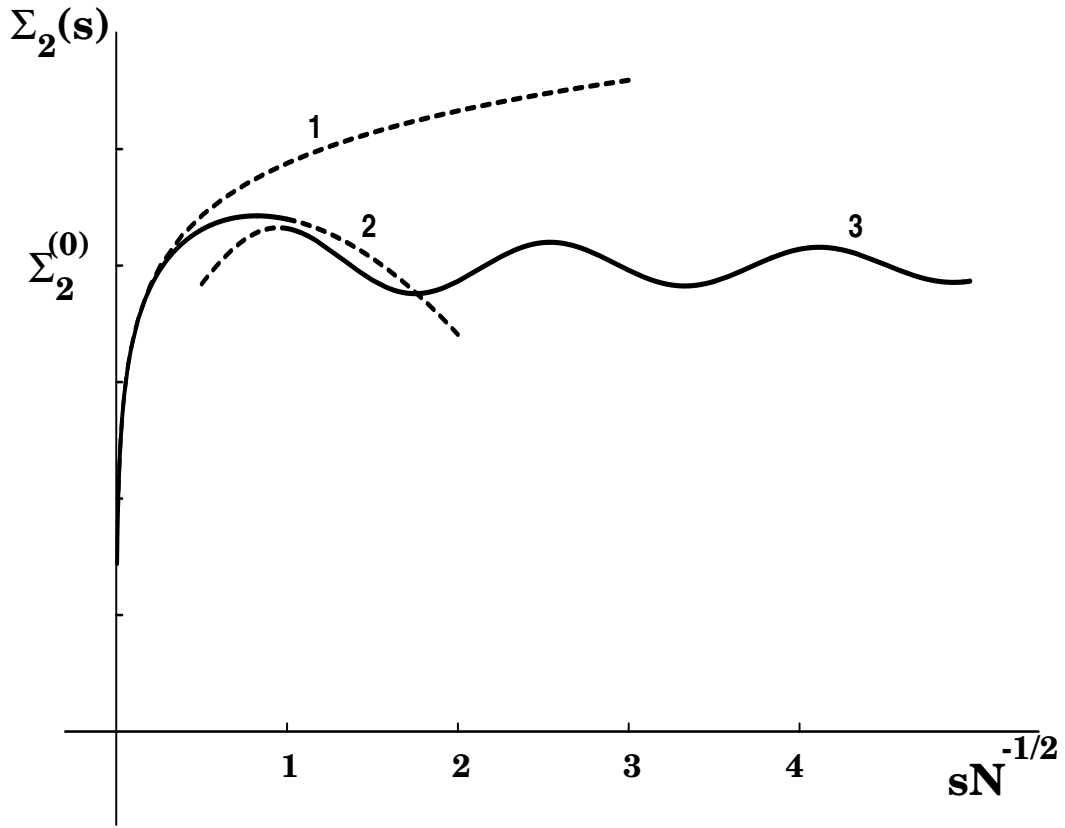}}}
\vspace{1cm}
\caption{Level number variance $\Sigma_2 (E)$ as a function of
energy; $s = E/\Delta$. Curve 1 shows the RMT result, while curves 2
and 3 correspond to asymptotic regimes of low
(\protect\ref{surf_lnv1}) and 
high (\protect\ref{surf_lnv2}) frequencies. The saturation value
$\Sigma_2^{(0)}$ is given in the text. From Ref.~\cite{bmm2}.} 
\label{surf}
\end{figure}

\subsubsection {Eigenfunction statistics.} 
\label{s5.3.4}

Let us recall that the system-specific contributions to fluctuations 
and correlations of the eigenfunction amplitudes were described in the
diffusive case in terms of the diffusion propagator $\Pi({\bf r},{\bf
r'})$, see Sec.~\ref{s4.4}, \ref{s4.5.3}. Its ballistic counterpart
$\Pi_{\rm B}$ is defined as follows
\begin{eqnarray}
\label{corrrev_ball}
& & \Pi_{\rm B} ({\bf r}_1, {\bf r}_2)
= \int {\rm d}{\bf n}_1 {\rm d}{\bf n}_2\, g({\bf r}_1, {\bf n}_1;
{\bf r}_2, 
{\bf n}_2), \nonumber\\
& & {\cal L} g({\bf r}_1, {\bf n_1}; {\bf r}_2, {\bf n}_2) =
 \left( \pi \nu \right)^{-1} \left[ \delta({\bf r}_1  - {\bf r}_2)
\delta({\bf n}_1  - {\bf n}_2) - V^{-1} \right].
\end{eqnarray} 
Equivalently,  the function $\Pi_{\rm B}({\bf r}_1 ,{\bf r}_2)$ can be
defined as
\begin{equation}
\label{corrrev_ball2}
\Pi_{\rm B}({\bf r}_1 ,{\bf r}_2)=\int_0^\infty {\rm d}t\int {\rm
d}{\bf n}_1 \,\tilde{g}({\bf r}_1 ,{\bf n}_1 ,t;{\bf r}_2)\ ,
\end{equation}
where $\tilde{g}$ is determined by the evolution equation
\begin{equation}
\label{corrrev_ball3}
\left({\partial\over\partial t}+v_F{\bf n_1 \nabla}_1 \right)
\tilde{g}({\bf r}_1 ,{\bf n}_1 ,t;{\bf r}_2)=0\ ,\qquad t>0
\end{equation}
with the boundary condition
\begin{equation}
\label{ball4}
\tilde{g}|_{t=0}=(\pi\nu)^{-1}[\delta({\bf r}_1 -{\bf r}_2)-V^{-1}].
\end{equation}
Equation (\ref{corrrev_ball}) is a
natural ``ballistic'' counterpart of Eq.(\ref{corrrev_diff}). 
Note, however, that $\Pi_{\rm B}({\bf r}_1 ,{\bf r}_2)$ contains a
contribution $\Pi_{\rm B}^{(0)}({\bf r}_1 ,{\bf r}_2)$
of the straight line motion from ${\bf r}_2$ to
${\bf r}_1 $ (equal to $1/(\pi p_{\rm F} |\bf r_1-\bf r_2|)$ in 2D and
to $1/2(p_{\rm F} |\bf r_1-\bf r_2|)^2$ in 3D), which is 
nothing but the smoothed version of the function 
$f_{\rm F}^2(|{\bf r}_1 -{\bf r}_2|)$. For this reason, 
$\Pi({\bf r}_1 ,{\bf r}_2)$ in
the formulas of Sec.~\ref{s4}  should 
be replaced in the ballistic case by 
$\Pi({\bf r}_1 ,{\bf r}_2)=
\Pi_{\rm B}({\bf r}_1 ,{\bf r}_2)-\Pi_{\rm B}^{(0)}
({\bf r}_1 ,{\bf r}_2)$. At large distances $|{\bf r}_1 -{\bf r}_2|\gg
\lambda_F$ the (smoothed) correlation function takes in the leading
approximation the form
\begin{equation}
\label{corrrev_ball1}
V^2\langle | \psi({\bf  r}_1)|^2 |\psi({\bf  r}_2)|^2
 = 1 + {2\over \beta} \Pi_{\rm B}({\bf r}_1 ,{\bf r}_2).
\end{equation}
For related results obtained in the semiclassical approach see 
Refs.~\cite{eckhardt95,srednicki98}.

Equation (\ref{corrrev_ball1}) shows that correlations in
eigenfunction amplitudes 
in remote points are determined by the classical dynamics in the
system. It is closely related to the phenomenon of scarring of
eigenfunctions by the classical orbits
\cite{heller91,agfish93}. 
Indeed, if ${\bf r}_1 $ and ${\bf r}_2$ belong to a short periodic
orbit, the function $\Pi_{\rm B}({\bf r}_1 ,{\bf r}_2)$ is positive,
so that  
the amplitudes $|\psi_k({\bf r}_1 )|^2$ and $|\psi_k({\bf r}_2)|^2$
are positively correlated. This is a reflection of the ``scars''
associated with this periodic orbit and a quantitative
characterization of their strength in the coordinate space. 
Note that this effect gets smaller
with increasing energy $E$ of eigenfunctions. Indeed, for a strongly
chaotic system and for $|{\bf r}_1 -{\bf r}_2|\sim L$ ($L$ being the
system size), we have in the 2D case
$\Pi_{\rm B}({\bf r}_1 ,{\bf r}_2)\sim\lambda_F/L$,
so that the magnitude of the correlations decreases as $E^{-1/2}$. 

For the model of a circular billiard with diffuse surface scattering
a direct calculation gives \cite{bmm2}
\begin{eqnarray} 
\Pi_{\rm B} ({\bf r}_1 , {\bf r}_2) & = & \Pi_1 ({\bf r}_1 , {\bf r}_2) +
\Pi_2 ({\bf r}_1 , {\bf r}_2), 
\end{eqnarray}
\begin{eqnarray} 
\label{surf_green}
\Pi_1 ({\bf r}_1 , {\bf r}_2) &=& \Pi_{\rm B}^{(0)}({\bf r}_1  
- {\bf r}_2)- 
V^{-1}\int {\rm d}^2{\bf r'}_1 \Pi_{\rm B}^{(0)}({\bf r'}_1  
- {\bf r}_2) \\ 
&&\hspace{-2cm}
-V^{-1}\int {\rm d}^2{\bf r'}_2\Pi_{\rm B}^{(0)}({\bf r}_1  
- {\bf r'}_2)
+V^{-2}\int {\rm d}^2{\bf r'}_1 {\rm d}^2{\bf r'}_2
\Pi_{\rm B}^{(0)}({\bf r'}_1  - {\bf r'}_2); \nonumber\\
\Pi_2 ({\bf r}_1 , {\bf r}_2) & = & \frac{1}{4\pi p_FR}
\sum_{k=1}^{\infty} \frac{4k^2 - 1}{4k^2} \left( \frac{r_1r_2}{R^2}
\right)^k \cos k \left( \theta_1 - \theta_2 \right) \nonumber
\end{eqnarray}
where $\Pi_{\rm B}^{(0)}({\bf r}) = 1/(\pi p_F|{\bf r}|)$, and
$(r,\theta)$ are the polar coordinates. This formula has
a clear interpretation. The function $\Pi_{\rm B}$ can be represented as a sum
over all paths leading from ${\bf r}_1 $ to ${\bf r}_2$, with
possible surface scattering in between. In particular, 
$\Pi_1$ corresponds to direct trajectories from ${\bf r}_1$ to
${\bf r}_2$ with no reflection from the surface, while  the
contribution $\Pi_2$ is due to the surface scattering. The first term
in the numerator $4k^2-1$ comes from trajectories with only one
surface reflection, while the second sums up contributions from
multiple reflections.

\subsection{$\sigma$-model for the kicked rotor.}
\label{s5.4}

Up to now we have considered autonomous systems, with the Hamiltonian
$\hat{H}$ having no time dependence. The supersymmetric $\sigma$-model
approach is, however, also applicable to periodically driven systems
($\hat{H}(t)$ periodic in time $t$), as we are going to discuss. The
standard system of the latter type is the quantum kicked rotor (QKR) 
\cite{casati79,chirikov91,izrailev90} defined by the Hamiltonian
\be
\label{kr.1}
\hat{H}={\hat{l}^2\over
2I}+\tilde{k}\cos\theta\sum_{m=-\infty}^\infty\delta(t-mT)\ .
\ee
Here $\hat{l}=-i\hbar\partial/\partial\theta$ is the angular momentum
operator conjugate to the angle variable $\theta$; $I$ is the moment
of inertia, $T$ the kick period, and $\tilde{k}$ the kick
strength. The 
classical version of this problem is characterized by a single
dimensionless parameter $K=\tilde{k}\tau/I$. For a sufficiently large
$K$ the motion becomes globally chaotic and the system exhibits
unbounded diffusion in the phase space (in the direction of the
angular momentum), the diffusion coefficient being
\be
\label{kr.2}
D\equiv\left.{\langle [l(t)-l(0)]^2\rangle\over
2t}\right|_{t\to\infty} 
\simeq {\tilde{k}^2\over 4T}\qquad {\rm for}\  K\gg 1.
\ee
In contrast to the classical problem, the quantum system depends
non-trivially on both parameters $\tilde{k}$ and $T$, since there are
two dimensionless combinations which can be formed:
$k=\tilde{k}/\hbar$ and $\tau=\hbar T/I$. The classical limit
corresponds thus to $k\to\infty$, $\tau\to 0$ at fixed $k\tau=K={\rm
const}$. After this short reminder of the classical-quantum
correspondence for the kicked rotor problem, we set $\hbar=1$ as in
the other parts of this article.

Numerical simulations of the QKR have shown that at long times the
classical diffusion is suppressed, the phenomenon known as ``dynamical
localization''. In Ref.~\cite{fishman82} an analogy between this
problem and the Anderson localization in 1D systems was drawn. 
Later studies revealed a close connection between the QKR problem and
the RBM ensemble (see Sec.~\ref{s2.5.2}). The evolution (Floquet)
operator of the QKR has in the angular momentum representation the
form
\be
\label{kr.3}
U_{ll'}=(-i)^{l-l'}\exp\left\{-i{\tau\over 2}l^2\right\}J_{l-l'}(k)\ ,
\ee
where $J_m(k)$ is the Bessel function. Since 
$J_m(k)\simeq(2\pi m)^{-1/2}(ek/2m)^m$ for $m\gg 1,k$, we see that the
matrix elements of $U$ are exponentially small for $|l-l'|\gg
k$. Therefore the matrix $U$ is indeed banded, with the bandwidth
$b\sim k\gg 1$. Furthermore, the second factor on the r.h.s. of
(\ref{kr.3}) mimics a generator of pseudorandom numbers (of absolute
value unity). Indeed, numerical simulations 
\cite{izrailev90,casati90,casati90a} have demonstrated that the
statistical properties of the two models (QKR and RBM) are very close
to each other. 

On the other hand, the RBM ensemble was shown to belong to the class
of quasi-1D disordered systems described by the 1D $\sigma$-model
\cite{fm91}, see also Sec.~\ref{s4.3.1}. The solution of the 1D
$\sigma$-model made possible a detailed analytical study  of
eigenfunctions of such systems \cite{fm94a}, see
Sec.~\ref{s4.3}. While the pseudo-RBM properties of the QKR evolution
operator $U_{nn'}$, as well as the numerical simulations,
suggested strongly that these results are also applicable to QKR, a
formal derivation of this fact has been missing until recently. This gap
was filled by Altland and Zirnbauer (AZ) \cite{alzir96} who achieved a
mapping of the QKR onto the 1D $\sigma$-model. The idea of their
calculation is essentially the same as the one described in
Sec.~\ref{s5.2.3}. Performing the averaging over the quasienergy
spectrum with making use of the transformation (\ref{b.11}) and
carrying out the semiclassical expansion, AZ obtained the 1D version of
the $\sigma$-model (\ref{e3.16}) (or its orthogonal-class counterpart
for unbroken time-reversal symmetry), with $D$ equal to the classical
diffusion constant of the rotor (\ref{kr.2}) and
$\nu=T/2\pi$. This is precisely what could have been expected from
the analogy with disordered wires or the RBM ensemble; the above value
of $\nu$ being the density of quasienergies $\omega_k$ corresponding to
the eigenvalues $e^{i\omega_k T}$ of the Floquet operator. 
This mapping allowed AZ  to conclude
that the statistical properties of the QKR are identical to those of
the quasi-1D disordered systems. In particular, the localization
length $L_{\rm loc}$ of the QKR governing the decay of a typical
eigenfunction, $|\psi^2(l)|_{\rm typ}\propto \exp(-|l-l_0|/L_{\rm
loc})$, is found to be $L_{\rm loc}=\beta\pi\nu D=(\beta/8)k^2$, in
agreement with numerical simulations of Shepelyansky 
\cite{shep86}.\footnote{Shepelyansky considered also another
localization length -- that of a steady state -- and concluded that it
is larger by a factor of 2. This is in disagreement with the 
field-theoretical results, which give the same value 
$L_{\rm loc}=(\beta/8)k^2$ for the both lengths. Presumably, the
discrepancy is due to insufficient numerical accuracy of evaluation of
the steady state asymptotic decay rate for $k^2\gg 1$ in
\cite{shep86}.} 

It should be noted, however, that the problem of sufficiency of the
energy averaging discussed in Sec.~\ref{s5.2.3} is equally applicable
here. In \cite{alzir-reply} AZ acknowledged that an additional
averaging over an ensemble of rotors having the same classical limit
({\it i.e.} of the type described in Sec.~\ref{s5.2.3})
was needed to justify the semiclassical expansion which is in the
heart of the derivation in \cite{alzir96}. 
In fact, the conclusion of the existence of such an implicit ensemble
averaging in 
\cite{alzir96} can also be supported by the following
argument. Following \cite{alzir96}, one can calculate, {\it e.g.}, the
distribution function of the quantity of the type (\ref{wv.10}), 
$v\propto|\psi_\alpha^2(l_1)\psi_\alpha^2(l_2)|$, which will have the
log-normal distribution (\ref{ijmpb_123}). In particular, the far tail
of this distribution is crucially important for identifying the 
localization length of a typical eigenstate on the basis of the
average value $\langle
v\rangle$ (which has a 4 times smaller decay rate). 
However, to be able to find the whole distribution, one
should average over an exponentially large 
[$\gg \exp(|l_1-l_2|/4L_{\rm loc})$] number of eigenfunctions, while
the energy averaging alone reduces effectively to an averaging over 
$\sim|l_1-l_2|$ eigenfunctions only and is thus insufficient. 

\subsection{Concluding remarks.}
\label{s5.5}

We finish these notes by comparing briefly the supersymmetric and the
semiclassic (periodic orbit) approach to the spectral statistics of
chaotic systems. The main advantage of the supersymmetry method is
that it allows one to get the RMT results in the leading
approximation. In contrast, the semiclassical approach is only
justified for times much shorter than the Heisenberg one,
$\tau/2\pi \ll 1$. Obtaining the GUE results within this approach
requires using an {\it ad hoc} regularization prescription
\cite{bogkeat,smila99}. Even the problem of calculating the
perturbative (in $\tau$) corrections to the leading behavior
$K(\tau)=\tau/\pi$ in GOE (for GUE the perturbative corrections are
identically zero) has not been solved semiclassically. Therefore, the
$\sigma$-model approach is the only known method allowing to obtain
the RMT results in a controlled way.

On the other hand, the ballistic $\sigma$-model approach is also not free
from problems. In particular,  construction of a regular expansion in
$1/g$ has not been achieved in this case (in contrast to the diffusive
$\sigma$-model). While the leading non-universal contributions
(discussed above) to the energy level and eigenfunction statistics
are only determined by the density relaxation modes $\phi_\mu$ and
their eigenvalues $\gamma_\mu$, the higher-order (in $1/g$) terms are
induced by interaction of these modes. In the case of the diffusive
$\sigma$-model, the corresponding contributions are known as weak
localization corrections and can be calculated systematically. At the
same time, recent attempts to perform such a calculation in the
ballistic case \cite{aleiner96,smith98} did not produce unambiguous
results, because of the singular nature of the expressions
obtained. This problem (as well as that of repetitions, see
Sec.~\ref{s5.2.5}) remains a challenge for future research.

\vspace{0.5cm}

{\bf Acknowledgment.}
Financial support by SFB 195 der Deutschen Forschungsgemeinschaft is
gratefully acknowledged.

\end{document}